\documentclass[twocolumn]{aastex631}

\usepackage{subfiles}
\usepackage[normalem]{ulem}
\usepackage{multirow}
\usepackage{enumitem}

\begin{document}

\title{The Eighteenth Data Release of the Sloan Digital Sky Surveys: \\ Targeting and First Spectra from SDSS-V}

\correspondingauthor{Gail Zasowski} 
\email{gail.zasowski@gmail.com} 
 
\author{Andr\'{e}s Almeida}
\affiliation{Department of Astronomy, University of Virginia, Charlottesville, VA 22904-4325, USA}

\author{Scott F. Anderson}
\affiliation{Department of Astronomy, University of Washington, Box 351580, Seattle, WA 98195, USA}

\author{Maria Argudo-Fern\'{a}ndez}
\affiliation{Instituto de F\'{i}sica, Pontificia Universidad Cat\'{o}lica de Valpara\'{i}so, Valpara\'{i}so, Chile}
\affiliation{Universidad de Granada (UGR), Departamento de F\'isica Te\'orica y del Cosmos, 18071, Granada, Spain}
\affiliation{Instituto Universitario Carlos I de F\'isica Te\'orica y Computacional, Universidad de Granada, 18071, Granada, Spain}

\author{Carles Badenes}
\affiliation{PITT PACC, Department of Physics and Astronomy, University of Pittsburgh, Pittsburgh, PA 15260, USA}

\author{Kat Barger}
\affiliation{Department of Physics \& Astronomy, Texas Christian University, Fort Worth, TX 76129, USA}

\author{Jorge K. Barrera-Ballesteros}
\affiliation{Instituto de Astronom\'{i}a, Universidad Nacional Aut\'{o}noma de M\'{e}xico, A.P. 70-264, 04510, Mexico, D.F., M\'{e}xico}

\author{Chad F. Bender}
\affiliation{Steward Observatory, University of Arizona, 933 North Cherry Avenue, Tucson, AZ 85721–0065, USA}

\author{Erika Benitez}
\affiliation{Instituto de Astronom\'{i}a, Universidad Nacional Aut\'{o}noma de M\'{e}xico, A.P. 70-264, 04510, Mexico, D.F., M\'{e}xico}

\author{Felipe Besser}
\affiliation{Las Campanas Observatory, Ra\'{u}l Bitr\'{a}n 1200, La Serena, Chile}

\author[0000-0001-5838-5212]{Jonathan C. Bird}
\affiliation{Department of Physics and Astronomy, Vanderbilt University, VU Station 1807, Nashville, TN 37235, USA}

% ...

%DocuLlama: Anne-Marie, Joel, Jordan, John D., Sean, Scott, Ani, Jose SG, Jennifer, Niall, Nathan, Diogo, Joleen, Nick T., Jose FT, Tom, William Zhang, Keith
%DocuLlama consultants: https://wiki.sdss.org/display/DATA/DocuLlama+2022
%Pramod

%\collaboration{20}{(AAS Journals Data Editors)}



\begin{abstract} %250 words

The eighteenth data release of the Sloan Digital Sky Surveys (SDSS) is the first one for SDSS-V, the fifth generation of the survey. SDSS-V comprises three primary scientific programs, or ``Mappers'': Milky Way Mapper (MWM), Black Hole Mapper (BHM), and Local Volume Mapper (LVM). This data release contains extensive targeting information for the two multi-object spectroscopy programs (MWM and BHM), including input catalogs and selection functions for their numerous scientific objectives. We describe the production of the targeting databases and their calibration- and scientifically-focused components. DR18 also includes $\sim$25,000 new SDSS spectra and supplemental information for X-ray sources identified by eROSITA in its eFEDS field. We present updates to some of the SDSS software pipelines and preview changes anticipated for DR19. We also describe three value-added catalogs (VACs) based on SDSS-IV data that have been published since DR17, and one VAC based on the SDSS-V data in the eFEDS field. 

\end{abstract}

%\tableofcontents



%\clearpage
%\tableofcontents
%\clearpage

\section{The first two decades of the Sloan Digital Sky Surveys}
\label{sec:intro}
This paper describes the eighteenth data release of the Sloan Digital Sky Survey (SDSS) and the first data release of SDSS-V.

Since its operations began in 1998, SDSS has been taking near-continuous observations of stars, galaxies, and quasars, and other objects, spanning from our solar system to the early days of the Universe. The first phase, SDSS-I \citep{York_2000_sdss}, imaged over 8000~deg$^2$ of the sky in the {\it ugriz} filters and collected optical spectra of more than 700,000 objects. SDSS-II completed the legacy imaging survey and added a dedicated supernova imaging survey \citep{Frieman_2008_sdss2supernovae} and a spectroscopic survey of $\sim$230,000 Milky Way stars \citep[SEGUE;][]{Yanny_09_SEGUE}. 

SDSS-III \citep{Eisenstein_11_sdss3overview} focused entirely on spectroscopy and comprised an extension of SEGUE \citep[SEGUE-2;][]{Rockosi_2022_segue2}; a radial-velocity exoplanet survey \citep[MARVELS;][]{Ge_2008_marvels}; a clustering survey of galaxies and intergalactic gas in the distant universe \citep[BOSS;][]{Dawson_2013_boss}, which required an upgrade to the optical spectrographs \citep{Smee_2013_bossspectrographs}; and an infrared survey of Milky Way and Local group stars \citep[APOGEE-1;][]{Majewski_2017_apogeeoverview}, which introduced infrared spectrographs to the suite of SDSS instrumentation \citep{Wilson_2019_apogeespectrographs}. SDSS-IV \citep{Blanton_2017_sdss4} included a significant expansion of the APOGEE survey (APOGEE-2; Majewski et al.\,in prep), including the deployment of a new APOGEE-S spectrograph for observations with the 2.5m DuPont telescope at Las Campanas Observatory (LCO), as well as an extension of BOSS observations to previously understudied redshifts \citep[eBOSS;][]{Dawson_2016_eboss}, and an optical IFU survey of the gas and stellar properties of low-redshift galaxies \citep[MaNGA;][]{Bundy_2015_manga}.

Continuing in this tradition, SDSS-V \citep[][Kollmeier et al.\,in prep.]{Kollmeier_2017_sdss5} constitutes 
a major innovation step in science scope and hardware. It comprises three primary scientific surveys, called ``mappers'': the Milky Way Mapper (MWM; Johnson et al.\,in prep.), the Black Hole Mapper (BHM; Anderson et al.\,in prep.), and the Local Volume Mapper (LVM; Drory et al.\,in prep.). See \S\ref{sec:status}, \S\ref{sec:mwm_target_cartons}, and \S\ref{sec:bhm_observations} for more details.

A hallmark of the SDSS family of surveys is the regular release of high-quality, well-documented, and ready-usable data to the entire world.  Beginning with the Early Data Release \citep{Stoughton_2002_sdssEDR}, SDSS helped usher in the era of ``open science'' through its regular data releases.  This practice has proven fruitful for the astronomical community, enabling a very broad scientific reach and impact.  The final data release of SDSS-IV was DR17, in December 2021 \citep{Abdurrouf_2021_sdssDR17}. As of November 2022, SDSS data (across all survey phases) have been cited in more than 11,000 refereed papers, with over 650,000 citations. \citet{Stalzer_2013_datadriven} named SDSS as one of the most influential data sets, even beyond astronomy and physics. SDSS data are also used in numerous educational contexts, from young schoolchildren to undergraduate and graduate students, especially through its {\it Voyages} and {\it SciServer} platforms\footnote{\url{https://voyages.sdss.org/}, \url{https://www.sciserver.org/outreach/}}.

One of the primary reasons for the widespread use of SDSS data is the collaboration's commitment to high-quality, user-friendly documentation to accompany each data release \citep[e.g.,][]{Weijmans_2019_docuDR}. 
SDSS-V continues that core practice with DR18, as summarized in this paper. In \S\ref{sec:status}, we briefly summarize SDSS-V's scientific and hardware components. In \S\ref{sec:scope}--\ref{sec:access}, we outline the scope of DR18 and how to access the information. \S\ref{sec:mos_targeting} describes the multi-object spectroscopic (MOS) databases and targeting software, while \S\ref{sec:mwm_target_cartons}, \S\ref{sec:bhm_observations}, and \S\ref{sec:open_fiber_programs} detail the MOS Milky Way Mapper, Black Hole Mapper, and ``open fiber'' programs, respectively. \S\ref{sec:spectra} describes the spectra that are released in DR18, and \S\ref{sec:software} focuses on new and modified software. \S\ref{sec:vacs} contains the Value-Added Catalogs released since DR17. Finally, \S\ref{sec:summary} summarizes all of this information and looks forward to the anticipated contents of DR19.

\section{SDSS-V: Status and Science Objectives}
\label{sec:status}
In this section, we provide a brief summary of the SDSS-V program at the time of this data release. Many of these elements have been described elsewhere \citep[e.g.,][]{Kollmeier_2017_sdss5} and will be described in greater detail in companion papers, including Kollmeier et al.\,(in prep).  We provide this information here as a standalone snapshot at this important survey milestone.

SDSS-V sets out to be the first astronomical survey to provide all-sky, multi-epoch spectroscopy in the optical and infrared. Its scientific goals and targets are grouped into three top-level ``Mapper'' programs. The Milky Way Mapper (MWM; \S\ref{sec:mwm_target_cartons}) is mapping the stellar populations and chemo-dynamics of the Milky Way to understand its evolution, and is probing stellar physics and stellar system architectures by collecting optical and infrared spectra of stars in the Milky Way and the Magellanic Clouds. The Black Hole Mapper (BHM; \S\ref{sec:bhm_observations}) is studying the physics of black hole growth through time-domain spectroscopy and is providing spectra for extragalactic X-ray-luminous sources from eROSITA \citep{Merloni2012,Predehl2021}, using optical spectra across the sky. The Local Volume Mapper (LVM) is examining gas emission, star formation, and stellar/interstellar energy exchange processes in the Milky Way and beyond, at unprecedented scales using ultra-wide-field optical IFU spectroscopy.  

MWM and BHM are designed to use the multi-object spectroscopy (MOS) infrastructure of SDSS-V, which includes the Sloan 2.5~m telescope at Apache Point Observatory (APO) in New Mexico, USA \citep{Gunn_2006_sloantelescope} and the du Pont 2.5~m telescope at Las Campanas Observatory (LCO) in Chile \citep{Bowen_1973_duPontTelescope}. Obtaining homogeneous, all-sky spectral survey data at both optical and infrared wavelengths requires near-identical sets of optical and near-infrared fiber spectrographs in both hemispheres. For SDSS-V, one of the optical BOSS spectrographs was moved from APO to LCO, so that each site is now equipped with a pair of APOGEE and BOSS instruments\footnote{{Details of APOGEE spectrographs: \url{https://www.sdss.org/instruments/apogee-spectrographs/} and details of BOSS spectrographs: \url{https://www.sdss.org/instruments/boss-spectrographs/}.} Within SDSS-V, one will frequently encounter references to ``BOSS spectra''. This term always means ``spectra obtained with one of the two BOSS spectrographs'', rather than ``spectra obtained as part of the SDSS-III BOSS (or SDSS-IV eBOSS) project''. }. The central new hardware component for MOS observations in SDSS-V is the implementation of a focal plane system (FPS) with robotic fiber positioners \citep{Pogge_2020_fps}, which replaces the plug-plate system and enables efficient, single-epoch pointings with $\sim 15$~minute exposure times.
The adoption of the FPS, in turn, required a new three-element corrector for the Sloan telescope \citep{Barkhouser_2022_sdss5corrector}. 

For the LVM, SDSS-V is building an entirely new facility at LCO to enable ultra wide-field IFU observations across $\sim$1,000~deg$^2$. The Local Volume Mapper Instrument (LVM-i) builds upon replicated instrument concepts from DESI \citep{Perruchot_2018_desispectrographs}
%that are the state of the art in very high multiplex spectroscopy  
and the successful IFU technology developed as part of SDSS-IV/MaNGA. To meet the survey science requirements, LVM-i will comprise a set of four 160~mm telescopes coupled via fiber IFUs to three spectrographs providing full coverage of the optical waveband, all housed in a new dedicated enclosure \citep{Herbst_2020_lvmi}.

\section{Scope of Data Release 18}
\label{sec:scope}
The main focus of DR18 is to lay the groundwork for future SDSS-V releases by presenting the updated access paths, data models, new targeting strategies, and other structures that will be used in DR19 and beyond {(Table~\ref{tab:dr18_contents})}.

DR18 consists primarily of targeting information (algorithms and databases) for the MWM and BHM programs (\S\ref{sec:mos_targeting}). These catalogs comprise 269.5~GB of data in ${\tt MOS\_TARGET\_DIR}$ (Table~\ref{tab:catalogs}). 
They also introduce the concept of a ``carton'', a new organizational unit for SDSS observations. A carton is a set of targets that results from a specific target selection algorithm, designed to advance certain science goals. Examples of cartons include white dwarfs selected from {\it Gaia}--SDSS catalogs for MWM, and reverberation mapping targets for BHM. Cartons will play a large role in understanding the SDSS-V targeting strategy, target selection algorithms, and eventual full sample. They are discussed in more detail in \S\ref{sec:mos_cartons}, and lists of MWM and BHM cartons are given in Tables \ref{tab:mwm_cartons} and \ref{tab:bhm_cartons}, respectively. Cartons are grouped into ``programs'', which are also columns in the \texttt{target} tables that refer to broader science cases whose goals will be met by targets from one or more cartons.

A small number of new BOSS spectra are being made available as part of eFEDS \citep{Brunner2022}, an eROSITA follow-up program (\S\ref{sec:efeds}). These add a volume of 301.9~GB to ${\tt BOSS\_SPECTRO\_REDUX}$, including the $\sim$25k spectra. Accompanying these spectra is a new value added catalog (VAC), which contains updated redshifts and classifications (see \S\ref{sec:vacs_dr18}). DR18 also contains new software routines and updates to existing SDSS software (\S\ref{sec:software}), particularly the BOSS spectral reduction package (\S\ref{sec:software_boss}).

Finally, all previous SDSS data are also available through the SDSS portals without any changes or additional processing, essentially summarized in DR17 \citep{Abdurrouf_2021_sdssDR17}. Newly available as part of DR17, coincident with DR18, are two new VACs based on DR17 data, as well as an earlier DR15-based VAC that has been updated with data from DR17. These VACs are described in greater detail in \S\ref{sec:vacs_dr17}.

%\begin{deluxetable*}{p{3cm}p{8cm}p{6cm}}[!hp]
\begin{deluxetable*}{lll}[!hp]
\rotate
\label{tab:dr18_contents}
\tabletypesize{\small}
\tablecaption{Summary of DR18 Contents}
\tablehead{\colhead{Item} & \colhead{Description} & \colhead{Data Access}}
\startdata
\multicolumn{3}{l}{Targeting Information\tablenotemark{1}} \\
\texttt{mos\_catalog} & Cross-match of input catalogs (\S\ref{sec:mos_targetdb}) & \url{https://skyserver.sdss.org/dr18} \\
\texttt{mos\_X} catalogs & Input catalogs (\S\ref{sec:mos_targetdb}, Table~\ref{tab:catalogs} for core catalogs) & \url{https://skyserver.sdss.org/dr18} \\
\texttt{mos\_target} & Objects meeting the selection criteria of one or more cartons (\S\ref{sec:mos_cartons}) & \url{https://skyserver.sdss.org/dr18} \\
\texttt{mos\_carton\_to\_target} & Table linking targets, cartons, and observational requirements (\S\ref{sec:mos_cartons}) & \url{https://skyserver.sdss.org/dr18} \\
\texttt{mos\_carton} & Table of SDSS-V cartons (\S\ref{sec:mos_targetdb}, \S\ref{sec:standards}, \S\ref{sec:mwm_target_cartons}, \S\ref{sec:bhm_observations}, \S\ref{sec:open_fiber_programs}) & \url{https://skyserver.sdss.org/dr18} \\
\hline 
Spectra & & \\
eFEDS spectra & Optical spectra of eROSITA-selected sources (\S\ref{sec:bhm:efeds} and \S\ref{sec:efeds}) & Directories and data models are in \S\ref{sec:efeds_data} \\
\hline
\multicolumn{3}{l}{Value-Added Catalogs (VACs)} \\
SDSS-IV VACs & Based on SDSS-IV data (\S\ref{sec:vacs_dr17}, Table~\ref{table:vac}) & \url{https://www.sdss.org/dr18/data_access/value-added-catalogs/} \\
SDSS-V VAC & Based on SDSS-V (eFEDS) data (\S\ref{sec:vacs_dr18}, Table~\ref{table:vac}) & \url{https://www.sdss.org/dr18/data_access/value-added-catalogs/} \\
\enddata %This throws an error when using the p{} column spec
\tablecomments{
\tablenotetext{1}{These are a subset of the total individual targeting tables available from the CAS. The full list can be seen in the CAS Schema Browser: \url{https://skyserver.sdss.org/dr18/MoreTools/browser}.}
}
\end{deluxetable*}

\section{Accessing the Data} 
\label{sec:access}

There are numerous ways to access the SDSS DR18 data products, summarized on the SDSS website\footnote{\url{https://www.sdss.org/dr18/data_access/}} and on the data release website\footnote{  \url{https://dr18.sdss.org}}. As in previous phases of SDSS, SDSS-V provides a searchable database for the Catalog Archive Server (CAS), using SkyServer\footnote{\url{http://skyserver.sdss.org}}, with both SQL and Jupyter notebook interfaces. The Science Archive Server (SAS) is well-suited for directly downloading flat files, including spectra.  Both the CAS and SAS are cumulative systems, with some data products replacing or extending data from previous releases, generally speaking.
Examples of data products that can be replacements (not in DR18) include spectroscopic reductions and their associated parameters and Value-Added Catalogs.  In these cases, only the latest version is included in the data release, but all previously released versions are available at their original locations.  An example of data products that are cumulative is the raw data transferred from the observatories for each night of observations.

The best way to access a particular data product will depend on the data product itself and the anticipated use case.  We recommended to use the CAS when accessing the MOS targeting data described in \S\ref{sec:mos_targetdb}, since this product was originally created as a relational database. The corresponding fits files on the SAS are primarily for archival purposes. Access to the eFEDS spectra is described in \S\ref{sec:spectra_efeds_access}; these spectra are also available for visual inspection in the Science Archive Webapp at \url{https://dr18.sdss.org/optical/}. 

Each type of data file on the SAS has an associated data model, which describes its format and content. SDSS-V is developing extensions to the existing data model products to include better accessibility in formats other than the current static \emph{html}, such as \emph{yaml}, \emph{json}, and backward compatibility with classic \emph{html}.  For DR18, all data models can be found at \url{https://data.sdss.org/datamodel/}.

The SDSS DR18 website has numerous tutorials and examples available to access and interact with both the MOS targeting data and the eFEDS spectra released in DR18. See \url{https://www.sdss.org/dr18/tutorials/} for more information.

\section{Multi-Object Spectroscopy Targeting}
\label{sec:mos_targeting}
The primary data products in DR18 are the underlying, cross-matched
catalog of objects targeted by SDSS-V and the list of targets. 
In this section, we describe how these data are organized into
cross-matched catalog database tables and target tables.
Throughout DR18, we use \texttt{0.5} to indicate the version of the catalog cross-matching process (\S\ref{sec:mos_targetdb}). Variants of the target selection of individual cartons (\S\ref{sec:mos_cartons}) using this cross-match are versioned as \texttt{0.5.X}, where ``\texttt{X}'' varies between cartons. A set of cartons and carton targeting versions is referred to as a ``targeting generation'', which is tracked similarly as \texttt{0.5.X}. Targeting generation \texttt{0.5.3} is the one released in DR18.

\subsection{Catalog Database Tables and Cross-Matching}
\label{sec:mos_targetdb}

Previous generations of SDSS relied almost exclusively on a single imaging catalog for targeting. In the optical, the SDSS imaging survey \citep{sdss_dr7,ubercal} formed the basis for targeting, while in the $H$-band, the 2MASS all-sky survey was used \citep{2mass}. For SDSS-V, this approach is no longer practical because of the need for all-sky, deep optical imaging and the desire to observe stars in {\it both} the optical and the infrared. Gaia imaging does not go deep enough in the extragalactic sky, while even the union of SDSS imaging, PanSTARRS imaging, and other wide-field imaging do not cover the full sky at the necessary wavelengths. 

Therefore, to ensure that each object in the sky of relevance to SDSS-V has a unique identifier, regardless of how many imaging catalogs it appears in, we created our own targeting database. All SDSS-V multi-object spectroscopy (MOS) targets (i.e., all targets for MWM and BHM) are stored in a suite of database tables,
which contains both a unified list of all sources within a set of crossmatched input catalogs and the targeting classifications of those sources. 
This section describes how the catalogs
are stored; \S\ref{sec:mos_cartons} describes
how the targeting classifications are stored.

\begin{figure*}[!hpbt]
\includegraphics[trim=2cm 0cm 2cm 0cm, angle=270, width=\textwidth]{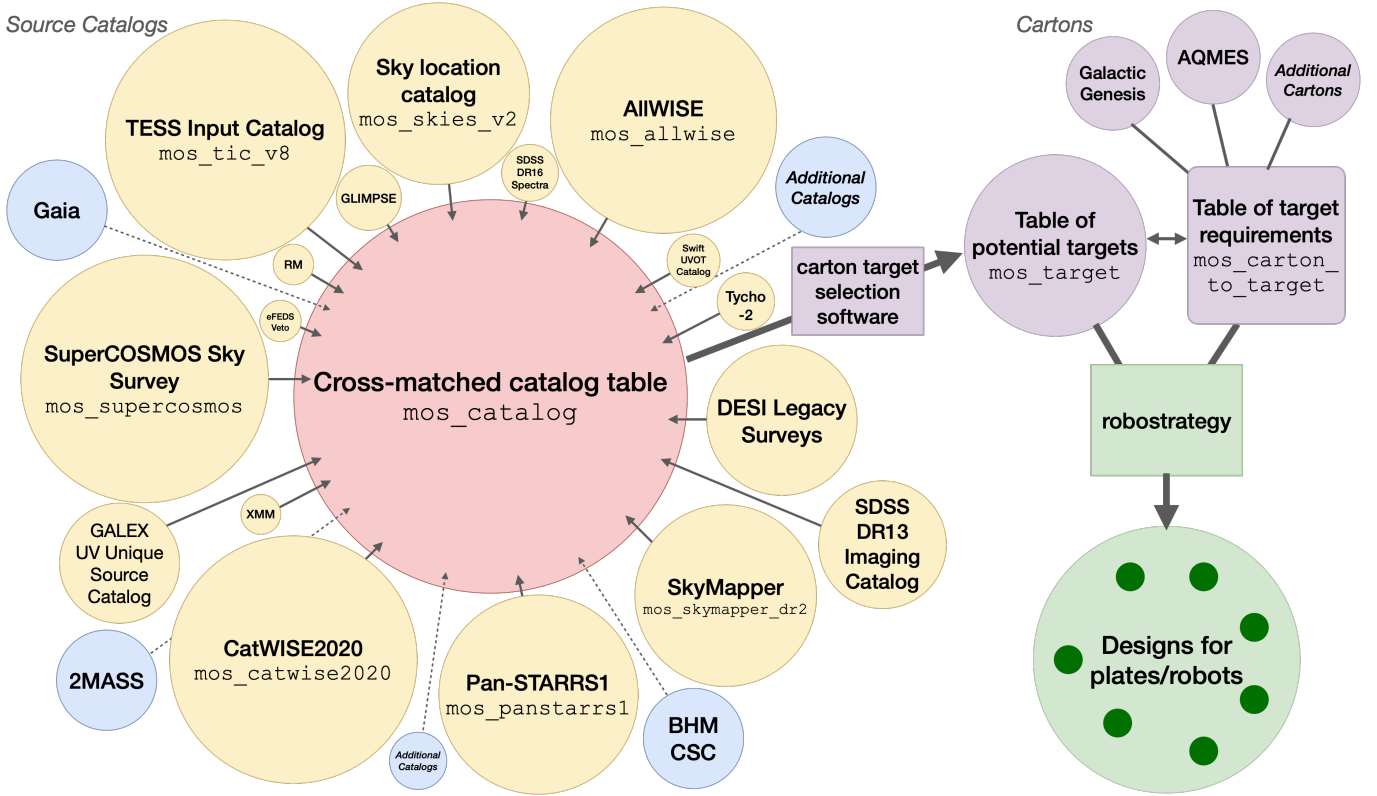} %trim=left botm right top
\caption{
Cartoon diagram of the relationships between several key database tables and of the workflow between sky sources and observed targets (\S\ref{sec:mos_targetdb}--\ref{sec:mos_cartons}). The objects colored yellow, blue, pink, and purple correspond to the same types of tables in Figures~\ref{fig:catalog_to_x}--\ref{fig:targetdb}. The green color has been repurposed to highlight the "observational" section of the workflow, where candidate targets are down-selected and grouped into observable designs (\S\ref{sec:mos_cartons}). {\it Left:} the pink \texttt{mos\_catalog} is built as a superset of the yellow core input catalogs (\S\ref{sec:mos_targetdb}, Table~\ref{tab:catalogs}), whose circles are approximately scaled here to the size of the catalog. The blue ``additional catalogs'' (\S\ref{sec:mos_targetdb}) provide supplementary information to sources in \texttt{mos\_catalog}. {\it Right:} the small purple circles represent the many individual cartons (\S\ref{sec:mos_cartons}), which specify both target selection criteria and observational requirements. The potential targets satisfying the selection criteria, along with the observational requirements (stored in \texttt{mos\_carton\_to\_target}) are used by the robostrategy code (Blanton et al., in prep) to produce observable designs, previously for plates (Covey et al., in prep) and now for the FPS robots.
}
\label{fig:database_cartoon}
\end{figure*}

Figure~\ref{fig:database_cartoon} shows a cartoon representation of the primary database tables and workflows presented in this section.
The {\tt mos\_catalog} table is the central unified list of sources.
It was created from {the set of spatially cross-matched core input
catalogs, which are listed in Table~\ref{tab:catalogs} and shown in yellow in Figures~\ref{fig:database_cartoon} and \ref{fig:catalog_to_tic_etc}}. We have a set of additional
catalogs used in targeting whose entries are associated with
entries in the spatially cross-matched catalogs; { these are the tables highlighted in blue in Figures~\ref{fig:database_cartoon} and \ref{fig:catalog_to_tic_etc}}.
The database in DR18 stores this set of catalogs and the associations
between objects within them.

\begin{figure}[!hpbt]
\includegraphics[width=0.45\textwidth]{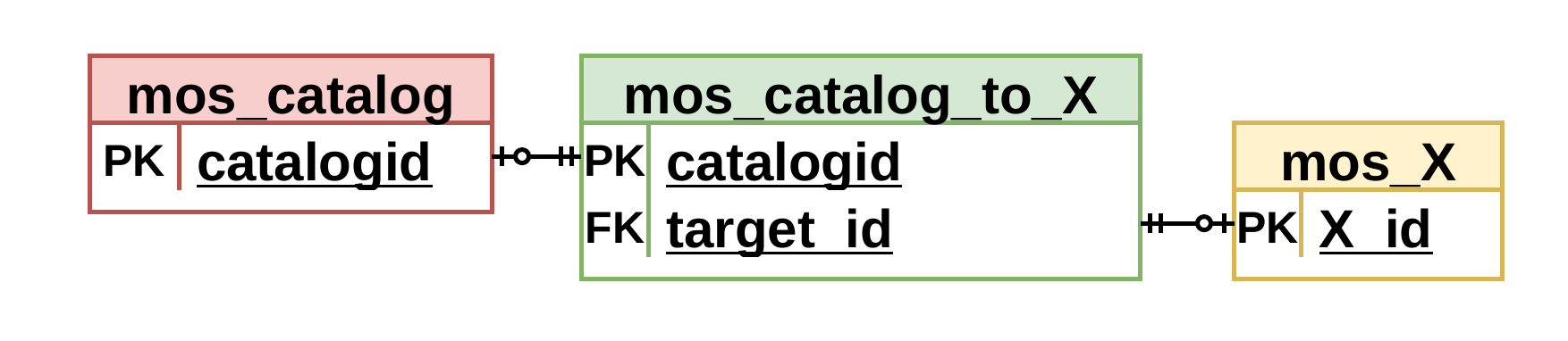}
\caption{
Relationships between the cross-matched {\tt mos\_catalog} table and the original source input catalogs (\S\ref{sec:mos_targetdb}).
%{\tt catalog} contains the full cross-matched list of sources.
There are 17 source input catalogs that are linked to { {\tt mos\_catalog}} through the structure shown above, with ``X'' representing the name of the original catalog (Table~\ref{tab:catalogs}).
}
\label{fig:catalog_to_x}
\end{figure}

The {\tt mos\_catalog} table contains only a minimum of information for each object: only the positional information and the {\tt catalogid} identifier, which is an internal ID used by the SDSS-V databases to uniquely identify each source.
All other information about the sources is stored in the
individual cross-matched and targeting catalogs.
Each {\tt catalogid} is associated with
one or more entries within these original catalogs. For 17
of the original input catalogs, the association between
{\tt mos\_catalog} and the original catalog is expressed as shown in Figure \ref{fig:catalog_to_x}, using a
join table {\tt mos\_catalog\_to\_X} for each input catalog { ``\texttt{mos\_X}''}.
{ As above}, the list of all original catalogs cross-matched in this fashion
is in Table \ref{tab:catalogs}, along with the primary
key used for each (i.e., the column
labeled ``{\tt X\_id}'' in Figure \ref{fig:catalog_to_x}).

\begin{deluxetable*}{lll}
\label{tab:catalogs}
\tablecaption{Cross-matched input catalogs for the catalog database (\S\ref{sec:mos_targetdb})}
\tablehead{\colhead{Table Name\tablenotemark{1}} &
\colhead{Catalog Description} & \colhead{Primary Key\tablenotemark{1}} }
\startdata
{\tt mos\_allwise} & AllWISE \citep{Cutri_2013_allwise} & {\tt cntr} \cr % {\tt https://wise2.ipac.caltech.edu/docs/release/allwise/}
{\tt mos\_bhm\_efeds\_veto } & eFEDS-related veto catalog & {\tt pk} \cr
{\tt mos\_bhm\_rm\_v0\_2 } & RM-related targeting catalog & {\tt pk} \cr
{\tt mos\_catwise2020} & CatWISE2020 \citep{Marocco2021} & {\tt source\_id} \cr  % https://catwise.github.io/
{\tt mos\_glimpse} & GLIMPSE I, II, 3D \citep{Churchwell_09_glimpses} & {\tt pk} \cr  % https://irsa.ipac.caltech.edu/data/SPITZER/GLIMPSE/
{\tt mos\_guvcat} & GALEX UV Unique Source Catalog \citep{Bianchi_2017_GALEXuvcat} & {\tt objid} \cr  %  https://archive.stsci.edu/hlsp/guvcat
{\tt mos\_legacy\_survey\_dr8} & DESI Legacy Surveys DR8 \citep{Dey_2019_DESIsurveys} & {\tt ls\_id} \cr  %  https://www.legacysurvey.org/dr8/
{\tt mos\_panstarrs1} & Pan-STARRS1 \citep{Flewelling_2020_PS1,Magnier_2020_PS1data} & {\tt catid\_objid} \cr  %  https://outerspace.stsci.edu/display/PANSTARRS/
{\tt mos\_sdss\_dr13\_photoobj\_primary} & SDSS DR13 Imaging Catalog \citep{Albareti2017_sdss_dr13} & {\tt objid} \cr  % https://www.sdss.org/dr13/imaging/
{\tt mos\_sdss\_dr16\_specobj} & SDSS DR16 Optical Spectra \citep{Ahumada2020_sdss_dr16} & {\tt specobjid} \cr  %  https://www.sdss.org/dr16/spectro/
{\tt mos\_skies\_v2} & Sky location catalog (\S\ref{sec:skies}) & {\tt pix\_32768} \cr  %
{\tt mos\_skymapper\_dr2} & SkyMapper DR2 \citep{Onken_2019_skymapperDR2} & {\tt object\_id} \cr  %  https://skymapper.anu.edu.au/data-release/
{\tt mos\_supercosmos} & SuperCOSMOS Sky Survey \citep{Hambly_2001_supercosmos} & {\tt objid} \cr  %  http://www-wfau.roe.ac.uk/sss/
{\tt mos\_tic\_v8} & TESS Input Catalog v8 \citep{Stassun_2019_TIC_v8} & {\tt id} \cr % https://tess.mit.edu/science/tess-input-catalogue/
{\tt mos\_tycho2} & Tycho-2 \citep{Hog_2000_tycho2} & {\tt designation} \cr  %  https://www.cosmos.esa.int/web/hipparcos/tycho-2
{\tt mos\_uvotssc1} & Swift UVOT Serendipitous Source Catalog \citep{Yershov_2014_xmmswift} & {\tt id} \cr  %  https://archive.stsci.edu/prepds/uvotssc/
{\tt mos\_xmm\_om\_suss\_4\_1} & XMM OM Serendipitous Ultraviolet Source Survey \citep{Page_2012_xmmUVsource} & {\tt pk} %  https://heasarc.gsfc.nasa.gov/W3Browse/all/xmmomsuob.html
\enddata
\tablenotetext{1}{ The Table Name and Primary Key are denoted as \texttt{mos\_X} and \texttt{id\_X}, respectively, in Figure~\ref{fig:catalog_to_x}.}
\end{deluxetable*}

The initial basis for \texttt{mos\_catalog} is the TESS Input Catalog v8 \citep[TIC;][]{Stassun_2019_TIC_v8}, which is all-sky and uses both Gaia~DR2 and 2MASS to identify objects\footnote{ We anticipate including Gaia DR3 data in the future.}. 
Then, cross-matching follows a three-phase procedure for each {of the subsequent input catalogs listed in Table~\ref{tab:catalogs}, which can include both point and extended sources}. 
First, we use existing literature cross-matches included with the 
catalog information to identify physical targets that already exist 
in the {\tt mos\_catalog} table. For those, a new entry is added to 
the appropriate {\tt mos\_catalog\_to\_X} table that references the 
matched {\tt catalogid} and the unique identifier in the input catalog. 

For targets in the input catalog that do not have an available association with 
{\tt mos\_catalog}, we perform a cone search around the coordinates of 
each target and find the associated {\tt mos\_catalog} entries within the search radius. We use 
a default 1~arcsec search radius, but this value is sometimes modified to match the spatial 
resolution of the input catalog. All the cone-search-matched {\tt catalogid} entries are 
associated with the input catalog target via the corresponding {\tt mos\_catalog\_to\_X} table. 
The match with the smallest on-sky separation is marked as the ``best'' match\footnote{This cross-matching approach is sensitive to the order in which input catalogs are ingested. In general, catalogs were processed in order of higher to lower spatial resolution, which minimizes the problem of mis-associations in the spatial cross-match phase due to mismatching resolutions.} by setting the {\tt 
best} column in the {\tt mos\_catalog\_to\_X} table to {\tt True}. Finally, 
targets in the input catalog that cannot be 
spatially cross-matched are considered new physical objects and added to 
{\tt mos\_catalog} and  {\tt mos\_catalog\_to\_X} with a new unique
{\tt catalogid}.

%This cross-matching approach is sensitive to the order in which input catalogs are ingested. In general, catalogs were processed in order of higher to lower spatial resolution, which minimizes the problem of mis-associations in the spatial cross-match phase due to mismatching resolutions.

For some of these 17 catalogs, the database contains further associations 
between them and the additional catalogs used in targeting.
Figure~\ref{fig:catalog_to_tic_etc} shows these other catalogs and
how they are associated in the database with the cross-matched
catalogs.

DR18 contains the catalog cross-matching version {\tt v0.5}, which was used
for targeting at the beginning of MOS observations in SDSS-V. {Due to the size of the databases involved,} the tables in the
released DR18 version only contain sources that were ultimately identified with
potential targets (i.e., are in one or more cartons). DR18 also excludes some X-ray catalogs from {\it eROSITA} that were used for targeting in {\tt v0.5} but not yet made public. 
Due to an error, DR18 also excludes tables associated with the TESS Objects of Interest, the {\it Spitzer}/MIPSGAL program \citep{Gutermuth15a}, and the {\it Gaia} ASAS-SN Classical Cepheid sample \citep{Inno2021}; these tables will be included in DR19.

\begin{figure*}[!hpbt]
\includegraphics[width=0.98\textwidth]{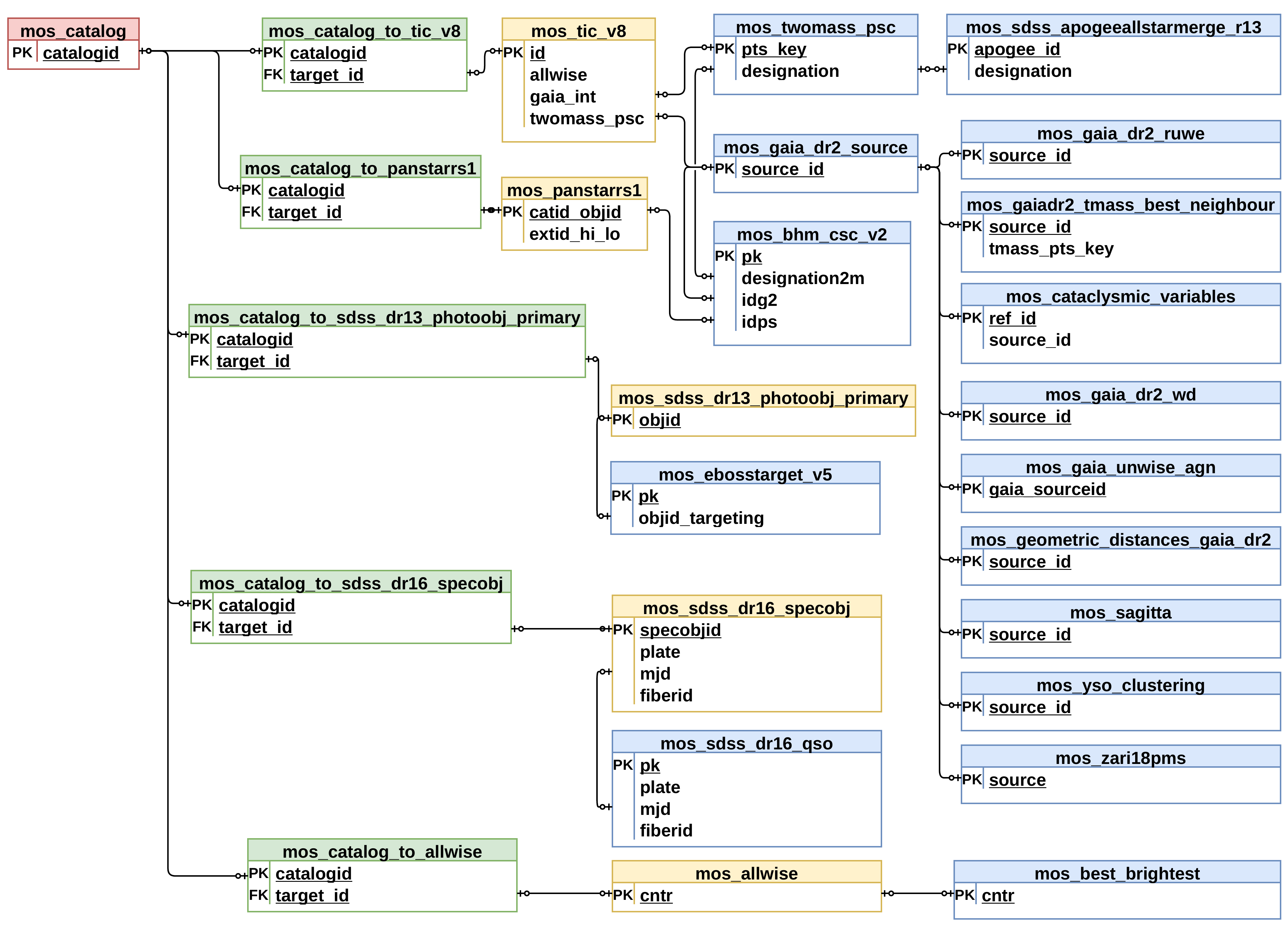}
\caption{\label{fig:catalog_to_tic_etc} Relationships between
cross-matched {\tt mos\_catalog} table (pink), original source catalogs (yellow)
and additional catalogs used for targeting (blue), for the cases
where additional such catalogs exist.
{\tt mos\_catalog} contains the full cross-matched list of
sources, and it is associated with the original source catalogs through
join tables (green) in the manner described in Figure~\ref{fig:catalog_to_x} and \S\ref{sec:mos_targetdb}.
Additional catalogs can be joined as shown to the original source catalogs,
and thereby to the {\tt mos\_catalog} table, as shown. For clarity, only the columns necessary
for table joins are shown in this diagram. 
}
\end{figure*}

\subsection{Target Database Tables and MOS Cartons}
\label{sec:mos_cartons}

Drawing on the catalog database {and the information stored in the linked catalogs} described above (\S\ref{sec:mos_targetdb}), the target selection software determines which objects should be targeted for spectroscopy using the selection criteria and other properties of the target cartons.
The target selection criteria are described in later sections, on the website, and in other program papers; here we only describe how the information is organized.

\begin{figure}[!hpbt]
\includegraphics[width=0.47\textwidth]{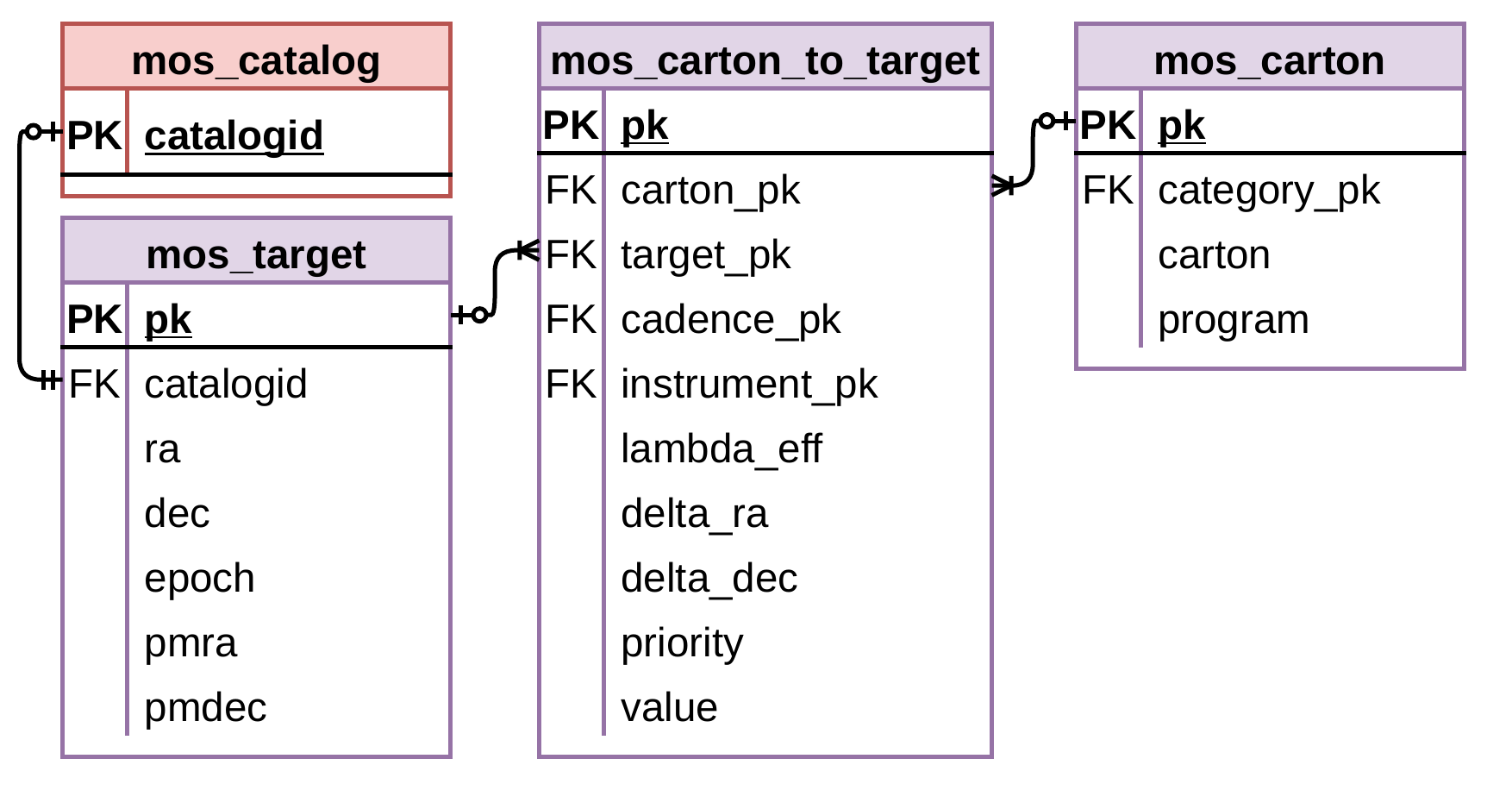}
\caption{\label{fig:targetdb} Association of {\tt catalogid}s with targeting
information. Each {\tt catalogid} that is selected by one or more
cartons is given an entry in the {\tt mos\_target} table. For each carton
that selects it, there is a {\tt mos\_carton\_to\_target} entry. Each
entry is associated with one target and one carton, and specifies the
cadence, instrument (BOSS or APOGEE), and other observing conditions.
Further tables in the database define the properties of the cadences,
the names of the instruments, and the names of the categories ({\tt science},
{\tt sky\_boss}, {\tt standard\_boss}, {\tt sky\_apogee}, and {\tt standard\_apogee}).}
\end{figure}

After the target selection software is run, the targeting database contains a set of targets
and a set of cartons.
Figure~\ref{fig:targetdb} shows how this information is stored for a generalized subset of the database tables.
The associations between targets and cartons are stored in
the {\tt mos\_carton\_to\_target} table. Each carton is associated with all of the candidate targets that satisfy the selection criteria for the carton. Each target is associated with one or more cartons.

{For example, consider a star that has a detection in Gaia~DR2. This star will appear in the TIC~v8, and thus will have an entry in \texttt{mos\_catalog}. If this star satisfies the selection criteria for {\it any} carton, it will also appear in \texttt{mos\_target}, with a corresponding entry in \texttt{mos\_carton\_to\_target}. If this star satisfies the criteria for multiple cartons, then multiple entries in \texttt{mos\_carton\_to\_target} will be associated with the star's single entry in \texttt{mos\_target}. Each \texttt{mos\_carton\_to\_target} entry is also associated with an entry in the \texttt{mos\_carton} table, which gives the name of the carton and other information.}

Each {\tt mos\_carton\_to\_target} entry has an associated set of entries --- \emph{cadence, value}, and \emph{priority} --- that are used in the process of determining the
survey strategy and fiber assignment process.
The \emph{cadence} describes how the target
should be observed (number of epochs, number of observations per epoch, and
observing conditions). The {\tt cadence\_pk} values allow
a join to a cadence table with the cadence name and other
parameters associated with it. 

The \emph{value} expresses how important the target is to the
overall objectives of the MOS program, which is used in the determination of 
the overall survey strategy.
The \emph{priority} expresses the order in which targets should be 
assigned to fibers during fiber assignment in a given pointing.
These quantities are more fully explained in the {\tt robostrategy} paper (Blanton et al.\,in prep).

Each carton also has an assigned \emph{category}, which can be one of {\tt science}, {\tt sky\_boss}, {\tt standard\_boss}, {\tt sky\_apogee}, or {\tt standard\_apogee}.
The first category indicates science targets, and the others are different types of calibration targets (\S\ref{sec:standards}--\ref{sec:skies}).

Among the science target cartons, there are those whose observations are required for the stated success of SDSS-V science, as laid out in the survey's science requirements document (``SRD''). These cartons tend to have descriptive names.
The cartons that arose from an initial call within the
SDSS-V collaboration for ``open fiber'' targets have carton names starting with {\tt opentargets}). There is also a set of filler targets for spare fibers. The released tables in DR18 do not unambiguously identify
which cartons are SRD, open, or filler (although the information can often be approximately guessed from the carton names), because this identification is associated with the fiber assignment results, which will not be released until DR19.

\subsection{Standard star cartons}
\label{sec:standards}

\subsubsection{APOGEE standard stars}
The standard stars for APOGEE observations are used primarily to correct telluric absorption by Earth's atmosphere. They are selected in a process very similar to that used for APOGEE-1 and APOGEE-2 observations, which aimed to select hot, blue stars (with few absorption lines of their own) distributed as evenly as possible across the field of view to allow for spatially-dependent telluric corrections \citep{Zasowski_2013_apogeetargeting,Zasowski_2017_apogee2targeting}.

The carton \texttt{ops\_std\_apogee} contains these calibration stars, which are drawn from the 2MASS PSC \citep{2mass}. They are restricted to have magnitude $7 < H < 11$, color $-0.25 < (J-K_s) < +0.5$, and magnitude uncertainties (JERR, HERR, KERR) of $\le$0.1~mag. In addition, we applied the requirements that the 2MASS read flag be equal to ``1'' or ``2'', the quality flag for $J$, $H$, and $K_s$ be equal to ``A'' or ``B'', the galaxy contamination flag be equal to ``0'', the confusion flag be equal to ``000'', the extkey ID \citep[linking to the 2MASS Extended Source Catalog;][]{Jarrett_2000_2massXSC} be \texttt{Null}, and the star lie at least 6$^{\prime\prime}$ from its nearest 2MASS neighbor. From this subset of potential standard stars, the five bluest sources are chosen from each HEALPix $\rm NSIDE=128$ pixel \citep[0.21~deg$^2$~pix$^{-1}$;][]{Gorski_2005_healpix} as telluric standard stars.
%Initially the carton used an upper limit for color of (J-K)<0.5, but this was removed for newer versions of the carton.

\subsubsection{BOSS spectrophotometric standard stars}

Spectrophotometric standard stars are used by the BOSS pipeline to calibrate the absolute and relative throughput of the instrument during science observations.
The strategy for selecting BOSS standards in SDSS-V builds upon past experience from the SDSS-III/IV BOSS and eBOSS projects \citep[see, e.g.,][]{Dawson_2013_boss,Dawson_2016_eboss}. Selecting standards for the SDSS-V program presents some new challenges: i) we need to select standard stars outside the footprint of the SDSS photometric imaging catalog, and ii) we need to reliably select standards along lines of sight that may be heavily reddened by Galactic dust. In order to satisfy the first challenge, we exploited wider area photometric and astrometric information from PanStarrs, {the DESI Legacy Surveys}, and Gaia. To mitigate the complications of extinction, we used the 3D reddening information provided by the TIC~v8 catalog \citep{Stassun_2019_TIC_v8}.

Below we briefly describe the target selection criteria for the various BOSS standards cartons used in early SDSS-V operations, which are released as part of DR18 (targeting generation \texttt{v0.5.3}).

The \texttt{ops\_std\_eboss} carton is identical to the spectrophotometric standards used by the eBOSS survey \citep{Dawson_2016_eboss}. These standards lie in the magnitude range $16 < r_{\mathrm{psf}} < 18$~AB and so are most suitable for use in dark time. Use of these standards maintains an important continuity with archival SDSS spectroscopy, which is especially important when investigating the long term variations of QSOs (\S\ref{sec:bhm_observations}).

The \texttt{ops\_std\_boss\_ps1dr2}, \texttt{ops\_std\_boss\_lsdr8}, and \texttt{ops\_std\_boss\_gdr2} standard star cartons are based on PanStarrs1-DR2, Legacy Survey DR8, and Gaia DR2 photometry, respectively. For each carton, we use the empirically determined location of eBOSS standards within the dereddened color space of the given photometric system to train our new selection locus. Additional cuts on data quality, and in the parallax vs magnitude plane (via Gaia~DR2), are applied to further clean outliers from the sample. Magnitude limits are applied such that these standards are appropriate for BOSS dark-time observations (approximately
$16 < r_{\mathrm{psf}} < 18$~AB, or $16 < G < 18$~Vega). Additionally, the \texttt{ops\_std\_boss\_gdr2} carton is limited to Galactic latitudes of $|b|>10^\circ$. We estimate SDSS $g_{\mathrm{psf}}, r_{\mathrm{psf}}, i_{\mathrm{psf}}, z_{\mathrm{psf}}$ magnitudes for each BOSS standard star target, transferring their native multi-band photometry into the SDSS system via color transforms derived from the \texttt{ops\_std\_eboss} sample.

Additional BOSS standard cartons were developed to facilitate all-sky observations during bright time: \texttt{ops\_std\_boss\_tic}, \texttt{ops\_std\_boss}, and \texttt{ops\_std\_boss\_red}. These cartons fill in the regions of the sky that are not covered by the BOSS standard cartons described above, and they include brighter standards, to mitigate the potential impact of cross-talk from brighter science targets in the design. The \texttt{ops\_std\_boss\_tic} carton consists of likely F stars, selected via T$_{\rm eff}$ and surface gravity cuts applied to TIC-V8 stellar parameters \citep{Stassun_2019_TIC_v8}. To provide uniform coverage over the sky, targets are sorted into a NSIDE=128 HEALpix grid (0.21~deg$^2$~pix$^{-1}$), and the 10 highest-gravity stars in each healpix are retained for the final carton. The \texttt{ops\_std\_boss} carton similarly targets F stars, but selected via cuts on Gaia parallax, color, and absolute magnitude.  Finally, the \texttt{ops\_std\_boss\_red} carton is intended to provide standards even in the most heavily extincted sections of the Galactic midplane. This carton uses cuts on observed and dereddened Gaia$+$2MASS colors, as well as parallax, proper motion and reduced proper motion criteria to reduce contamination from non-F stars whose observed colors can be dereddened into the relevant areas of color-magnitude space. 

\subsection{Sky fiber locations}
\label{sec:skies}

The data reduction pipelines for BOSS and APOGEE both require a number of fibers to be placed at ``sky'' locations, which do not contain light from astrophysical sources and thus allow for the correction of the observed spectra for contamination by emission processes in Earth's atmosphere. The number and ``quality'' of sky fibers per configuration is dependent on the type of observation to be performed. For example, for SDSS-V observations targeting faint extragalactic populations in dark time, we reserve 20\% of the BOSS fibers for sky observations, and those locations must be empty of astrophysical sources down to the magnitude limit of the deepest wide area imaging that we currently have available. In contrast, observations of bright targets in bright time generally require fewer sky fibers, and can sometimes tolerate mild contamination from faint astrophysical sources. However, even with such reduced criteria, suitable sky fiber locations can be challenging to find in the densest Galactic plane fields.

With these constraints in mind we have designed a hierarchy of sky fiber cartons, which collectively satisfy all SDSS-V FPS observational requirements. BOSS observations use three cartons: \texttt{ops\_sky\_boss\_best}, \texttt{ops\_sky\_boss\_good}, and \texttt{ops\_sky\_boss\_fallback}, and APOGEE observations draw on two cartons: \texttt{ops\_sky\_apogee\_best}, and \texttt{ops\_sky\_apogee\_good}.

First, the sky is divided into HEALPix $\rm NSIDE=32$ pixel ``tiles'', comprising roughly 3.4~deg$^2$~pix$^{-1}$, which is approximately the area spanned by a single FPS configuration. Each tile is further divided into HEALpix $\rm NSIDE=32768$ pixels (41\,arcsec$^2$~pix$^{-1}$), the centers of which are considered candidate sky locations\footnote{For comparison, BOSS and APOGEE-N fibers cover $\sim$3.14~arcsec$^2$ on the sky, and APOGEE-S fibers $\sim$1.32~arcsec$^2$.}. Each candidate sky location is then compared to a set of input catalogs (given below) and labeled \emph{valid} (with respect to each comparison catalog, separately) if i) their $\rm NSIDE=32768$ HEALpixel contains no astrophysical objects in that catalog, and ii) they lie further than a magnitude-dependent separation from the nearest potentially contaminating astrophysical source. This minimum separation is computed as $s^* = s + \frac{(m_{\mathrm{thr}}-m)^{\beta}}{a}$, where $s=5$~arcsec is a minimum radius floor and $m_{\mathrm{thr}}$ is a fixed brightness threshold (14~mag). The $a=0.15$ and $\beta=1.5$ parameters control the scaling of the exclusion radius with brightness, and are set to conservative values. We consider the $H$, $G$, $V_{\mathrm{total}}$, $r_{\mathrm{psf}}$, and $r_{\mathrm{model}}$ magnitudes for objects from the 2MASS PSC, Gaia DR2, Tycho2, Pan-STARRS1 DR2, and Legacy Survey DR8 catalogs, respectively \citep{2mass,GaiaCollab_2018_gaiaDR2,Hog_2000_tycho2,Flewelling_2020_PS1,Dey_2019_DESIsurveys}. Candidate sky locations lying near 2MASS extended sources \citep{Jarrett_2000_2massXSC} are considered to be valid if they lie outside the ``extrapolation/total radius'' of the source. All of these candidate locations that are valid in one or more catalogs are stored in the \texttt{catalogdb} table \texttt{skies\_v2} and serve as inputs for the cartons below.

The highest quality sky carton for BOSS (i.e., suitable for darkest observations) is \texttt{ops\_sky\_boss\_best}. For this carton, we require locations to be valid in each of the 2MASS (point- and extended sources), Tycho2, and Gaia~DR2 catalogs. Furthermore, selected sky locations must be within the Pan-STARRS1 or Legacy Survey DR8 survey footprints, and must not be contaminated by sources from either catalog. Somewhat less restrictive is the \texttt{ops\_sky\_boss\_good} carton, which drops the requirement to avoid Pan-STARRS1 and Legacy Survey DR8 sources, and so extends the list of available sky locations to the entire sphere, excepting an area of the Galactic bulge and the inner parts of the Magellanic Clouds. In order to provide sky fiber locations in those remaining regions, the \texttt{ops\_sky\_boss\_fallback} carton loosens the radial exclusion criteria, preferring locations that lie furthest from Gaia~DR2 sources but requiring only that the sky location lies more than 3$^{\prime\prime}$ from a Gaia~DR2 or Pan-STARRS1 source, more than 5$^{\prime\prime}$ from any 2MASS source, and more than 15$^{\prime\prime}$ from any Tycho2 star.

For APOGEE observations, the highest quality \texttt{ops\_sky\_apogee\_best} carton requires sky locations to be valid in each of the 2MASS (point- and extended sources), Tycho2, and Gaia~DR2 catalogs. The lower-priority \texttt{ops\_sky\_apogee\_good} carton is dedicated to finding the remaining least-bad sky locations in dense regions. These positions have a maximum nearest neighbor brightness of $H=10$, and candidates lying farthest from 2MASS contaminants are preferred.

\section{Milky Way Mapper: Science Programs and Target Selection}
\label{sec:mwm_target_cartons}
The Milky Way can serve as a ``model organism'' for understanding  the physical processes that shape galaxies over an enormous rage of temporal and physical scales in the context of hierarchical cosmogony. The Milky Way Mapper (MWM) takes advantage of our unique perspective within the Milky Way to create an unprecedented high-resolution map of our Galaxy’s stellar populations. MWM is observing stars formed in the dawn of the Galaxy through the present day, thoroughly sampling the H-R diagram and chemical abundance space. Through multi-epoch, multi-wavelength observations of single stars and stellar systems, MWM seeks to understand the evolution of the luminous constituents of the Galaxy and detect unseen compact objects and substellar objects. 

MWM focuses on four connected themes:
\begin{itemize} \itemsep -2pt
\item {\bf Galactic Archaeology}: 
We can reconstruct the deep history of the Milky Way through observing the number, masses, composition, ages, and motions of its stars and the structures that they create. Luminous red giants cover a large range of ages, from $\sim$1 Gyr and older.
However, samples of red giants are insensitive to the history of intermediate-mass stars older than $\sim$1 Gyr. Such stars have turned into stellar remnants, mostly white dwarfs. 
Other sites of Galactic archaeology studied in open fiber programs (\S\ref{sec:open_fiber_programs}) in targeting generation \texttt{v0.5.3} include the Galactic halo and the Magellanic Clouds.

\item {\bf The Young Galaxy}:
The programs under Galactic Archaeology investigate the cumulative history of star formation, chemical enrichment, and radial migration of the Milky Way, as told through evolved stars across a range of masses. Young stars (here $\sim 5-100$ Myrs) fill in the early phases of low-mass stellar evolution and the full life cycle of massive stars. They provide a detailed and complete snapshot of the youngest generation in the Milky Way. In MWM, we systematically target low-mass young stars (YSOs) within $\sim 500~$pc and \emph{all} luminous, hot stars throughout the Galactic disk. Dust is relevant to not only the current structure of the Milky Way, but also how metals are incorporated into the ISM and thence into the next generation of stars. MWM takes advantage of dust and extinction indicators along lines of sight to its observed stars to map the dust and its properties in the disk.

\item {\bf The High Energy Galaxy}:
Galactic sources of X-ray emission include compact object accretors, young stellar objects, and flaring stars, all objects of intense interest for the other themes here as well. The eROSITA mission is measuring X-ray fluxes for millions of point sources in the Galaxy. However, to identify and explore the physical nature of such objects, optical and/or infrared spectra rich in absorption and emission features are critical.

\item {\bf Stellar Physics and Stellar Systems}:
Undergirding all of MWM's Galactic explorations are the properties of stars and stellar systems.  We begin these investigation close to home --- in the solar neighborhood, the only place in the universe where it is practicable to obtain a spectroscopic census of stars down to the hydrogen burning limit. 

To reliably predict the age, evolutionary state, nucleosynthesis, internal mixing, and end state of stars, we need to understand their structure. Astereoseismology --- the study of how waves propagate through stellar interiors --- provides a powerful tool for this work, especially when combined with stellar and dynamical parameters from spectra. MWM is targeting oscillating red giant and hot stars. In the latter case, we are focusing on OBA stars in massive eclipsing binaries.

Most stars orbit other stars. MWM is unique among current large spectroscopic surveys in targeting a vast range of stellar types for time-domain radial velocity observations, including YSOs, OB stars, WDs, and red giants. Particular attention is also paid to compact binaries, which are binary systems where at least one component is a WD, neutron star, or black hole, and to stars observed at high cadence for planet transits. By increasing the number of epochs, MWM is increasing the probability of orbit reconstruction with the goal of probing long baselines ($>$8 years) and the brown dwarf desert. 

\end{itemize}

These top-level science goals are achieved by observing an anticipated $\sim$5 million stars, with the different target categories structured in a set of target cartons. Each carton has a well-defined selection function to enable subsequent population modeling \citep{Rix2021}. These cartons are summarized in Table~\ref{tab:mwm_cartons}, with complete selection functions provided in Appendix~\ref{sec:mwm_cartons_appendix} and on the DR18 website\footnote{\url{https://www.sdss.org/dr18/mwm/programs/cartons/}}. It is important to note that these target selections  are the {\it input} into the observational design code (\S\ref{sec:software}). The final targets that will actually be observed are an as-yet-unknown subset of these selections. See Johnson et al.\,(in prep) for a complete description of MWM's goals and targeting.

DR18 contains substantial targeting information for MWM, including the input catalogs used to generate potential target lists, the selection functions for the numerous cartons, and the data models for the anticipated output files. Users anticipating the first spectroscopic data release in DR19 can use this information to prepare for analysis of the DR19 data.
We confine our discussion here to the \texttt{v0.5.3} targeting scheme, deferring presentation of cartons added in subsequent targeting schemes (e.g., for validating stellar parameters, observing {\it Gaia} binaries, and other expansions and improvements) to the future MWM overview and DR19 papers. 
Thus, the subsections below enumerate the specific programs, and their constituent cartons, designed under the \texttt{v0.5.3} targeting strategy to address the scientific goals above.

\subsection{Galactic Genesis}
\label{sec:mwm:gg}

Galactic Genesis is the flagship program of the MWM, with near-infrared (APOGEE) observations of over 4 million stars planned. Galactic Genesis's central science goal to obtain a fine sampling of the chemo-orbital distribution of stars across the entire radial extent of the Milky Way, probing the Galactic plane, the central regions of our Galaxy, and the far side of the disk.  To do this, Galactic Genesis targets luminous red giants, above the red clump, where the multi-million sample size with dust-penetrating APOGEE observations provides a decisive advantage over other surveys.

All of the Galactic Genesis targets are contained in the \texttt{mwm\_galactic\_core} carton.

\subsection{White Dwarfs}
\label{sec:mwm:wds
}
White dwarfs (WDs) are tracers of Galactic star formation, progenitors of type Ia supernovae, important end products of both single and binary star evolution, and important cosmic physics laboratories (e.g., for the formation of strong magnetic fields). 
%Their pristine photospheres can show signs of tiny amounts of recent contamination from infalling cometary or planetary material. 
{\it Gaia} has recently produced enormous, well defined samples of WDs ($\sim$250,000 objects). SDSS-V is enlarging the subset of these WDs that have complementary spectroscopy (especially in the southern hemisphere) and also providing multi-epoch observations for empirical input on the binary WD merging channels towards SNe~Ia and other remnant products \citep[e.g.,][]{Chandra_2021_WDbinary}.

All of the MWM WDs targeted for this program are contained in the \texttt{mwm\_wd\_core} carton.

\subsection{Solar Neighborhood Census}
\label{sec:mwm:snc}

 MWM's Solar Neighborhood Census (SNC) program targets a volume-limited sample of stars with the goals of cataloging low-luminosity stellar populations. While not technically ``complete'', the SNC will provide high-quality, two-epoch BOSS or APOGEE spectra of $>10^5$ stars within 100~pc and within 250~pc. The most complete census will be within 100~pc. However, there are few stars hotter than K types that close to the Sun, so a sample of stars out to 250~pc away is also targeted to enable a reliable match to higher luminosity populations.

The SNC targets span four cartons, all beginning with \texttt{mwm\_snc}: \texttt{mwm\_snc\_100pc\_apogee}, \texttt{mwm\_snc\_100pc\_boss}, \texttt{mwm\_snc\_250pc\_apogee}, \texttt{mwm\_snc\_250pc\_boss}, where the latter part of the carton names indicate the volume size and instrument.

\subsection{Young Stellar Objects (YSOs)}
\label{sec:mwm:yso}

The MWM YSO program targets the pre-main-sequence phase of low-mass stars. These objects are key to understanding the early phases of low-mass stellar evolution, on what orbits stars are born, and how they disperse from their clustered birth sites to become field stars. The YSO program selects objects via their {position in the color-magnitude diagram (CMD)}, optical/IR SED, or variability. APOGEE spectra then provide stellar parameters and velocities, while BOSS spectra provide indicators of youth, such as Li absorption and H$\alpha$ emission. Kounkel et al.\,(submitted) presents a complete discussion of the target selection rationale and on-sky validation.

Cartons that fall under the YSO program have names beginning with \texttt{mwm\_yso}, a third label indicating their selection method, and a fourth label indicating the instrument used (either \texttt{apogee} or \texttt{boss}). The shorthand labels for the selection methods are \texttt{cluster}, \texttt{cmz}, \texttt{disk}, \texttt{embedded}, \texttt{nebula}, \texttt{pms}, and \texttt{variable}. For example, YSO targets selected based on their variability and observed with the APOGEE instrument are contained in the \texttt{mwm\_yso\_variable\_apogee} carton. The full selection functions for these methods are detailed in Appendix~\ref{sec:mwm_cartons_appendix}.

\subsection{OB Stars}
\label{sec:mwm:ob}

Massive stars on the main sequence are unambiguously young because of their short hydrogen-burning lifetimes. They are excellent tracers of the recent star formation throughout the disk \citep[e.g.,][]{Zari2021} and they are most likely luminous companions to black holes and neutron stars, objects that were once even more massive stars. These hot stars are also the dominant ionizers of the interstellar medium. The MWM OB program targets all stars brighter than $G=16$ that are also likely to be hotter than $\sim$8000~K. 

The massive stars in this program fall into two cartons: \texttt{mwm\_ob\_cepheids}, targeting known Cepheid stars \citep{Inno2021}, and \texttt{mwm\_ob\_core}, targeting a larger sample of OBA stars across the MW and Magellanic Clouds.

\subsection{Galactic eROSITA Sources}
\label{sec:mwm:xray}

In addition to the spectroscopic follow-up of high-redshift X-ray sources from eROSITA (\S\ref{sec:bhm_xray_followup}), the Galactic eROSITA sources program in MWM is targeting likely Milky Way X-ray sources, including accreting compact objects, YSOs, and flare stars. The optical and IR counterparts of these sources were identified by the eROSITA Stars and Compact Objects working groups and are being observed with the APOGEE and/or BOSS instrument(s), depending on the $H$-band or $G$-band magnitude, for source characterization.

The Galactic eROSITA program targets comprise three cartons: \texttt{mwm\_erosita\_stars}, for likely stellar coronal emitters, and \texttt{mwm\_erosita\_compact\_gen} and \texttt{mwm\_erosita\_compact\_var} for likely compact object accretors, using two different methods for finding the likeliest optical counterpart.

\subsection{Massive Eclipsing Binaries}
\label{sec:mwm:eb}

The main-sequence structure of massive stars, including the size of their convective cores, is intimately linked to their (eventual) remnant mass and other properties. The Massive Eclipsing Binaries (EBs) program uses asteroseismic and photometric data to identify the likeliest double-lined spectroscopic EBs with pulsational variability, among OBA stellar types. 

These targets are contained in the \texttt{manual\_mwm\_tess\_ob} carton.

\subsection{Binary Systems}
\label{sec:mwm:binaries}

The APOGEE-1 and -2 surveys contributed to binary studies both through statistical studies of large samples \citep[e.g.,][]{apw2020,Mazzola_2020_closebinaries} and through reconstructing orbits \citep[e.g.,][]{Washington_2021_symbioticstars}. In MWM, the orbital size and mass range spanned by targets in the OB Stars and YSO programs will be increased by serendipitous observations of previously-observed stars that meet the Galactic Genesis criteria (see above). The Binary Systems program is further enhancing this sample of stellar and substellar companions of stars across the HR diagram by investing fibers to extending time baselines of red giants, subgiants, and M~dwarfs. This program exclusively uses the APOGEE spectrograph because of its superior spectral resolution and improved RV precision. 

Targets in this program fall in one of two cartons: \texttt{mwm\_rv\_long\_fps}, for stars {\it with} existing multi-epoch observations from APOGEE, and \texttt{mwm\_rv\_short\_fps}, for stars without these earlier observations.

\subsection{Compact Binaries}
\label{sec:mwm:cb}
 Spectra of compact binary systems are frequently marked by non-stellar emission -- such as X-ray, UV, or H$\alpha$ flux -- that can probe mass transfer or other non-thermal processes. The Compact Binaries program is observing a large number of likely compact binary systems, identified through different combinations of CMD position, UV excess, and previous association with a cataclysmic variable (from the AAVSO\footnote{\url{https://www.aavso.org/cataclysmic-variables}}).

These different combinations of selections are reflected in the many cartons contained in the Compact Binaries program. All of the carton names begin with \texttt{mwm\_cb} or \texttt{manual\_nsbh}, followed by additional labels indicating the selection method and (in some cases) the instrument used for the MWM observations. The shorthand labels for the selection methods are \texttt{300pc}, \texttt{cvcandidates}, \texttt{gaiagalex}, and \texttt{uvex[1-5]}. The full selection functions for these methods are detailed in Appendix~\ref{sec:mwm_cartons_appendix}.

\subsection{Planet Hosts}
\label{sec:mwm:planethosts}
Understanding the conditions that impact a star's likelihood of hosting planets, and the properties of the planetary orbits and the planets themselves, is essential for understanding planet formation and solar system evolution. The Planet Hosts program targets TESS stars {\it with} and {\it without} associated TESS Objects of Interest (TOIs). These will be observed with the APOGEE instrument to provide stellar parameters and detailed stellar abundances.

The single carton in this program is called \texttt{mwm\_tess\_planet}.

\subsection{Asteroseismic Red Giants}
\label{sec:mwm:asteroseismic}

For stars on the red giant branch (RGB), asteroseismology arguably provides the gold standard of stellar mass and surface gravity measurements, but maximizing its power requires additional input from spectroscopy (such as stellar metallicity). The goal of the Asteroseismic Red Giants program is to obtain spectra for stars with asteroseismic signals, especially those observed by TESS. The start of SDSS-V overlapped with the nominal mission of the TESS satellite, so this program targets bright RGB stars throughout most of the sky (avoiding the Galactic plane), under the expectation that some large fraction of them would eventually have detectable oscillations. This simple selection function will enhance the legacy value of this program's sample.

The single carton in this program is called \texttt{mwm\_tessrgb\_core}.

\subsection{Dust}
\label{sec:mwm:dust}
Interstellar dust presents both a challenge and an opportunity in studying the structure and history of our Galaxy. For example, rigorous modeling of the Milky Way's structure, based on stellar observations, requires (approximate) knowledge of the 3D extinction distribution, as seen from the Sun. 
%Constructing such a map requires both good distances and precise effective temperatures for extensive sets of stars. 
Such a map then permits dust-corrected estimates of stellar distances for stars with spectroscopic luminosity estimates but poor parallaxes, but also valuable information on the 3D structure of the cold interstellar medium. MWM's Dust program is obtaining spectra for mapping distribution of the density and properties of dust \citep[such as $R_V$; e.g.,][]{Schlafly_2016_OIRextinctionlaw}. Those spectra will of also provide information on the column density and kinematics of the ISM via the diffuse interstellar bands \citep[e.g..][]{Zasowski_2019_highDdust}. The targets for this program are RGB stars selected to ``fill in'' the regions of low extinction avoided by the Galactic Genesis program, to achieve an even sampling of the Milky Way's dust with the total MWM sample.

The single carton in this program is called \texttt{mwm\_dust\_core}.

\subsection{Science Validation}
\label{sec:mwm:scival}

The only carton in this program in the \texttt{v0.5.3} targeting generation is \texttt{mwm\_legacy\_ir2opt}, which is a filler carton to obtain BOSS spectra of stars observed in APOGEE-1 and -2. The stellar parameters from the higher-resolution APOGEE spectrum of a given star are used to provide labels for the BOSS spectrum, facilitating data-driven modeling and cross-calibration of the BOSS spectra. Future targeting generations will include additional cartons targeting stars with known fundamental parameters and stars targeted by other large spectroscopic surveys.

\begin{deluxetable*}{llll}
\label{tab:mwm_cartons}
\tablecaption{Milky Way Mapper Cartons}
%\tablehead{\colhead{\multirow{2}{*}{Name}} & \colhead{\multirow{2}{*}{Selection Summary$^1$}} & \colhead{Number Observed} & \colhead{Approx Number} \\ [-0.2cm]
% \colhead{} & \colhead{} & \colhead{So Far$^2$} & \colhead{in DR19}}
\tablehead{\colhead{\multirow{2}{*}{Carton Name}} & \colhead{\multirow{2}{*}{Selection Summary\tablenotemark{1}}} & 
\colhead{\multirow{2}{*}{Instrument}} & \colhead{Available} \\ [-0.2cm]
\colhead{} & \colhead{} & \colhead{} & \colhead{Targets\tablenotemark{2}}}

\startdata
manual\_mwm\_tess\_ob & OB(AF)-type pulsating stars in eclipsing binaries & APOGEE & 364 \\ 
manual\_nsbh\_apogee & Compact NS/BH binaries APOGEE & APOGEE & 75 \\ 
manual\_nsbh\_boss & Compact NS/BH binaries BOSS & BOSS & 249 \\ 
mwm\_cb\_300pc\_apogee & Compact binaries within 300pc from Gaia+GALEX $G<16$ & APOGEE & 17382 \\ 
mwm\_cb\_300pc\_boss & Compact binaries within 300pc from Gaia+GALEX $G>=16$ & BOSS & 52267 \\ 
mwm\_cb\_cvcandidates\_apogee & Compact binaries from literature $G<16$ & APOGEE & 61 \\ 
mwm\_cb\_cvcandidates\_boss & Compact binaries from literature $G>=16$ & BOSS & 5167 \\ 
mwm\_cb\_gaiagalex\_apogee & Compact binaries from Gaia+GALEX $G<16$ & APOGEE & 16287 \\ 
mwm\_cb\_gaiagalex\_boss & Compact binaries from Gaia+GALEX $G>=16$ & BOSS & 261570 \\ 
mwm\_cb\_uvex1 & Compact binaries from Gaia+GALEX FUV+NUV for BOSS identification & BOSS & 110447 \\ 
mwm\_cb\_uvex2 & Compact binaries from Gaia+GALEX NUV only for BOSS identification & BOSS & 278961 \\ 
mwm\_cb\_uvex3 & Compact binaries from Gaia+XMM-Newton OM for BOSS identification & BOSS & 25032 \\ 
mwm\_cb\_uvex4 & Compact binaries from Gaia+Swift UVOT for BOSS identification & BOSS & 21036 \\ 
mwm\_cb\_uvex5 & Compact binaries from Gaia+GALEX RV targets for APOGEE & APOGEE & 8641 \\ 
mwm\_dust\_core & Sample of bright, nearby, mid-plane giants complementary with Galactic Genesis & APOGEE & 61926 \\ 
mwm\_erosita\_compact\_gen & eROSITA/Gaia compact binary candidates for BOSS spectroscopy & BOSS & 77038 \\ 
mwm\_erosita\_compact\_var & eROSITA/Gaia compact binary candidates, Gaia variability, BOSS identifiation & BOSS & 16835 \\ 
mwm\_erosita\_stars & eROSITA/Gaia coronal emitters & BOSS & 89994 \\ 
mwm\_galactic\_core & Luminous Giants across the sky based on $G-H$ color cuts & APOGEE & 5459267 \\ 
mwm\_legacy\_ir2opt & Follow-up of APOGEE-1/2 targets with BOSS -- Filler Carton & BOSS & 209893 \\ 
mwm\_ob\_cepheids & Cepheids from \citet{Inno2021} with BOSS & BOSS & 2357 \\ 
mwm\_ob\_core & OB(A) stars in the Milky Way and Magellanic clouds with BOSS  & BOSS & 603854 \\ 
mwm\_rv\_long\_fps & FPS Radial Velocity Monitoring -- Long Temporal Baseline & APOGEE & 111233 \\ 
mwm\_rv\_short\_fps & FPS Radial Velocity Monitoring -- Short Temporal Baseline & APOGEE & 6283604 \\ 
mwm\_snc\_100pc\_apogee & Volume limited census of the solar neighbourhood $G<16$ & APOGEE & 114173 \\ 
mwm\_snc\_100pc\_boss & Volume limited census of the solar neighbourhood $G>=16$ & BOSS & 322222 \\ 
mwm\_snc\_250pc\_apogee & Volume limited census of the solar neighbourhood, extension to earlier types $G<16$ & APOGEE & 404618 \\ 
mwm\_snc\_250pc\_boss & Volume limited census of the solar neighbourhood, extension to earlier types $G>=16$ & BOSS & 417431 \\ 
mwm\_tess\_planet & TESS TOI and 2 minute Cadence Follow-up  & APOGEE & 223764 \\ 
mwm\_tessrgb\_core & TESS Observed Red Giant Astroseismology Follow-up & APOGEE & 1004714 \\ 
mwm\_wd\_core & White dwarfs selected from Gaia + SDSS-V \citep{GentileFusillo2019} & BOSS & 192534 \\ 
mwm\_yso\_cluster\_apogee & Candidates identified through phase space clustering & APOGEE & 45461 \\ 
mwm\_yso\_cluster\_boss & Candidates identified through phase space clustering & BOSS & 59065 \\ 
mwm\_yso\_cmz\_apogee & Candidates located in Central Molecular Zone+inner Galactic disk & APOGEE & 13170 \\ 
mwm\_yso\_disk\_apogee & Optically bright candidates identified through infrared excess & APOGEE & 28832 \\ 
mwm\_yso\_disk\_boss & Optically bright candidates identified through infrared excess & BOSS & 37478 \\ 
mwm\_yso\_embedded\_apogee & Optically faint candidates identified through infrared excess & APOGEE & 11086 \\ 
mwm\_yso\_nebula\_apogee & Candidates found towards areas of high nebulosity & APOGEE & 1112 \\ 
mwm\_yso\_pms\_apogee & Low mass candidates located above main sequence on HR diagram & APOGEE & 76361 \\ 
mwm\_yso\_pms\_boss & Low mass candidates located above main sequence on HR diagram & BOSS & 73214 \\ 
mwm\_yso\_variable\_apogee & Low mass candidates identified through Gaia variability & APOGEE & 52691 \\ 
mwm\_yso\_variable\_boss & Low mass candidates identified through Gaia variability & BOSS & 47758 \\ 
\enddata
\tablecomments{
\tablenotetext{1}{For more complete selection details, see Appendix~\ref{sec:mwm_cartons_appendix} and online documentation: \url{https://www.sdss.org/dr18/mwm/programs/cartons/}.} 
\tablenotetext{2}{This column is the number of targets that satisfy the carton selection function in the \texttt{target} database. The number of targets ultimately observed for each carton will be smaller than this value.}
}
\end{deluxetable*}

\section{Black Hole Mapper: Science Programs and Target Selection} 
\label{sec:bhm_observations}
The Black Hole Mapper (BHM) sets out to better understand the growth of supermassive black holes at the centers of galaxies, through both time-domain spectroscopy of quasars and other AGN, and the first large-area optical spectroscopic follow-up of the newly available eROSITA X-ray survey.  In its quasar time-domain program, SDSS-V will provide orders of magnitude advances in both time baseline and sample sizes. The BHM time-domain core programs (\S\ref{sec:bhm:time-domain}) will target about $\sim$10$^{4.5}$ previously known quasars with multiple additional optical spectral epochs. SDSS-V will also provide optical counterpart spectra for $\sim$10$^{5.5}$ X-ray sources (\S\ref{sec:bhm_xray_followup}), especially recently-discovered X-ray sources from SRG/eROSITA \citep{Predehl2021}.

In DR18, BHM is releasing two categories of data: i) initial catalogs of candidate targets for its main science programs, which may provide guidance to the community in planning for future SDSS-V data releases, and ii) optical BOSS spectra for $\sim10^4$ candidate counterparts to eROSITA X-ray sources from the eROSITA Final Equatorial Depth Survey field \citep[eFEDS, \S\ref{sec:bhm:efeds};][]{Brunner2022}.
In the subsections below, we provide a high-level summary of those BHM science programs that directly flow down from the driving science goals of the project (i.e., the ``core'' programs); these are summarized in Table \ref{tab:bhm_cartons}. Further details of the target selection criteria for the full set of BHM target cartons released in DR18 can be found in Appendix~\ref{sec:bhm_cartons_appendix} and on the SDSS DR18 website\footnote{\url{https://www.sdss.org/dr18/bhm/programs/cartons/}}. As in the MWM section above, these targeting selections provide {\it input} into the observational design software (\S\ref{sec:software}). The actual targets that are finally observed are a subset of these selections, determined by the survey optimization software.

\subsection{Spectral time-domain programs} 
\label{sec:bhm:time-domain}
The time-domain BHM core programs will target about $\sim$10$^{4.5}$ previously known quasars with multiple additional optical spectral epochs.
These spectral time-domain programs aim to sample a broad range of timescales and cadences, ranging from days to decades (when all SDSS data are combined), as different aspects of quasar physics result in variability on very different time scales. SDSS-V studies black hole masses via reverberation mapping, possible SMBH binarity, time-resolved accretion events, broad line region (BLR) dynamics, and Broad Absorption Line Quasi-Stellar Object (BALQSO) outflows. The SDSS-V quasar time-domain programs build on the results and experience of earlier generations of SDSS, e.g., the SDSS-III/IV reverberation mapping project \citep[RM;][]{Shen2015} and the Time Domain Spectroscopic Survey \citep[TDSS;][]{Macleod2018}, to enable long time baselines.

\subsubsection{All-Quasar Multi-Epoch Spectroscopy}
\label{sec:bhm:aqmes}
The All-Quasar Multi-Epoch Spectroscopy (AQMES) program is core to the BHM science, with different tiers in survey area and number of epochs. It includes cartons that are aimed at wide areas and low cadence, and cartons aimed at a more modest area with higher cadence (see Table~\ref{tab:bhm_cartons}). Together these two tiers add new epochs in SDSS-V for $\sim$22,000 quasars that already have at least one previous epoch of spectroscopy from SDSS-I--IV. These AQMES targets are selected from the SDSS DR16 quasar catalog \citep{Lyke2020} and are readily observable from the Apache Point Observatory.

For $\sim$20,000 known quasars,
%to be targeted under the `AQMES-Wide' core program, 
the primary associated targeting carton is \texttt{bhm\_aqmes\_wide2}. 
This carton aims to add $\sim$2 additional epochs of SDSS-V optical spectroscopy for each of these known SDSS quasars. When combined with their archival SDSS optical spectra these data will sample $\sim$1--25~yr timescales (observer frame). The primary science goals include probing the BLR dynamics of the most massive black holes, constraining the statistics of changing look quasars, and charting broad absorption line (BAL) disappearance/emergence. 
This is a relatively wide area, but low-cadence time-domain tier, encompassing $\sim$2000--3000 deg$^2$ of the sky.

For $\sim$2000 quasars,
%to be targeted under the `AQMES-Medium' core program, 
the \texttt{bhm\_aqmes\_medium} carton (Table \ref{tab:bhm_cartons})
aims to add $\sim$10 optical spectral epochs in SDSS-V, probing down to 1-month to 1-year timescales (in addition to the longer baseline timescales enabled with archival SDSS spectroscopy). 
The primary science goals of AQMES-Medium are to trace out BLR structural and dynamical changes, including BLR changes in modest mass SHMBs. 
This is a medium time-domain tier in area and cadence, encompassing $\sim$200-300~deg$^2$.

\subsubsection{Reverberation Mapping}
\label{sec:bhm:rm}
The Reverberation Mapping (RM) program is core to the BHM science of measuring black hole masses.

For $\sim$1000 quasars in 4--5 dedicated fields to be targeted under the BHM RM program, a set of cartons (with names starting with \texttt{bhm\_rm}, Table~\ref{tab:bhm_cartons}) aim to obtain optical repeat spectra with a high cadence of up to $\sim$174 epochs, which sample down to (observer frame) timescales of days to weeks. The time lags between the continuum and BLR emission, plus line velocity widths, yield virial estimates of the black hole mass, advancing RM measures yet further to a broad range of luminosity and redshift. This is a small-area but high-cadence time-domain tier, encompassing a total of $\sim$30 deg$^2$.

\subsection{Spectroscopic follow-up of X-ray sources}
\label{sec:bhm_xray_followup}
In addition to the time-domain science projects described above, BHM is carrying out a 
program of spectroscopic characterization of counterparts to an unprecedentedly large sample of X-ray sources. In almost all cases this will be via optical spectroscopy. The broad science emphasis is to chart the astrophysics, and the growth and evolution, of SMBHs. The X-ray selection provides a probe of accretion onto SMBHs that is significantly less sensitive to intervening absorption, and that generally selects a broader range of AGN luminosities, than purely optical-based selections. 
BHM features a large core program aimed at eROSITA X-ray sources (\S\ref{sec:bhm:spiders}) and a substantial complementary program following up {\it Chandra} archival X-ray source catalogs (\S\ref{sec:bhm:csc}).

\subsubsection{Spectroscopic Identification of eROSITA Sources}
\label{sec:bhm:spiders}
The SPectroscopic Identification of ERosita Sources (SPIDERS) program is core to BHM science and will help reveal the connections between statistical samples of X-ray emitting quasars/AGN and clusters of galaxies, and the large mass structures that they trace. The BHM SPIDERS program expands greatly on the SDSS-IV SPIDERS program \citep[e.g.,][]{Clerc2016,Dwelly2017,Comparat2020}.

The main goal of the SPIDERS program (sometimes referred to as the ``(Southern) Hemisphere'' survey) is to provide complete and homogeneous optical spectroscopic follow-up of $\sim$10$^{5.5}$ X-ray sources 
detected by eROSITA, across $\sim$10,000~deg$^2$ at high Galactic latitude (nominally $|b|>15$~deg), within the German or ``DE'’ half of the sky. Ultimately, SPIDERS, via its wide-area survey, aims to obtain the optical spectroscopic identifications and redshifts, evolution, and astrophysics of $\sim$250,000 counterparts to X-ray point-like sources. It is anticipated that AGN/quasar identifications will form the large majority of the targets, but with a significant minority of X-ray emitting stars (from compact binaries to coronal emitters). 
Eventually, SPIDERS will use X-ray sources from eROSITA’s first 1.5~years of survey operation (termed ``eRASS:3'' because it contains data from three passes over the whole sky), which were completed in June 2021. 
The SPIDERS ``AGN'' project cartons have names that start with \texttt{bhm\_spiders\_agn} (Table~\ref{tab:bhm_cartons}).
However, note that the initial SPIDERS targeting information in \texttt{v0.5.3} released in DR18 is derived from the somewhat shallower first 6~months of eROSITA-DE observations (``eRASS:1''). 

An additional SPIDERS project is targeting $\sim$10$^4$ X-ray emitting clusters of galaxies ($>5\times 10^4$ galaxy targets), selected from a combination of X-ray imaging and the eROSITA red-sequence Matched-filter Probabilistic Percolation galaxy cluster finding algorithm \citep[eROMaPPeR;][]{Rykoff2014,IderChitham2020} applied to multi-band wide+deep optical/IR imaging catalogs. For the SPIDERS clusters cartons (with names including \texttt{bhm\_spiders\_clusters}), the new SDSS-V spectroscopy will provide redshifts and confirmations for the candidate clusters, as well as constraints on the velocity dispersion of the member galaxies. See \citet{Bulbul_2022_erositagroups} for additional discussion on the complexity of selecting galaxy clusters in X-ray data sets.

A related pilot survey that began in SDSS-IV, and extended into early SDSS-V, has already largely completed BOSS spectroscopy for $\sim$10$^4$ candidate counterparts in the eROSITA-eFEDS mini-survey field, a $\sim$140~deg$^2$ region of the sky. The optical spectra in eFEDS are a centerpiece of the DR18 data release, and are detailed further in \S\ref{sec:efeds}.

\subsubsection{Chandra Source Catalog}
\label{sec:bhm:csc}
The Chandra Source Catalog (CSC) program is a more modest, complementary program of mainly optical spectroscopy of X-ray counterparts (but including some IR APOGEE spectra as well). The SDSS-V/CSC program produces identifications, redshifts, and other properties of tens of thousands of X-ray source counterparts selected from the Chandra Source Catalog\footnote{\url{https://cxc.cfa.harvard.edu/csc/}} \citep{Evans2010}; again many are likely to be verified as AGN in their SDSS-V spectra. The CSC sample is expected to probe fainter X-ray fluxes than the SPIDERS Hemisphere samples.
The CSC target cartons released as part of DR18 are based on CSC2.0. In the future we plan to switch to the more recent CSC2.1 catalog in order to increase the available pool of targets.

\begin{deluxetable*}{llccc}
\label{tab:bhm_cartons}
\tablecaption{Black Hole Mapper Core Cartons$^1$}
\tablehead{\colhead{\multirow{2}{*}{Carton Name}} & \colhead{\multirow{2}{*}{Simplified Selection\tablenotemark{1}}} 
& \colhead{No. Epochs\tablenotemark{2}} &  \colhead{No. Targets} \\ [-0.2cm]
\colhead{} &  
& \colhead{Anticipated} & \colhead{Anticipated}}
\startdata
\texttt{bhm\_aqmes\_wide2} & \multirow{2}{*}{SDSS DR16Q, $16<i_{\mathrm{psf}}<19.1$ AB} & $\ge 2$ & $\sim$20,000 \\
\texttt{bhm\_aqmes\_med}   & & $\ge 10$ & $\sim$2000 \\
\hline
\texttt{bhm\_rm\_known\_spec} & & \multirow{4}{*}{100--174} & \multirow{4}{*}{$\sim$1400} \\
\texttt{bhm\_rm\_core} & \multirow{1}{*}{Confirmed and candidate QSOs/AGN} & & & \\
\texttt{bhm\_rm\_var} & \multirow{1}{*}{$15<i_{\mathrm{psf}}<21.7$ AB (or $16 < G < 21.7$ Vega)} & & & \\
\texttt{bhm\_rm\_ancillary} & & & & \\
\hline
\texttt{bhm\_spiders\_agn\_lsdr8} & eROSITA pointlike; $r_{\mathrm{fiber}}<22.5$ or $z_{\mathrm{fiber}}<21$ AB & \multirow{2}{*}{1} & \multirow{2}{*}{$\sim$250,000} \\
\texttt{bhm\_spiders\_agn\_ps1dr2} & $F_{0.5-2\mathrm{keV}}>2\times 10^{-14}$ erg\,s$^{-1}$\,cm$^{-2}$&  &  \\
\hline
\texttt{bhm\_spiders\_clusters\_lsdr8} & eROSITA + red sequence finder & \multirow{2}{*}{1} & \multirow{2}{*}{$\sim$50,000} \\
\texttt{bhm\_spiders\_clusters\_ps1dr2} & $r_{\mathrm{fiber}} < 21$ or $z_{\mathrm{fiber}} < 20$ AB & & & \\
\enddata
\tablecomments{
\tablenotetext{1}{For more complete selection details, see Appendix~\ref{sec:bhm_cartons_appendix}  or online documentation: \url{https://www.sdss.org/dr18/bhm/programs/cartons/}.}
\tablenotetext{2}{Each BHM epoch typically relies on a 1-2~hr BOSS exposure.}
}
\end{deluxetable*}

\subsection{The SDSS-V/eFEDS Mini-survey}
\label{sec:bhm:efeds}
The SDSS-V/eFEDS survey 
is a pathfinder component of the BHM survey program that exploits early performance validation observations from the SRG/eROSITA X-ray telescope in the eFEDS field \citep[][]{Brunner2022}. 
The eFEDS X-ray footprint comprises 140~deg$^2$ centered near $(\alpha, \delta) = (9^h, +1^\circ)$, 
encapsulating the ``GAMA09'' Field \citep{Driver2009}. We used plate observations at APO 
to collect optical BOSS spectroscopy for counterparts to both point-like and extended X-ray sources \citep{Salvato2022,Klein2022}. 
From past
experience \citep[e.g.,][]{Menzel2016,LaMassa2019, Clerc2016, Clerc2020}, one can expect these X-ray populations to be numerically dominated by AGN and clusters of galaxies, but we also expect a substantial minority to be associated with stellar coronal emitters or accreting compact objects.

Below we give a summary of the observational goals,
target selection, and survey strategy for the SDSS-V/eFEDS
observations. Details of the plate design and the scope and quality of the observed dataset can be found in \S\ref{sec:efeds}.

\subsubsection{SDSS-V/eFEDS Observational goals}

The primary observational goal for this program was to achieve near-complete and reliable
redshifts and classifications for optical counterparts to X-ray sources that were detected
as part of the eROSITA/eFEDS performance verification survey,
particularly for counterparts in the magnitude range $16 < r < 22$~AB. 
On one hand, this goal was constrained by the
number of hours of dark observing time available to the
project, a general desire to minimize the total number of drilled plug
plates, and the overall capabilities of the plate system to
place fibers on naturally clustered targets.
On the other hand, this goal of high completeness was assisted by the wealth of previous
spectroscopic survey data in the eFEDS field, both from previous SDSS
generations and from other telescopes and instruments. Our strategy was
therefore to prioritize targets that did not have existing high
quality spectroscopic observations.

\subsubsection{Target selection for SDSS-V/eFEDS plates}
\label{sec:efeds_targeting}

The generation of targeting used for the eFEDS plates, during the Dec 2020--May 2021 run of plate operations predates the global \texttt{v0.5.3} targeting information that is being released in DR18. 
The parent catalogs from which the eFEDS targets were selected were derived
from early reductions of the eROSITA/eFEDS X-ray dataset, and are based on early
attempts to match those X-ray sources to longer-wavelength
counterparts provided by several supporting optical/IR catalogs,
including the DESI Legacy Survey \citep[DR8;][]{Dey_2019_DESIsurveys}, 
SDSS DR13 \citep{Albareti2017_sdss_dr13}, and the Hyper SuprimeCam Subaru Strategic Project \citep[HSC-SSP DR2;][]{Aihara2019}. 
The eFEDS science target selection process concentrates on AGN candidates
(one target carton) and galaxy cluster candidates (four cartons):

\begin{description}

\item[\texttt{bhm\_spiders\_agn-efeds}] 
A carton that contains candidate AGN targets found in the
eROSITA/eFEDS X-ray survey field. This carton provides optical
counterparts to point-like (unresolved) X-ray sources detected in
early reductions (eROSITA Science Analysis Software, eSASS, version ``c940/V2T'') 
of the eROSITA performance validation
survey data in the eFEDS field. The sample is expected to contain a
mixture of QSOs, AGN, stars and compact objects. The X-ray sources
have been cross-matched by the eROSITA-DE team to the Legacy Survey\footnote{\url{https://legacysurvey.org/dr8}}
optical/IR counterparts \citep{Salvato2022}.

\item[\texttt{bhm\_spiders\_clusters-efeds-ls-redmapper}]
A carton that contains galaxy cluster targets found in the
eROSITA/eFEDS X-ray survey field. The carton provides a list of
galaxies that are candidate members of clusters selected from early
reductions (eSASS version ``c940'') of the eROSITA performance verification survey data. 
The parent sample of galaxy clusters and their member
galaxies have been selected via a joint analysis of X-ray and
(multiple) optical/IR datasets using the eROMaPPeR red-sequence finder
algorithm \citep{Rykoff2014, IderChitham2020}. This particular carton relies on optical/IR data from
the DESI Legacy Surveys \citep{Dey_2019_DESIsurveys}.

\item[\texttt{bhm\_spiders\_clusters-efeds-sdss-redmapper}]
Similar to the bhm\_spiders\_clusters-efeds-ls-redmapper carton,
except that the red sequence finder algorithm is run on the SDSS DR13
photometric catalogue \citep{Albareti2017_sdss_dr13}.

\item[\texttt{bhm\_spiders\_clusters-efeds-hsc-redmapper}]
Similar to the bhm\_spiders\_clusters-efeds-ls-redmapper carton,
except that the red sequence finder algorithm is run on the HSC-SSP DR2 photometric
catalogue \citep{Aihara2019}.

\item[\texttt{bhm\_spiders\_clusters-efeds-erosita}]
This carton includes manually-identified counterparts to X-ray
extended sources that were not also selected by the eROMaPPeR
algorithm, when applied to any of the DESI Legacy Survey, SDSS--DR13, or
HSC-SSP datasets.
\end{description}

Additional details on these cartons can be found in Appendix~\ref{sec:bhm_cartons_appendix} and at \url{https://www.sdss.org/dr18/bhm/programs/cartons/}.

\section{MOS Open-Fiber Programs} 
\label{sec:open_fiber_programs}

The hardware and the survey strategy for SDSS-V's MOS programs were designed around a set of core science cases that define this part of the survey. The underlying target densities in these science cartons vary dramatically across the sky, from star-rich Baade's Window towards the Galactic bulge to sparse high galactic latitudes. In addition, the FPS is subject to geometric constraints in placing fibers within a pointing's field of view: in a given individual exposure, fibers cannot be placed closer than $\sim$50$^{\prime\prime}$ to each other, and --- given the limited patrol radius of each robot --- the mean fiber density across the field of view cannot be changed.
These constraints imply that after survey optimization for SDSS-V's core science, many robots with either BOSS or APOGEE fibers remain unallocated. After simulating the entire survey, over three million anticipated fiber-exposures remained available.

In light of this, SDSS-V set up an ``open fiber'' program, based on multiple collaboration-wide calls for ideas and proposals, that serves three purposes. First, and most immediately, to make sure that as few fibers as possible go scientifically unused.  Second, to provide a mechanism to broaden the science case beyond the initial, defining science cases. And third, to provide a mechanism, through repeated open fiber proposal calls, to incorporate scientifically exciting targets that only become known or available as the survey goes on.

In 2020, SDSS-V issued a first call for open fiber proposals from the consortium, whose results have entered the targeting documented here in various ways. Boundary conditions for these proposals were that i) the science goals could be reached with single 15~minute exposures using either APOGEE or BOSS; ii) the targeting information was freely available to all consortium members, and a target selection function could be specified; iii) the existing core science goals were complemented or enhanced; and iv) the science goals had proponents within the SDSS-V consortium.  

The project received nearly 30 proposals in late 2020. Much of the proposed science fell into one of three broad categories: 1) more extensive surveying of AGNs and X-ray sources across the sky; 2) broader stellar physics, in particular binary star physics, across the CMD; and 3) studying the Milky Way's stellar halo and metal-poor population, which (intentionally) had not been a focus of the MWM Galactic Genesis program (\S\ref{sec:mwm_target_cartons}).  These programs were reviewed in early 2021 by the SDSS-V technical advisory group and the project leadership, who called on three consortium members not involved in the proposals to help with the review. The resulting target cartons are listed in Table~\ref{tab:open_fiber_cartons}. 
{For the \texttt{v0.5.X} versions of the targeting, including the \texttt{v0.5.3} generation released in DR18, we adopted these cartons as-is. In future versions of the target selection, a number of these cartons may be consolidated and incorporated into the versioned science programs of the BHM and MWM surveys.}

\begin{deluxetable*}{llll}
\label{tab:open_fiber_cartons}
\tablecaption{Open Fiber Cartons}
\tablehead{\colhead{\multirow{2}{*}{Carton Name}} & \colhead{\multirow{2}{*}{Selection Summary\tablenotemark{1}}} & 
\colhead{\multirow{2}{*}{Instrument}} & \colhead{Available} \\ [-0.2cm]
\colhead{} & \colhead{} & \colhead{} & \colhead{Targets\tablenotemark{2}}}

\startdata
openfibertargets\_nov2020\_3 & Binaries - Bright Eclipsing Binaries & BOSS & 34689 \\ 
openfibertargets\_nov2020\_5 & Galactic Halo - $\rm [Fe/H] < -3$ Candidates & BOSS & 39099 \\ 
openfibertargets\_nov2020\_6a & Galactic Halo - SkyMapper BHB & BOSS & 50611 \\ 
openfibertargets\_nov2020\_6b & Galactic Halo - SkyMapper RGB Metal-Poor & BOSS & 340436 \\ 
openfibertargets\_nov2020\_6c & Galactic Halo - SkyMapper Dwarf Metal-Poor & BOSS & 890043 \\ 
openfibertargets\_nov2020\_8 & Young/Massive Stars - Young Galactic Disk & BOSS/APOGEE & 704638 \\ 
openfibertargets\_nov2020\_9 & Young/Massive Stars - F Star Physics & BOSS/APOGEE & 1197354 \\ 
openfibertargets\_nov2020\_10 & Binaries - RVs in chemically anomalous stars & BOSS/APOGEE & 917 \\ 
openfibertargets\_nov2020\_11 & All QSOs/AGN - SDSS-I/II Quasars observed only once & BOSS & 39684 \\ 
openfibertargets\_nov2020\_12 & Binaries - Close Binary Stars & BOSS/APOGEE & 33577 \\ 
openfibertargets\_nov2020\_14 & Binaries - MS-WD Binaries & BOSS & 10798 \\ 
openfibertargets\_nov2020\_15 & Open Cluster Survey & BOSS/APOGEE & 177185 \\ 
openfibertargets\_nov2020\_17 & Young/Massive Stars - Census of YSOs $<$ 500pc & BOSS & 32891 \\ 
openfibertargets\_nov2020\_18 & All QSOs/AGN - Hard X-Ray Sources & BOSS & 10012 \\ 
openfibertargets\_nov2020\_19a & TESS 2min cadence targets - a & BOSS/APOGEE & 224773 \\ 
openfibertargets\_nov2020\_19b & TESS 2min cadence targets - b & BOSS/APOGEE & 1605 \\ 
openfibertargets\_nov2020\_19c & TESS 2min cadence targets - c & BOSS/APOGEE & 1615 \\ 
openfibertargets\_nov2020\_22 & Binaries - Ellipsoidal Binaries & APOGEE & 28408 \\ 
openfibertargets\_nov2020\_24 & Chemodynamics in the Solar Neighborhood & BOSS & 1323442 \\ 
openfibertargets\_nov2020\_25 & Galactic Halo - Local Halo & BOSS/APOGEE & 293741 \\ 
openfibertargets\_nov2020\_26 & All QSOs/AGN - Quasar Winds & BOSS & 2105 \\ 
openfibertargets\_nov2020\_27 & All QSOs/AGN - All Bright Quasars & BOSS & 814963 \\ 
openfibertargets\_nov2020\_28a & Galactic Halo - Distant K Giants & BOSS & 73627 \\ 
openfibertargets\_nov2020\_28b & Galactic Halo - Distant BHBs & BOSS & 49900 \\ 
openfibertargets\_nov2020\_28c & Galactic Halo - Distant RRLs & BOSS & 51234 \\ 
openfibertargets\_nov2020\_29 & Extra-tidal Globular Cluster Stars & BOSS/APOGEE & 107688 \\ 
openfibertargets\_nov2020\_30 & All QSOs/AGN - JWST North Ecliptic Pole Time Domain & BOSS & 237 \\ 
openfibertargets\_nov2020\_31 & RAVE Cross Calibration & APOGEE & 62275 \\ 
openfibertargets\_nov2020\_32 & Binaries - Wide MS-WD Binaries & BOSS/APOGEE & 34230 \\ 
openfibertargets\_nov2020\_33 & All QSOs/AGN - Variability Selected AGN and Blazars & BOSS & 22250 \\ 
openfibertargets\_nov2020\_34a & CMD tiling - nonvariables & BOSS & 8701166 \\ 
openfibertargets\_nov2020\_34b & CMD tiling - variables & BOSS & 643546 \\ 
openfibertargets\_nov2020\_35a & Galactic Halo - Metal-poor giants (infrared) & BOSS & 345721 \\ 
openfibertargets\_nov2020\_35b & Galactic Halo - Metal-poor giants (SkyMapper) & BOSS & 52696 \\ 
openfibertargets\_nov2020\_35c & Galactic Halo - Metal-poor giants (SAGES) & BOSS & 7521 \\ 
openfibertargets\_nov2020\_46 & Binaries - LAMOST multi-epoch & BOSS & 663835 \\ 
openfibertargets\_nov2020\_47a & SDSS-V Magellanic Genesis: RGB & BOSS & 278987 \\ 
openfibertargets\_nov2020\_47b & SDSS-V Magellanic Genesis: AGB & APOGEE & 27731 \\
openfibertargets\_nov2020\_47c & SDSS-V Magellanic Genesis: Massive & BOSS & 1000 \\ 
openfibertargets\_nov2020\_47d & SDSS-V Magellanic Genesis: Massive & APOGEE & 1000 \\ 
openfibertargets\_nov2020\_47e & SDSS-V Magellanic Genesis: Symbiotic & APOGEE & 24 \\ 
openfibertargets\_nov2020\_1000 & Stellar Clusters & BOSS/APOGEE & 102172 \\ 
openfibertargets\_nov2020\_1001a & Binaries - Wide Binaries - Merged 21/36a & BOSS/APOGEE & 81303 \\ 
openfibertargets\_nov2020\_1001b & Binaries - Wide Binaries - Merged 36/36b & BOSS & 10545 \\ 
\enddata

\tablecomments{\tablenotetext{1}{See online documentation for expanded selection details: \url{https://www.sdss.org/dr18/targeting/open_fiber_programs/}} \tablenotetext{2}{\textbf{Available targets} is the number of targets that satisfy the carton selection function in the targeting database. The number of targets ultimately observed for each carton will be smaller than this value.}} 
\end{deluxetable*}

\section{Spectra}
\label{sec:spectra}

This section describes the spectra being released in DR18, particularly the eFEDS survey and its data products.

\subsection{The SDSS-V/eFEDS mini-survey} 
\label{sec:efeds}
The scientific objectives and carton design of the eFEDS program can be found in \S\ref{sec:bhm:efeds}. Below we describe the observational design details and the final data set.

\subsubsection{Tiling and plate design}

Due to COVID-19-induced delays in the completion of the FPS systems, SDSS-V started observations at APO using the existing fiber plugplate system, including the eFEDS observations. Modifications to the existing system included a new joint BOSS+APOGEE configuration, in which the focal plane was populated with 500
fibers feeding a BOSS optical spectrograph and 300
fibers feeding an APOGEE IR spectrograph. New
procedures and software were put in place to manage target selection,
fiber assignment and plate design for the single season of SDSS-V data that
were obtained in this mode. These procedures will be described in more detail in
a future publication (Covey et al.\,in prep). 
Here we provide a summary of special considerations that
apply specifically to the SDSS-V/eFEDS plates.

\begin{figure*}[!hptb]
\begin{center}
\includegraphics[width=\textwidth]{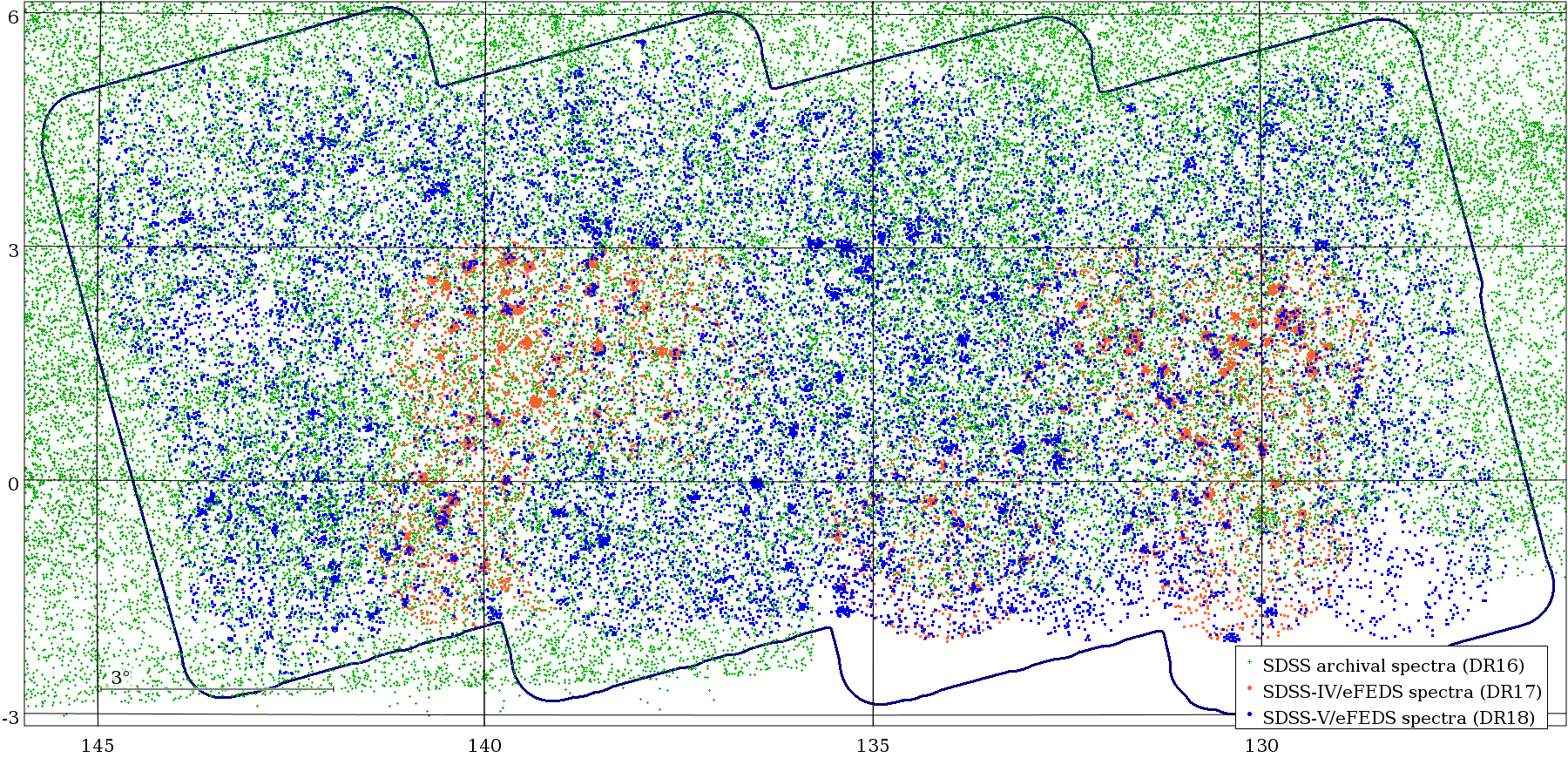}
\caption{
SDSS spectroscopic coverage in the eFEDS field (\S\ref{sec:efeds}), color-coded by the SDSS phase when the spectrum was taken. Green points %(and green $\sim$3$^\circ$ diameter circles) 
show the locations of individual SDSS/BOSS spectra %(and plates) 
made available in DR16 or before; orange corresponds to SDSS-IV/eFEDS BOSS spectroscopy released in DR17, and blue shows the new optical BOSS spectroscopy being released in DR18. X-ray-selected galaxy clusters, specifically targeted in DR17 and DR18, are readily visible as compact concentrations of points. The approximate bound of the {\it eROSITA} X-ray data is shown with the thick black line. The southwest (lower right) corner is empty of DR16 spectroscopy because it lies outside the nominal SDSS imaging footprint. The map is shown on a tangential projection, in Equatorial coordinates, with units of degrees.
}
\label{fig:efeds_target_map}
\end{center}
\end{figure*}

Each plate-based observing run comprises a set of plates created to observe a set
of targets from a given field (a region of the sky) and drilled
to observe around a specific hour angle (to account for atmospheric
refraction effects). The targets on a given plate were assigned using a
specific combination of cartons (\S\ref{sec:efeds_targeting}).
For each plate, we created a prioritized list of cartons (including cartons associated with 
calibration targets such as skies and spectrophotometric standards, \S\ref{sec:mos_cartons}) to fill the fibers for each spectrograph. 
%This means that to fill APOGEE and BOSS fibers we covered the corresponding list of cartons in priority order until all fibers for each instrument were filled. 
Individual fiber filling rules were shared by multiple plates in the same observing run or even across
different runs; different fiber filling rules are distinguished by shifting in priority of a carton with respect to another, or an update on the version of the carton that was used. Each combination of fiber
filling rule and unique set of targets was identified with a \texttt{designID}.
Figure \ref{fig:efeds_target_map} shows the sky locations of the SDSS-V/eFEDS sources
in relation to the other data sets available in the region.

SDSS-V/eFEDS plates were designed within two independent
iterations of the tiling and plate design process. The first of these
plate runs (\texttt{2020.11.a.bhm-mwm}) consisted of 31 initial plates
spanning most of the $\sim$140\,deg$^2$ eFEDS field, with an inner region
(roughly corresponding to the GAMA09 field) 
targeted with two plates per sky position. The large fractional
overlap between the eFEDS plates required special attention to ensure
proper control over which targets were prioritized in which plates, as the standard SDSS \texttt{platedesign} software treats all plates
independently. 
This complexity, and constraints on the fiber reach of the plate fibers, 
resulted in only 20 of the \texttt{2020.11.a.bhm-mwm} plates being available for observation.
A second eFEDS plate run (\texttt{2021.01.a.bhm-mwm}) was designed to
recover eFEDS targets from the unobservable plate designs and to take
advantage of updated projections of the available observing
time and plug-plate manufacturing resources. 
A new heuristic tiling algorithm resulted in all 17 plates from this run being observed.

For each of these 37 SDSS-V/eFEDS plates, we reserved at least 80 BOSS
fibers per plate for skies and 20 BOSS fibers for spectrophotometric
standard stars, leaving up to 400 BOSS fibers for science targets\footnote{The SDSS-V/eFEDS plates also included up to 300 APOGEE targets per plate;
these will be presented in a future SDSS data release (\S\ref{sec:other_spectra}).}. 
We applied a bright limit for science
targets of 
%${\rm psfmag}_{g,r,i} > 16.5$\,AB, 
$g_{\rm psf}, r_{\rm psf}, i_{\rm psf} > 16.5$\,AB, 
and we avoided placing fibers near brighter stars.
These bright limits are imposed to allow observation of the faint end of the target population 
(Figure~\ref{fig:efeds_histos}) without significant degradation from on-chip cross-talk between
neighbouring BOSS fiber traces.

\subsubsection{Spectroscopic observations, reductions, and data quality}
\label{sec:efeds_data}

The SDSS-V/eFEDS observing strategy prioritized coverage over depth, given
the somewhat uncertain amount of good quality observing time that
would be available whilst the eFEDS field was visible from APO. By the
end of the visibility window (Dec 2020--May 2021), all 37 eFEDS plates had achieved a
total fiducial $g$-band signal-to-noise ratio squared (${\rm SNR^2_g}$) of at least 7.5. However, most of the
plates are significantly deeper than that, with 33/37 of the eFEDS
plates having ${\rm SNR^2_g} > 10$ \citep[comparable to the typical exposure depth in BOSS/eBOSS;][]{Dawson_2016_eboss}, and more than half the plates having ${\rm SNR^2_g} > 20$. 
The fiducial SNR$^2$ in the $i$-band is typically twice that
measured in the $g$-band.
An updated version of the BOSS
\texttt{idlspec2d} data reduction pipeline \citep[][see \S\ref{sec:software_boss} for updates]{Bolton2012} was used to
generate a set of spectral data products per \texttt{PLATE-MJD} combination,
with individual 1D spectra for each combination of
\texttt{PLATE-MJD-CATALOGID}.

Many science targets were observed on more
than one \texttt{PLATE-MJD} combination. 
Therefore, a specialized co-adding algorithm (within
\texttt{idlspec2d}) was additionally implemented for these eFEDS plates, collating
spectroscopic data across plates and MJDs, in order to increase SNR in
the 1D spectra (\S\ref{sec:software_boss}). This routine creates a single
stacked spectrum per astrophysical object, using all
suitable SDSS-V BOSS data. 
A total of 13269 specially coadded spectra from eFEDS science targets are released in DR18, along with 2608 skies, 671 spectrophotometric standards, and several hundred unvetted spectra not intended for scientific use (\S\ref{sec:other_spectra}).

Redshifts, classifications, and quality flags were computed
automatically on the specially coadded spectra using the
\texttt{spec1d} template fitting pipeline \citep{Bolton2012}. A warning flag was set
for 669/13269 science spectra, indicating a problem with the data or
the fit. Of the 12600 science spectra without warnings, the main
pipeline gives source classifications of `QSO' for 6204, `GALAXY' for
4782 and `STAR' for 1614 spectra. Pipeline redshifts for non-flagged
`QSO' and `GALAXY' classified spectra span the range $0.0 < z < 4.5$,
with 90\% of these objects falling in the range $0.14 < z < 2.55$
(median 0.55).

\begin{figure}
\begin{center}
\includegraphics[width=8.5cm]{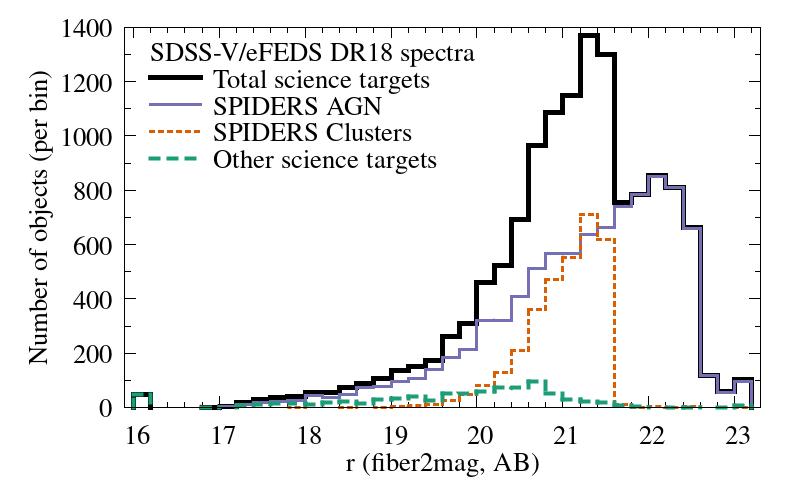}
\includegraphics[width=8.5cm]{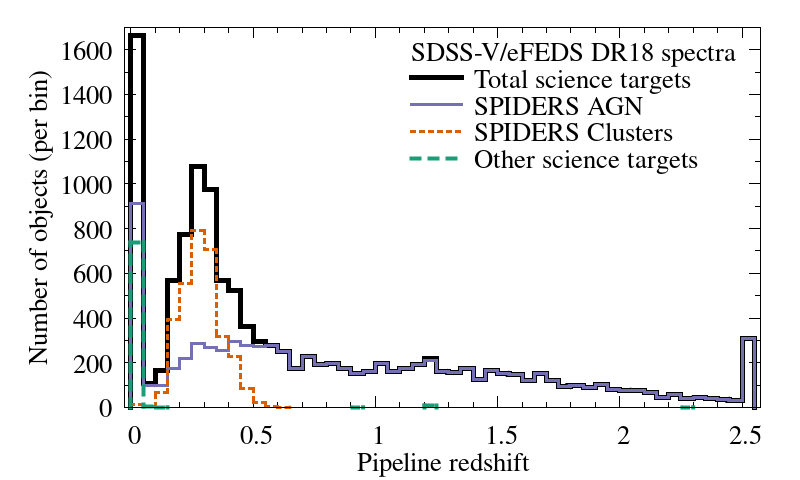}
\caption{
\textit{Upper panel:} Optical magnitude distribution of BOSS science targets in the SDSS-V/eFEDS plates (\S\ref{sec:efeds}). Note that target magnitudes in SDSS-V are synthesized from a range of input photometric systems; in this case we show the equivalent of an $r$-band `fiber2mag' (i.e., the light falling into a 2\,arcsec diameter fiber under nominal APO seeing conditions). \textit{Lower panel:} Distribution of pipeline-determined redshifts for science targets in the SDSS-V/eFEDS plate reductions (\S\ref{sec:efeds_data}). Spectra with redshift fitting warnings have been removed. Redshifts outside the range of the plots are included in the extremal bins. Target selection for SPIDERS AGN and Clusters is described in \S\ref{sec:efeds_targeting}, and for other targets in \S\ref{sec:other_spectra}.
}
\label{fig:efeds_histos}
\end{center}
\end{figure}

Figure \ref{fig:efeds_histos} shows the derived redshift distribution for good quality (\texttt{ZWARNING}=0) spectra obtained from the 
SDSS-V/eFEDS plates, grouped by their primary target selection criteria. The observed redshift distributions are 
broadly in line with expectations. SPIDERS AGN targets span a wide range of redshifts ($0<z<4.5$, median 0.8), including 
a small fraction (10.0\%) that are revealed by spectroscopy to be stellar in nature. SPIDERS galaxy cluster targets 
are found at relatively lower redshift ($0.1<z<0.6$, median 0.28).

In order to increase the completeness and reliability of the
redshifts and classifications derived from these spectra, we
carried out visual inspections of a large fraction of the
SDSS-V/eFEDS spectra. Those inspections have been combined with
earlier SDSS spectroscopic redshift information and with non-SDSS redshift
information gathered from the literature. Please see Merloni et al., (in
prep.), and the associated Value Added Catalog\footnote{\url{https://www.sdss.org/dr18/data_access/value-added-catalogs/}} (\S\ref{sec:vacs_dr18}) for further details of the eFEDS spectroscopic compilation.

This data set has enabled in-depth study of the clustering of X-ray selected AGN, resulting in new constraints on the 
AGN halo occupation distribution \citep[][]{Comparat2023}.

\subsubsection{Accessing eFEDS Data}
\label{sec:spectra_efeds_access}
To download the eFEDS spectra, go to \url{https://dr18.sdss.org/sas/dr18/spectro/boss/redux/eFEDS/}; the associated data models are at \url{https://data.sdss.org/datamodel/files/BOSS_SPECTRO_REDUX/RUN2D/} and subdirectories. Note that these spectra were generated using the specialized co-adding scheme described in \S\ref{sec:efeds_data}. The ``standard'' coadds can be found at \url{https://dr18.sdss.org/sas/dr18/spectro/boss/redux/v6_0_4/}. To download the eFEDS VAC of improved redshifts (\S\ref{sec:vacs_dr18}), visit the DR18 VAC website: \url{https://www.sdss.org/dr18/data_access/value-added-catalogs/}.

\subsection{Other Spectra}
\label{sec:other_spectra}
Users of DR18 may come across science spectra (i.e., not standard stars or sky targets) 
that were included in the eFEDS plates (\S\ref{sec:efeds}) but that are not associated with 
eROSITA/eFEDS target cartons.  These data are included in DR18 for logistical 
reasons but are not intended for scientific exploitation (e.g., they have not undergone any quality assurance testing). 
They can be identified by one of the following \texttt{FirstCarton} labels in the database:
\begin{itemize} \itemsep -2pt
    \item \texttt{mwm\_cb\_cvcandidates} %MWM\_CB\_CVCANDIDATES
    \item \texttt{mwm\_cb\_gaiagalex} %MWM\_CB\_GAIAGALEX
    \item \texttt{mwm\_cb\_uvex1} %MWM\_CB\_UVEX1
    \item \texttt{mwm\_cb\_uvex2} %MWM\_CB\_UVEX2
    \item \texttt{mwm\_cb\_uvex3} %MWM\_CB\_UVEX3
    \item \texttt{mwm\_cb\_uvex4} %MWM\_CB\_UVEX4
    \item \texttt{mwm\_cb\_uvex5} %MWM\_CB\_UVEX5
    \item \texttt{mwm\_halo\_bb} %MWM\_HALO\_BB
    \item \texttt{mwm\_halo\_sm} %MWM\_HALO\_SM
    \item \texttt{mwm\_snc\_100pc} %MWM\_SNC\_100PC
    \item \texttt{mwm\_wd} %MWM\_WD
\end{itemize}

Users can expect improved and vetted versions of these spectra in DR19.

\section{Software} 
\label{sec:software}
SDSS operations, data processing, and analysis rely on a large suite of software components. A limited number of these, new or updated for SDSS-V, have already been released or are being released as part of DR18.

\subsection{Updates to the BOSS Spectral Reduction Algorithms}
\label{sec:software_boss}
The optical BOSS data released in DR18 (\S\ref{sec:efeds}) are processed with version {\tt v6\_0\_4} of the BOSS pipeline software \texttt{idlspec2d} \citep[][Morrison et al.\,in prep]{Bolton2012,Dawson_2013_boss}. This is the first official version of \texttt{idlspec2d} to be used for SDSS-V, with a few changes since the last full public release (\texttt{v5\_13\_2}) of SDSS-IV/eBOSS. All SDSS-V versions of \texttt{idlspec2d} are available for download from the SDSS GitHub, with the version described here available at \url{https://github.com/sdss/idlspec2d/releases/tag/v6_0_4}.

One of the most significant operational changes to the pipeline is a modification to reduce 500 fibers from a single BOSS spectrograph, rather then a combined reduction of 1000 fibers from the twin BOSS spectrographs. This change was made to support the move of the second BOSS spectrograph to LCO.

A number of smaller changes were implemented into the pipeline to both improve ongoing reductions and facilitate future reductions of FPS/BOSS data.
First, to improve the radial velocity precision of the extracted spectra, a refined arc lamp linelist \citep[developed for the SDSS-IV MANGA project;][]{Bundy_2015_manga} replaced the list from SDSS-IV, and skylines are utilized as a second-level calibrator. This second step required an adjustment to the outlier exclusion algorithm (designed to exclude cosmic rays) to preserve the peaks of the skylines by increasing the outlier rejection from $4\sigma$ to $50\sigma$. This change, in some rare cases, might leave the remnant of a cosmic ray in the extracted spectra.

Second, we modified the coadding scheme of targets between exposure frames and the red/blue cameras; individual spectra are now grouped according to the RA,Dec coordinates of the fiber(s) (or the SDSS-V CatalogID, if one exists), rather than on the combination of BOSS \texttt{PLATE-MJD-FIBERID}.
The coadding scheme was also split into a two step process, in which the red and blue data for each exposure are combined, and then the all-exposures red and blue data are combined to produce coadds for the full observation epoch.
These changes were implemented to improve the signal-to-noise ratio of the final reduced spectra.

Third, \texttt{v6\_0\_4} was modified to produce a special version of the DR18 SDSS-V/eFEDS spectra (\S\ref{sec:efeds_data}), in addition to the existing standard pipeline results. This new combination includes every exposure of each target, regardless of the plate on which it was observed, to produce a set of spectra with maximum signal-to-noise. Both the standard pipeline, \texttt{v6\_0\_4}, and the \texttt{eFEDS} special coadds are available through the SAS\footnote{\url{https://dr18.sdss.org/sas/dr18/spectro/boss/redux}}; and the CAS using SkyServer\footnote{\url{http://casjobs.sdss.org/casjobs/}}.  Future data releases will include variations of this special coadding routine, with the spectra included in each coadd chosen to fulfil the science requirements of the individual SDSS-V science programs.

Fourth, in addition to the changes in the reduction and coadding schemes, some modifications were made to the \texttt{spec1d} analysis part of \texttt{idlspec2d}. Utilizing 849 SDSS-IV RM quasars \citep{Shen2019} and the Weighted \texttt{empca} package \citep{Bailey2012,Bailey_2016_empca}, a new set of QSO PCA templates are used for PCA reconstruction and redshift estimation of SDSS-V quasars. These new PCA templates are implemented with 10 eigenvectors instead of the 4 used in SDSS-I through SDSS-IV. To further support the MWM targets on the reduced eFEDs plates, on the observed MWM plates (not released in DR18 but documented here for completeness), and in future FPS observations, the SFD model for dust extinction \citep{Schlegel1998, Schlafly2011} for MWM plates was replaced by the 3D Bayestar2015 dust map \citep{Green2015} to properly account for differential dust extinction within the Milky Way. The package \texttt{pyXCSAO} \citep{marina_kounkel_2022_6998993}, a python replication of the \texttt{xcsao} package \citep{Kurtz1992,Mink1998,Tonry1979}, was added as part of the \texttt{spec1d} analysis pipeline for cross-correlating a spectrum against template spectra of known velocities. While this step is run for all targets, the results are only valid for stellar targets.

Finally, \texttt{idlspec2d}'s internal Python dependencies were updated to Python3, with the deprecation of Python~2.7.
All of changes described here improve either data quality or spectroscopic classification success rates, or support the changes to the BOSS spectrograph for the SDSS-V plate program and FPS observations.

\subsection{Migration to GitHub}

For SDSS-V, all of the SDSS software has been moved to a GitHub Organization\footnote{\url{https://github.com/sdss}}, from the previous Subversion private repository. With the exception of a few repositories containing proprietary code or credentials, all the code is publicly available under a BSD 3-Clause license. This includes most of the operational software, telescope and instrument control software, and data reduction pipelines.

\subsection{MOS Software for DR19}

The migration from plug-plates to robotic fiber positioners in SDSS-V required the development of a suite of new software for targeting and operations. Although this software was not used for the acquisition of DR18's spectra (\S\ref{sec:efeds}), it is currently in use for FPS operations and will become relevant for DR19 and future data releases. Here we provide a short summary of FPS targeting and operations as a preview for those data releases. Further details are given in \citet{SanchezGallego2020}.

Combining the complex set of observing constraints and epoch cadences defined by the various target selection cartons into a set to observable FPS configurations requires the use of an specialized algorithm. {\tt robostrategy} is the pipeline that takes the outputs of \texttt{target selection} and determines {\it how} the targets should be observed. It determines the cadence with which each field will be observed, which includes the number of epochs, the number of observations (``designs'') per epoch, and the desired observing conditions for each design (including, e.g., the sky brightness conditions and hour angle). At the beginning of each night of observations, the software {\tt roboscheduler} selects the optimal list of designs to be observed based on cadence requirements and observing conditions, as well as on previous observations and achieved completion.

One major challenge of the SDSS robotic fiber positioner system derives from the fact that the patrol radius of each positioner overlaps with its neighboring ones. For a given set of target-to-positioner assignments defined in a {\tt robostrategy} design, an algorithm named {\tt Kaiju} \citep{Sayres_2021_kaiju} provides a deterministic trajectory for each robotic positioner from a starting ``folded'' state, in which all robots are identically oriented, to the desired FPS configuration. {\tt Kaiju} works following a reverse-solve method in which it is considered simpler to calculate collision-free trajectories for each robot from a complex, deployed state, to a lattice-like folded configuration. To create a path between two observable configurations {\tt Kaiju} calculates the reverse trajectories from each configuration to the folded state, and then applies the reverse trajectory. Following this approach {\tt Kaiju} is able to determine deadlock-free trajectories between any two physically reachable robot configurations in over 99\% of cases \citep{Sayres_2021_kaiju}.

\section{Value-Added Catalogs}
\label{sec:vacs}

In addition to its fundamental data products, SDSS regularly includes ``Value Added Catalogs'' as part of its data releases\footnote{\url{https://www.sdss.org/dr18/data_access/value-added-catalogs/}}.  These catalogs contain curated quantities from the data, usually to enable a specific scientific projects, but recognized as having value above and beyond the particular project for which the catalog was initially constructed.  

\begin{deluxetable*}{lll}
\label{tab:vacs_dr17}
\tablecaption{New or Updated Value Added Catalogs}
\label{table:vac}
\tablehead{\colhead{Name} & \colhead{Reference(s)}}
\startdata
\multicolumn{2}{l}{\bf SDSS-IV APOGEE-2} \\ 
BACCHUS Analysis of Weak Lines in APOGEE Spectra & \citet{Masseron_2016_bacchus}, \citet{Hayes_2022_bawlas} \\
\tableline
\multicolumn{2}{l}{\bf SDSS-IV MaNGA} \\ 
MaNGA Dwarf Galaxy Sample (MaNDala) & \citet{CanoDiaz_2022_mangadwarfs} \\
MaNGA Visual Morphology & \citet{VazquezMata_2022_MANGAmorphVAC} \\
\tableline
\multicolumn{2}{l}{\bf SDSS-V Black Hole Mapper} \\ 
eFEDS Spectral Compilation VAC & \S\ref{sec:vacs_dr18}, Merloni et al. (in prep.) \\
\enddata
\end{deluxetable*}

\subsection{SDSS-IV VACs}
\label{sec:vacs_dr17}
Table~\ref{tab:vacs_dr17} includes the following VACs that rely on SDSS-IV data but are updated or published for the first time after DR17:

{\it BACCHUS Analysis of Weak Lines in APOGEE Spectra (BAWLAS)}: Weak-lined species in APOGEE spectra are challenging to measure and in some cases, cannot be done with the standard ASPCAP pipeline. 
This VAC contains approximately 120,000 high-SNR giants from APOGEE DR17, for which Na, P, S, V, Cu, Ce, and Nd abundances and $^{12}$C/$^{13}$C isotopic ratios have been derived with a specialized analysis \citep{Hayes_2022_bawlas}. An updated version of the code BACCHUS \citep[Brussels Automatic Code for Characterizing High accUracy Spectra;][]{Masseron_2016_bacchus} was used in the derivation, along with ASPCAP fundamental stellar parameters but also a dedicated set of individual line quality flags and a new prescription for identifying upper limits. \citet{Hayes_2022_bawlas} show how these newly derived abundances compare with literature values and provide examples of scientific projects that can be done with APOGEE and the new catalog, ranging from nuclear physics to Galactic chemical evolution and Milky Way population studies. 

{\it MaNGA Dwarf Galaxy Sample (MaNDala):}
This VAC presents properties for a sample of 125 dwarf galaxies ($M_{\star}<10^{9} M_{\odot}$) observed with MaNGA \citep{CanoDiaz_2022_mangadwarfs}, including newly derived photometric results, such as surface brightness, color, and position angle. Also measured are photometric profiles along with S\'ersic fits and new estimations of effective radii for all of the galaxies, using DECaLS DR9 $g,r,z$ images \citep{Dey_2019_DESIsurveys}. Analysis of the MaNGA data provides estimates of $M_{\star}$, SFRs, metallicities, ages, and other properties, 
using MaNGA Pipe3D data products \citep{Sanchez_2016_pipe3d,Sanchez_2018_mangaagn}. Details for the MaNDala sample and analysis are given in \citet{CanoDiaz_2022_mangadwarfs}.

{\it MaNGA Visual Morphology}: This is an update of a VAC, originally released in DR17 \citep[Section~5.5.2 in][]{Abdurrouf_2021_sdssDR17}, that contains visual morphological classifications for MaNGA galaxies based on SDSS and DECaLs \citep{Dey_2019_DESIsurveys} images. See \citet{VazquezMata_2022_MANGAmorphVAC} for details.

\subsection{SDSS-V VACs}
\label{sec:vacs_dr18}
The only VAC in DR18 derived from SDSS-V data is that providing supplementary properties for the eFEDS spectroscopic data set (\S\ref{sec:bhm:efeds}). 

This VAC provides updated redshift and classification information for optical counterparts to X-ray sources detected in the eFEDS field \citep[][]{Brunner2022,Salvato2022}. We update the spectroscopic redshift and classifications \citep[with respect to][]{Salvato2022} using a large spectroscopic compilation dominated by SDSS optical spectroscopy. Most importantly, we include new information from the 37 dedicated SDSS-V/eFEDS plates, described in detail in \S\ref{sec:efeds}. We combine automated redshifts and classifications, derived from the BOSS \texttt{idlspec1d} pipeline (\S\ref{sec:software_boss}), with an extensive set of visual inspections, which together increase the reliability and completeness of the spectroscopic coverage. 

The VAC includes three separate catalogs:
\begin{enumerate}[label=\roman*)] \itemsep -2pt
\item \texttt{eFEDS\_Main\_speccomp} --- an update of redshift and classification information for the eFEDS Main (0.2--2.3~keV selection) source counterparts catalog; 
\item \texttt{eFEDS\_Hard\_speccomp} --- an update of redshift and classification information for the eFEDS Hard band (2.3--5~keV selection) source counterparts catalog; and 
\item \texttt{eFEDS\_SDSSV\_spec\_results} --- a catalog of spectroscopic redshifts and classifications derived solely from the SDSS-V/eFEDS plate data set, supplemented by the results of an extensive visual inspection process. 
\end{enumerate}
A full description of this VAC will be provided in Merloni et al.\, (in prep).

\section{Summary and DR19 Preview}
\label{sec:summary}

In this paper, we have described the contents of the first data release of SDSS-V, which is also the eighteenth data release of the SDSS family of surveys. DR18 contains extensive targeting information for the Milky Way Mapper (\S\ref{sec:mwm_target_cartons}) and Black Hole Mapper (\S\ref{sec:bhm_observations}), both compiled target catalogs and the selection algorithms for their numerous scientific programs. Nearly $\sim$25,000 new spectra for X-ray sources in the eROSITA eFEDS field are also made available, along with substantial supplementary information, including improved redshift and classifications based on visual inspections. Significant improvements to the BOSS data spectral reduction pipelines were also made, in support of these and other optical spectra.

The next SDSS-V data release will be DR19, anticipated in 2024. This will contain the first MWM spectra and derived stellar parameters \& abundances, and the first spectra and derived properties from the primary BHM programs. Updates to the targeting catalogs presented in this paper will also be made available, along with updated simulations that predict the number of candidate targets in a carton that are likely be to observed. 
Additional tools for accessing and exploring all of the data are also anticipated to be released.

The capabilities and flexibility of SDSS-V make it unique among recent, ongoing, or imminent surveys, with well-vetted hardware, software, and collaboration infrastructure serving as the foundation for new innovations in all three areas. SDSS-V looks forward to expanding the SDSS legacy with high-quality, optical IFU data of 1000~deg$^2$ and optical $+$ infrared spectra of millions of sources across the entire sky.

%\begin{acknowledgements}
\section*{Acknowledgements}

The documentation workshop for DR18 (``DocuLlama'') was held virtually in September 2022, organized by Anne-Marie Weijmans and Gail Zasowski. This event was the main venue for the documentation of DR18, including significant progress on this paper and the website, and was attended by Scott Anderson, Joel Brownstein, Joleen Carlberg, Niall Deacon, Nathan De Lee, John Donor, Tom Dwelly, Keith Hawkins, Jennifer Johnson, Sean Morrison, Jordan Raddick, Jos\'{e} S\'{a}nchez-Gallego, Diogo Souto, Taylor Spoo, Ani Thakar, Nick Troup, Anne-Marie Weijmans, Gail Zasowski, William Zhang, three llamas, and an elderly goat named Nibblets.

Funding for the Sloan Digital Sky Survey V has been provided by the Alfred P. Sloan Foundation, the Heising-Simons Foundation, the National Science Foundation, and the Participating Institutions. SDSS acknowledges support and resources from the Center for High-Performance Computing at the University of Utah. The SDSS web site is \url{www.sdss.org}.

SDSS is managed by the Astrophysical Research Consortium for the Participating Institutions of the SDSS Collaboration, including the Carnegie Institution for Science, Chilean National Time Allocation Committee (CNTAC) ratified researchers, the Gotham Participation Group, Harvard University, The Johns Hopkins University, L'Ecole polytechnique f\'{e}d\'{e}rale de Lausanne (EPFL), Leibniz-Institut f\"{u}r Astrophysik Potsdam (AIP), Max-Planck-Institut f\"{u}r Astronomie (MPIA Heidelberg), Max-Planck-Institut f\"{u}r Extraterrestrische Physik (MPE), Nanjing University, National Astronomical Observatories of China (NAOC), New Mexico State University, The Ohio State University, Pennsylvania State University, Smithsonian Astrophysical Observatory, Space Telescope Science Institute (STScI), the Stellar Astrophysics Participation Group, Universidad Nacional Aut\'{o}noma de M\'{e}xico, University of Arizona, University of Colorado Boulder, University of Illinois at Urbana-Champaign, University of Toronto, University of Utah, University of Virginia, and Yale University.

%Full acknowledgements for Pan-STARRS1, 2MASS, Spitzer, Gaia, Legacy Surveys, and other facilities will be included in the published journal version.

%Pan-STARRS1
The Pan-STARRS1 Surveys (PS1) and the PS1 public science archive have been made possible through contributions by the Institute for Astronomy, the University of Hawaii, the Pan-STARRS Project Office, the Max-Planck Society and its participating institutes, the Max Planck Institute for Astronomy, Heidelberg and the Max Planck Institute for Extraterrestrial Physics, Garching, The Johns Hopkins University, Durham University, the University of Edinburgh, the Queen's University Belfast, the Harvard-Smithsonian Center for Astrophysics, the Las Cumbres Observatory Global Telescope Network Incorporated, the National Central University of Taiwan, the Space Telescope Science Institute, the National Aeronautics and Space Administration under Grant No. NNX08AR22G issued through the Planetary Science Division of the NASA Science Mission Directorate, the National Science Foundation Grant No. AST-1238877, the University of Maryland, Eotvos Lorand University (ELTE), the Los Alamos National Laboratory, and the Gordon and Betty Moore Foundation.

%2MASS
This publication makes use of data products from the Two Micron All Sky Survey, which is a joint project of the University of Massachusetts and the Infrared Processing and Analysis Center/California Institute of Technology, funded by the National Aeronautics and Space Administration and the National Science Foundation.

%Spitzer (GLIMPSE)
This work is based in part on observations made with the Spitzer Space Telescope, which was operated by the Jet Propulsion Laboratory, California Institute of Technology under a contract with NASA.

%WISE
This publication makes use of data products from the Wide-field Infrared Survey Explorer, which is a joint project of the University of California, Los Angeles, and the Jet Propulsion Laboratory/California Institute of Technology, funded by the National Aeronautics and Space Administration.

%Gaia
This work presents results from the European Space Agency (ESA) space mission Gaia. Gaia data are being processed by the Gaia Data Processing and Analysis Consortium (DPAC). Funding for the DPAC is provided by national institutions, in particular the institutions participating in the Gaia MultiLateral Agreement (MLA). The Gaia mission website is \url{https://www.cosmos.esa.int/gaia}. The Gaia archive website is \url{https://archives.esac.esa.int/gaia}.

%Legacy Surveys
The Legacy Surveys consist of three individual and complementary projects: the Dark Energy Camera Legacy Survey (DECaLS; Proposal ID \#2014B-0404; PIs: David Schlegel and Arjun Dey), the Beijing-Arizona Sky Survey (BASS; NOAO Prop. ID \#2015A-0801; PIs: Zhou Xu and Xiaohui Fan), and the Mayall z-band Legacy Survey (MzLS; Prop. ID \#2016A-0453; PI: Arjun Dey). DECaLS, BASS and MzLS together include data obtained, respectively, at the Blanco telescope, Cerro Tololo Inter-American Observatory, NSF’s NOIRLab; the Bok telescope, Steward Observatory, University of Arizona; and the Mayall telescope, Kitt Peak National Observatory, NOIRLab. Pipeline processing and analyses of the data were supported by NOIRLab and the Lawrence Berkeley National Laboratory (LBNL). The Legacy Surveys project is honored to be permitted to conduct astronomical research on Iolkam Du’ag (Kitt Peak), a mountain with particular significance to the Tohono O’odham Nation.

NOIRLab is operated by the Association of Universities for Research in Astronomy (AURA) under a cooperative agreement with the National Science Foundation. LBNL is managed by the Regents of the University of California under contract to the U.S. Department of Energy.

This project used data obtained with the Dark Energy Camera (DECam), which was constructed by the Dark Energy Survey (DES) collaboration. Funding for the DES Projects has been provided by the U.S. Department of Energy, the U.S. National Science Foundation, the Ministry of Science and Education of Spain, the Science and Technology Facilities Council of the United Kingdom, the Higher Education Funding Council for England, the National Center for Supercomputing Applications at the University of Illinois at Urbana-Champaign, the Kavli Institute of Cosmological Physics at the University of Chicago, Center for Cosmology and Astro-Particle Physics at the Ohio State University, the Mitchell Institute for Fundamental Physics and Astronomy at Texas A\&M University, Financiadora de Estudos e Projetos, Fundacao Carlos Chagas Filho de Amparo, Financiadora de Estudos e Projetos, Fundacao Carlos Chagas Filho de Amparo a Pesquisa do Estado do Rio de Janeiro, Conselho Nacional de Desenvolvimento Cientifico e Tecnologico and the Ministerio da Ciencia, Tecnologia e Inovacao, the Deutsche Forschungsgemeinschaft and the Collaborating Institutions in the Dark Energy Survey. The Collaborating Institutions are Argonne National Laboratory, the University of California at Santa Cruz, the University of Cambridge, Centro de Investigaciones Energeticas, Medioambientales y Tecnologicas-Madrid, the University of Chicago, University College London, the DES-Brazil Consortium, the University of Edinburgh, the Eidgenossische Technische Hochschule (ETH) Zurich, Fermi National Accelerator Laboratory, the University of Illinois at Urbana-Champaign, the Institut de Ciencies de l’Espai (IEEC/CSIC), the Institut de Fisica d’Altes Energies, Lawrence Berkeley National Laboratory, the Ludwig Maximilians Universitat Munchen and the associated Excellence Cluster Universe, the University of Michigan, NSF’s NOIRLab, the University of Nottingham, the Ohio State University, the University of Pennsylvania, the University of Portsmouth, SLAC National Accelerator Laboratory, Stanford University, the University of Sussex, and Texas A\&M University.

BASS is a key project of the Telescope Access Program (TAP), which has been funded by the National Astronomical Observatories of China, the Chinese Academy of Sciences (the Strategic Priority Research Program “The Emergence of Cosmological Structures” Grant \#XDB09000000), and the Special Fund for Astronomy from the Ministry of Finance. The BASS is also supported by the External Cooperation Program of Chinese Academy of Sciences (Grant \#114A11KYSB20160057), and Chinese National Natural Science Foundation (Grant \#12120101003, \#11433005).

The Legacy Survey team makes use of data products from the Near-Earth Object Wide-field Infrared Survey Explorer (NEOWISE), which is a project of the Jet Propulsion Laboratory/California Institute of Technology. NEOWISE is funded by the National Aeronautics and Space Administration.

The Legacy Surveys imaging of the DESI footprint is supported by the Director, Office of Science, Office of High Energy Physics of the U.S. Department of Energy under Contract No. DE-AC02-05CH1123, by the National Energy Research Scientific Computing Center, a DOE Office of Science User Facility under the same contract; and by the U.S. National Science Foundation, Division of Astronomical Sciences under Contract No. AST-0950945 to NOAO.

%SkyMapper
The national facility capability for SkyMapper has been funded through ARC LIEF grant LE130100104 from the Australian Research Council, awarded to the University of Sydney, the Australian National University, Swinburne University of Technology, the University of Queensland, the University of Western Australia, the University of Melbourne, Curtin University of Technology, Monash University and the Australian Astronomical Observatory. SkyMapper is owned and operated by The Australian National University's Research School of Astronomy and Astrophysics. The survey data were processed and provided by the SkyMapper Team at ANU. The SkyMapper node of the All-Sky Virtual Observatory (ASVO) is hosted at the National Computational Infrastructure (NCI). Development and support of the SkyMapper node of the ASVO has been funded in part by Astronomy Australia Limited (AAL) and the Australian Government through the Commonwealth's Education Investment Fund (EIF) and National Collaborative Research Infrastructure Strategy (NCRIS), particularly the National eResearch Collaboration Tools and Resources (NeCTAR) and the Australian National Data Service Projects (ANDS).

%TESS
This paper includes data collected by the TESS mission. Funding for the TESS mission is provided by the NASA's Science Mission Directorate.

%Swift
We acknowledge the use of public data from the Swift data archive.

%XMM-Newton
Based on observations obtained with XMM-Newton, an ESA science mission with instruments and contributions directly funded by ESA Member States and NASA.

%ADS
This research has made use of NASA's Astrophysics Data System Bibliographic Services.

%\end{acknowledgements}

\facilities{Du Pont (APOGEE), Sloan, Spitzer, WISE, 2MASS, Gaia, GALEX, PS1, TESS, Swift, CXO, XMM, eROSITA}

%A handy "cheat sheet" that provides the necessary \latex\ to produce 17 different types of tables is available at \url{http://journals.aas.org/authors/aastex/aasguide.html#table_cheat_sheet}.

%\section{Using Chinese, Japanese, and Korean characters}

%Authors have the option to include names in Chinese, Japanese, or Korean (CJK) characters in addition to the English name. The names will be displayed in parentheses after the English name. The way to do this in AASTeX is to use the CJK package available at \url{https://ctan.org/pkg/cjk?lang=en}. Further details on how to implement this and solutions for common problems, please go to \url{https://journals.aas.org/nonroman/}.


\bibliography{ms}{}
\bibliographystyle{aasjournal}


\appendix

\section{Details of MWM \texttt{v0.5.3} target cartons}
\label{sec:mwm_cartons_appendix}

In this Appendix we provide the detailed criteria used to select the MWM target cartons that are being released as part of SDSS DR18. We remind the reader that inclusion in a carton means that the target fits the carton selection criteria, not that it is guaranteed to be observed. 

For each target carton, we provide the following information: 
The {\bf Description of selection criteria} provides a short summary of the carton selection in  human-readable terms.  It also includes a list of the limits in color, magnitude, parallax, and other quantities applied to the carton's target candidates\footnote{We note that 
%the MWM cartons have selection functions of varying complexity, and 
the selection descriptions here describe the cuts made by the SDSS team. In some cases, targets were drawn from external catalogs, which may have undergone additional selection criteria prior to inclusion in the MWM survey. We refer the reader to the cited papers for more details on those selections.}. {\bf Data Sources} gives the catalogs from which these quantities are drawn. In {\bf Target priority options}, we indicate which priority is given to targets in this carton for observing; smaller priorities are more likely to be assigned fibers. The {\bf Cadence options} describes which exposure time requirement(s) are assigned to sources in the carton. See Table~\ref{tab:mwm_cartons} for each carton's instrument (BOSS or APOGEE) and the number of candidate targets that meet the carton's selection criteria.

\begin{center}\rule{0.5\linewidth}{0.5pt}\end{center}

\subsection{manual\_mwm\_tess\_ob}\label{manual_mwm_tess_ob_plan}
\begin{description}[nosep]
    \item[Description of selection criteria] A system selected from the \citet{Ijspeert_2021} catalog of massive eclipsing binaries in TESS. Systems chosen were in the {TESS Continuous Viewing Zones (CVZs)} and had TESS lightcurves that show pronounced eclipses and intrinsic variability. 
\item[Data Sources] \citet{Ijspeert_2021}
\item[Target priority options] 2200
\item[Cadence options]  bright\_8x1, bright\_8x2, bright\_8x4
\end{description}

\begin{center}\rule{0.5\linewidth}{0.5pt}\end{center}

%manual_nsbh_apogee_plan
\subsection{manual\_nsbh\_apogee}\label{manual_nsbh_apogee_plan}
\begin{description}[nosep]
    \item[Description of selection criteria] List of targets drawn from public and private catalogs of binary systems with known or suspected black holes or neutron stars, such as X-ray binaries and pulsars. Must have a valid 2MASS $H$ magnitude.
   
\item[Data Sources] 2MASS PSC (H), Gaia DR2 {(G)}
\item[Target priority options] 1400
\item[Cadence options] bright\_1x1
\end{description}

\begin{center}\rule{0.5\linewidth}{0.5pt}\end{center}

%manual_nsbh_boss_plan
\subsection{manual\_nsbh\_boss}\label{manual_nsbh_boss_plan}
\begin{description}[nosep]
    \item[Description of selection criteria] List of targets drawn from public and private catalogs of binary systems with known or suspected black holes or neutron stars, such as X-ray binaries and pulsars. Must have a valid Gaia DR2 $G$ magnitude.
%    \begin{itemize}[nosep]
%        \item H and G magnitude limits
%        \item Detected in Gaia DR2
%    \end{itemize}
\item[Data Sources] 2MASS PSC (H), Gaia DR2 (G)
\item[Target priority options] 1400
\item[Cadence options] bright\_1x1, dark\_1x2 
\end{description}

\begin{center}\rule{0.5\linewidth}{0.5pt}\end{center}

%mwm_cb_300pc_apogee_plan
\subsection{mwm\_cb\_300pc\_apogee}\label{mwm_cb_300pc_apogee_plan}
\begin{description}[nosep]
    \item[Description of selection criteria] Bright compact binary candidates within 300 pc.
    \begin{itemize}[nosep]
        \item $(FUV - 5\log_{10}(r_{\rm est}/10)) > 14 (FUV-NUV) - 46$
        \item $r_{\rm est} < 300$ pc
        \item $H < 11$
    \end{itemize}
\item[Data Sources] \citet{Bailer-Jones_dr2} distance catalog ($r_{\rm est}$), 2MASS PSC (H), GALEX (FUV, NUV)
\item[Target priority options] 1400
\item[Cadence options]  bright\_1x1
\end{description}

\begin{center}\rule{0.5\linewidth}{0.5pt}\end{center}

%mwm_cb_300pc_boss_plan
\subsection{mwm\_cb\_300pc\_boss}\label{mwm_cb_300pc_boss_plan}
\begin{description}[nosep]
     \item[Description of selection criteria] Faint compact binary candidates within 300 pc.
    \begin{itemize}[nosep]
        \item $(FUV - 5\log_{10}(r_{\rm est}/10)) > 14 (FUV-NUV) - 46$
        \item $r_{\rm est} < 300$ pc
    \end{itemize}
\item[Data Sources]\citet{Bailer-Jones_dr2} distance catalog ($r_{\rm est}$), GALEX (FUV, NUV)
\item[Target priority options] 1400
\item[Cadence options]  bright\_2x1, dark\_2x1
\end{description}

\begin{center}\rule{0.5\linewidth}{0.5pt}\end{center}

%mwm_cb_cvcandidates_apogee_plan
\subsection{mwm\_cb\_cvcandidates\_apogee}\label{mwm_cb_cvcandidates_apogee_plan}
\begin{description}[nosep]
    \item[Description of selection criteria] Bright AAVSO cataclysmic variables
    \begin{itemize}[nosep]
        \item Target in \texttt{mos\_cataclysmic\_variables} table 
        \item $H < 11$
    \end{itemize}
\item[Data Sources] 2MASS PSC (H)
\item[Target priority options] 1400
\item[Cadence options] bright\_1x1
\end{description}

\begin{center}\rule{0.5\linewidth}{0.5pt}\end{center}

%mwm_cb_cvcandidates_boss_plan
\subsection{mwm\_cb\_cvcandidates\_boss}\label{mwm_cb_cvcandidates_boss_plan}
\begin{description}[nosep]
    \item[Description of selection criteria] Faint AAVSO cataclysmic variables
    \begin{itemize}[nosep]
        \item Target in \texttt{mos\_cataclysmic\_variables} table 
        \item $H \geq 11$
    \end{itemize}
\item[Data Sources] 2MASS PSC (H)
\item[Target priority options] 1400
\item[Cadence options] bright\_2x1, dark\_2x1
\end{description}

\begin{center}\rule{0.5\linewidth}{0.5pt}\end{center}

%mwm_cb_gaiagalex_apogee_plan
\subsection{mwm\_cb\_gaiagalex\_apogee}\label{mwm_cb_gaiagalex_apogee_plan}
\begin{description}[nosep]
    \item[Description of selection criteria] Bright compact binary candidates using Gaia $G$ and GALEX FUV color cut
    \begin{itemize}[nosep]
        \item $\varpi / \sigma_\varpi  > 3$
        \item $G < 20$
         \item $FUV + 5 \log1_{10}(\varpi/ 1000) + 5 > 1.5 + 1.28 (FUV – G)$
         \item $H < 11$
    \end{itemize}
\item[Data Sources]  Gaia DR2 (G, $\varpi$, $\sigma_\varpi$), 2MASS PSC (H), GALEX (FUV)
\item[Target priority options] 1400
\item[Cadence options]  bright\_1x1
\end{description}

\begin{center}\rule{0.5\linewidth}{0.5pt}\end{center}

%mwm_cb_gaiagalex_boss_plan
\subsection{mwm\_cb\_gaiagalex\_boss}\label{mwm_cb_gaiagalex_boss_plan}
\begin{description}[nosep]
    \item[Description of selection criteria] Faint compact binary candidates using Gaia $G$ and GALEX FUV color cut
    \begin{itemize}[nosep]
        \item $\varpi / \sigma_\varpi  > 3$
        \item $G < 20$
         \item $FUV + 5 \log_{10}(\varpi/ 1000) + 5 > 1.5 + 1.28 (FUV – G)$
         \item $H \geq 11$
    \end{itemize}
\item[Data Sources]  Gaia DR2 (G, $\varpi$, $\sigma_\varpi$), 2MASS PSC (H), GALEX (FUV)
\item[Target priority options] 1400
\item[Cadence options] bright\_2x1, dark\_2x1
\end{description}

\begin{center}\rule{0.5\linewidth}{0.5pt}\end{center}

%mwm_cb_uvex1_plan
\subsection{mwm\_cb\_uvex1}\label{mwm_cb_uvex1_plan}
\begin{description}[nosep]
    \item[Description of selection criteria] Gaia and GALEX cross-match, keeping the nearest match within 5\arcsec\ and removing objects with small proper motion and parallax. Color cuts utilize both FUV and NUV magnitudes. ``AB'' indicates magnitudes transformed to AB magnitudes, and $M_G$ is Gaia absolute magnitude calculated with $r_{\rm est}$.
    \begin{itemize}[nosep]
        \item $r_{\rm lo} \leq 1500$
        \item g\_vis\_per $> 5$
        \item Neither of the following two conditions are met
        \begin{itemize}[nosep] %logpmdpm = $ \log_{10}(\mu/\sigma_\mu) $
            \item $(\log_{10}(\mu/\sigma_\mu) < 0.301)$ AND $(\varpi/\sigma_\varpi > -1.4996\log_{10}(\mu/\sigma_\mu) – 4.05)$ AND $(\varpi/\sigma_\varpi < 1.4995\log_{10}(\mu/\sigma_\mu) + 4.05)$
            \item $(\log_{10}(\mu/\sigma_\mu) – 0.301)^2/0.39794^2 + (\varpi/\sigma_\varpi)^2/4.5^2 \leq 1$
        \end{itemize}
        \item $NUV > -100$
        \item $FUV > -100$
        \item $\sigma_{NUV} <0 .2$
        \item $\sigma_{FUV} <0 .2$
        \item $M_G > 4.09$ OR $M_G > 4.5457 (G_{\rm BP} – G_{\rm RP}) + 4.0457$
        \item $M_G > -1.11749253\times 10^{-3}(FUV-NUV)^3 + 1.53748615\times 10^{-2}(FUV-NUV)^2 + 3.66419895\times 10^{-1} (FUV-NUV) + 2.20026639)$
        \item $(FUV-G_{\rm AB}) < 6.08$ OR $[(FUV-G_{\rm AB}) < 11.82(G_{\rm BP,AB} – G_{\rm RP,AB}) + 2.58$ AND $(FUV-G_{\rm AB}) < -0.79(G_{\rm BP,AB} – G_{\rm RP,AB} + 9.21)]$
    \end{itemize}
\item[Data Sources]  Gaia DR2 ($M_G$, $G_{\rm BP}$, $G_{\rm RP}$, $\mu$, $\sigma_mu$, $\varpi$, $\sigma_\varpi$, $r_{\rm lo}$, g\_vis\_per), \citet{Bailer-Jones_dr2} distance catalog ($r_{\rm est}$), GALEX (NUV, FUV, $\sigma_{NUV}$, $\sigma_{FUV}$)
\item[Target priority options] 1400
\item[Cadence options] bright\_1x1, dark\_1x2, dark\_1x3
\end{description}

\begin{center}\rule{0.5\linewidth}{0.5pt}\end{center}

%mwm_cb_uvex2_plan
\subsection{mwm\_cb\_uvex2}\label{mwm_cb_uvex2_plan}
\begin{description}[nosep]
    \item[Description of selection criteria] Gaia and GALEX cross-match, keeping the nearest match within 5\arcsec\ and removing objects with small proper motion and parallax. Color cuts utilize only NUV magnitudes. ``AB'' indicates magnitudes transformed to AB magnitudes, and $M_G$ is Gaia absolute magnitude calculated with $r_{\rm est}$.
    \begin{itemize}[nosep]
        \item $r_{\rm lo} \leq 1500$
        \item g\_vis\_per $> 5$
        \item Neither of the following two conditions are met
        \begin{itemize}[nosep] %logpmdpm = $ \log_{10}(\mu/\sigma_\mu) $
            \item $(\log_{10}(\mu/\sigma_\mu) < 0.301)$ AND $(\varpi/\sigma_\varpi > -1.4996\log_{10}(\mu/\sigma_\mu) – 4.05)$ AND $(\varpi/\sigma_\varpi < 1.4995\log_{10}(\mu/\sigma_\mu) + 4.05)$
            \item $(\log_{10}(\mu/\sigma_\mu) – 0.301)^2/0.39794^2 + (\varpi/\sigma_\varpi)^2/4.5^2 \leq 1$
        \end{itemize}
        \item $NUV > -100$
        \item $\sigma_{NUV} <0 .2$
        \item $M_G > 4.09$ OR $M_G > 4.5457 (G_{\rm BP} – G_{\rm RP}) + 4.0457$
        \item $(NUV – G_{\rm AB} < 2.25)$ OR $[(NUV – G_{\rm AB} < 6.725(G_{\rm BP,AB} – G_{\rm RP,AB)} – 1.735)$ AND $(NUV – G_{\rm AB} < -0.983(G_{\rm BP,AB} – G_{\rm RP,AB} + 8.24)]$
    \end{itemize}
\item[Data Sources]  Gaia DR2 ($M_G$, $G_{AB}$, $G_{\rm BP}$, $G_{\rm BP,AB}$, $G_{\rm RP}$, $G_{\rm RP,AB}$, $\mu$, $\sigma_mu$, $\varpi$, $\sigma_\varpi$, g\_vis\_per), \citet{Bailer-Jones_dr2} distance catalog ($r_{\rm lo}$, $r_{\rm est}$), GALEX (NUV, $\sigma_{NUV}$)
\item[Target priority options] 1400
\item[Cadence options] bright\_1x1, dark\_1x2, dark\_1x3
\end{description}

\begin{center}\rule{0.5\linewidth}{0.5pt}\end{center}

%mwm_cb_uvex3_plan
\subsection{mwm\_cb\_uvex3}\label{mwm_cb_uvex3_plan}
\begin{description}[nosep]
    \item[Description of selection criteria] Gaia and XMM-Newton Optical Monitor SUSS Catalog cross-match,  keeping the nearest match with 3\arcsec\ and removing objects with small proper motion and parallax. Color and quality cuts utilize the XMM UVM2 band.
    \begin{itemize}[nosep]
        \item $r_{\rm lo} \leq 1500$
        \item g\_vis\_per $> 5$
        \item Neither of the following two conditions are met
        \begin{itemize}[nosep] %logpmdpm = $ \log_{10}(\mu/\sigma_\mu) $
            \item $(\log_{10}(\mu/\sigma_\mu) < 0.301)$ AND $(\varpi/\sigma_\varpi > -1.4996\log_{10}(\mu/\sigma_\mu) – 4.05)$ AND $(\varpi/\sigma_\varpi < 1.4995\log_{10}(\mu/\sigma_\mu) + 4.05)$
            \item $(\log_{10}(\mu/\sigma_\mu) – 0.301)^2/0.39794^2 + (\varpi/\sigma_\varpi)^2/4.5^2 \leq 1$
        \end{itemize}
        \item The following conditions are all NOT met
        \begin{itemize}[nosep]
             \item qflag bit 1, 7, 8, or 9 is set
             \item qflag bit 2, or 3 is set AND UVM2$_{\rm signif} < 10$
        \end{itemize}
    \item  $M_G > 4.09$ OR $M_G > 4.5457 (G_{\rm BP} – G_{\rm RP}) + 4.0457$
    \item ${\rm UVM2_{AB}} – G_{\rm AB} < 2.25$ OR $[({\rm UVM2_{AB}} – G_{\rm AB} < 6)$ AND $({\rm UVM2_{AB}} – G_{\rm AB} < 5.57377 (G_{\rm BP,AB} – G_{\rm RP,AB}) + 0.2049)]$
    \end{itemize}
\item[Data Sources] Gaia DR2 ($M_G$, $G_{AB}$, $G_{\rm BP}$, $G_{\rm BP,AB}$, $G_{\rm RP}$, $G_{\rm RP,AB}$, $\mu$, $\sigma_mu$, $\varpi$, $\sigma_\varpi$, g\_vis\_per), \citet{Bailer-Jones_dr2} distance catalog ($r_{\rm lo}$, $r_{\rm est}$), XMM OM SUSS v4.1(${\rm UVM2_{AB}}$, UVM2$_{\rm signif}$, qflag)
\item[Target priority options] 1400
\item[Cadence options]  bright\_1x1, dark\_1x2, dark\_1x3 
\end{description}

\begin{center}\rule{0.5\linewidth}{0.5pt}\end{center}

%mwm_cb_uvex4_plan
\subsection{mwm\_cb\_uvex4}\label{mwm_cb_uvex4_plan}
\begin{description}[nosep]
    \item[Description of selection criteria] Gaia and Swift UVOT Catalog cross-match,  keeping the nearest match with 3\arcsec\ and removing objects with small proper motion and parallax. Color cuts utilize the UVW2 band, and quality cuts utilize both UVW2 and UVW1
    \begin{itemize}[nosep]
        \item $r_{\rm lo} \leq 1500$
        \item g\_vis\_per $> 5$
        \item ${\rm UVW1_{AB}} > -100$
         \item ${\rm UVW2_{AB}} > -100$
        \item Neither of the following two conditions are met
        \begin{itemize}[nosep] %logpmdpm = $ \log_{10}(\mu/\sigma_\mu) $
            \item $(\log_{10}(\mu/\sigma_\mu) < 0.301)$ AND $(\varpi/\sigma_\varpi > -1.4996\log_{10}(\mu/\sigma_\mu) – 4.05)$ AND $(\varpi/\sigma_\varpi < 1.4995\log_{10}(\mu/\sigma_\mu) + 4.05)$
            \item $(\log_{10}(\mu/\sigma_\mu) – 0.301)^2/0.39794^2 + (\varpi/\sigma_\varpi)^2/4.5^2 \leq 1$
        \end{itemize}
        \item The following conditions are all NOT met
        \begin{itemize}[nosep]
             \item qflag bit 0 or 6 is set
             \item qflag bit 1, 2, or 5 is set AND UVW2$_{\rm signif} < 10$
        \end{itemize}
    \item  $M_G > 4.09$ OR $M_G > 4.5457 (G_{\rm BP} – G_{\rm RP}) + 4.0457$
        \item ${\rm UVW2_{AB}} – G_{\rm AB} < 2.25$ OR $[({\rm UVW2_{AB}} – G_{\rm AB} < 6)$ AND $({\rm UVW2_{AB}} – G_{\rm AB} < 5.57377((G_{\rm BP,AB} – G_{\rm RP,AB}) + 0.2049)]$
    \end{itemize}
\item[Data Sources] Gaia DR2 ($M_G$, $G_{AB}$, $G_{\rm BP}$, $G_{\rm BP,AB}$, $G_{\rm RP}$, $G_{\rm RP,AB}$, $\mu$, $\sigma_mu$, $\varpi$, $\sigma_\varpi$, g\_vis\_per), \citet{Bailer-Jones_dr2} distance catalog ($r_{\rm lo}$, $r_{\rm est}$), Swift UVOT(${\rm UVW1_{AB}}$, ${\rm UVW2_{AB}}$, UVW2$_{\rm signif}$, qflag2)
\item[Target priority options] 1400
\item[Cadence options] bright\_1x1, dark\_1x2, dark\_1x3 
\end{description}

\begin{center}\rule{0.5\linewidth}{0.5pt}\end{center}

%mwm_cb_uvex5_plan
\subsection{mwm\_cb\_uvex5}\label{mwm_cb_uvex5_plan}
\begin{description}[nosep]
    \item[Description of selection criteria] Gaia and GALEX cross-match, keeping the nearest match within 5\arcsec\ and removing objects with small proper motion and parallax. Only objects with FUV and NUV detections are kept, but the color cuts utilize only optical colors and magnitudes. Here, the expected Gaia Main Sequence locus (GMS) is defined by GMS = $0.00206868742x^6 + 0.0401594518x^5– 0.842512410x^4 + 4.89384979x^3 – 12.3826637x^2 + 17.0197205x – 3.19987835$, where $x=G-G_{\rm RP}$. $M_G$ is the Gaia absolute magnitude calculated with $r_{\rm est}$.
    \begin{itemize}[nosep]
        \item $r_{\rm lo} \leq 1500$
        \item g\_vis\_per $> 5$
        \item Neither of the following two conditions are met
        \begin{itemize}[nosep] %logpmdpm = $ \log_{10}(\mu/\sigma_\mu) $
            \item $(\log_{10}(\mu/\sigma_\mu) < 0.301)$ AND $(\varpi/\sigma_\varpi > -1.4996\log_{10}(\mu/\sigma_\mu) – 4.05)$ AND $(\varpi/\sigma_\varpi < 1.4995\log_{10}(\mu/\sigma_\mu) + 4.05)$
            \item $(\log_{10}(\mu/\sigma_\mu) – 0.301)^2/0.39794^2 + (\varpi/\sigma_\varpi)^2/4.5^2 \leq 1$
        \end{itemize}
        \item $NUV > -100$
        \item $FUV > -100$
        \item $\sigma_{NUV} <0 .2$
        \item $\sigma_{FUV} <0 .2$
        \item $M_G > 4.0866$
        \item $H < 15$
        \item $|{\rm GMS}-M_G| \leq 0.5$
        \item $r_{\rm est} < 0.51 \times 10^{0.2291G}$
    \end{itemize}
\item[Data Sources] Gaia DR2 ($M_G$, $G_{\rm RP}$, $\mu$, $\sigma_mu$, $\varpi$, $\sigma_\varpi$, $r_{\rm lo}$, g\_vis\_per), \citet{Bailer-Jones_dr2} distance catalog ($r_{\rm est}$), GALEX (NUV, FUV, $\sigma_{NUV}$, $\sigma_{FUV}$),  2MASS PSC (H)
\item[Target priority options] 1400
\item[Cadence options] bright\_1x1, dark\_1x2, dark\_1x3
\end{description}

\begin{center}\rule{0.5\linewidth}{0.5pt}\end{center}

%mwm_dust_core_plan
\subsection{mwm\_dust\_core}\label{mwm_dust_core_plan}
\begin{description}[nosep]
    \item[Description of selection criteria] The dust carton selects bright, nearby, midplane giants with the same quality cuts as Galactic Genesis (GG)  to supplement GG targets in regions of high reddening. The Galaxy within 5 kpc and with $|z| < 0.2$ kpc is divided into 100 pc$^3$ regions (``voxels'') and the number of GG targets in each voxel are counted. In voxels with fewer than 10 GG stars, $10-n_{\rm GG}$ stars are selected for inclusion in this carton.
    \begin{itemize}[nosep]
        \item $\sigma_\varpi /\varpi < 0.2$
        \item $ 1/\varpi<$ 5 kpc
        \item $|z| < 0.2$ kpc {(using $\varpi$-based distances)}
        \item $M_K < 2.6$, adopting $A_K=0.918(H-4.5)$
        \item $(J-K)_0 > 0.5$, where  $E(J-K)=1.5 A_K$
        \item $H < 11.2$
         \item gal\_contam == 0
         \item cc\_flg == 0
        \item 0 $<$ rd\_flag $<= 3$
         \item ph\_qual flag is A or B
    \end{itemize}
\item[Data Sources]  Gaia DR2 ($\varpi$, $\sigma_\varpi$), 2MASS PSC (J, H, K, gal\_contam, cc\_flag, rd\_flag, ph\_qual)
\item[Target priority options] 2720
\item[Cadence options] bright\_1x1
\end{description}

\begin{center}\rule{0.5\linewidth}{0.5pt}\end{center}

%mwm_erosita_compact_gen_plan
\subsection{mwm\_erosita\_compact\_gen}\label{mwm_erosita_compact_gen_plan}
\begin{description}[nosep]
    \item[Description of selection criteria] Optical counterpart in Gaia DR2 to a candidate compact binary detected by eROSITA, chosen as the possible counterpart with the smallest angular separation
    \begin{itemize}[nosep]
    \item $G > 16$
    \item $L > 8$ in at least one of the three eROSITA energy bands
    \item Detections in three eROSITA energy bands
    \item $\log{\left( f_x/f_{opt}\right)} < 2.7$
    \item $\rm radec\_err > 0$ and $\rm s/radec\_err < 2.1$, where s is the separation between the Gaia and eROSITA sources
    \item Neither of the following two conditions are met
        \begin{itemize}[nosep] %logpmdpm = $ \log_{10}(\mu/\sigma_\mu) $
            \item $(\log_{10}(\mu/\sigma_\mu) < 0.301)$ AND $(\varpi/\sigma_\varpi > -1.4996\log_{10}(\mu/\sigma_\mu) – 4.05)$ AND $(\varpi/\sigma_\varpi < 1.4995\log_{10}(\mu/\sigma_\mu) + 4.05)$
            \item $(\log_{10}(\mu/\sigma_\mu) – 0.301)^2/0.39794^2 + (\varpi/\sigma_\varpi)^2/4.5^2 \leq 1$
        \end{itemize}
    \item Is the optical source that meets the above criteria with the smallest separation s
    \end{itemize}
\item[Data Sources] eROSITA ($f_x$, $L$, radec\_err), Gaia DR2 ($G$, $\mu$, $\sigma_mu$, $\varpi$, $\sigma_\varpi$, $f_{opt}$ (from $G$))
\item[Target priority options] 1910, 2400
\item[Cadence options] bright\_1x1, dark\_1x2, dark\_1x3
\end{description}

\begin{center}\rule{0.5\linewidth}{0.5pt}\end{center}

%mwm_erosita_compact_var_plan
\subsection{mwm\_erosita\_compact\_var}\label{mwm_erosita_compact_var_plan}
\begin{description}[nosep]
    \item[Description of selection criteria] Optical counterpart to a candidate compact binary detected by eROSITA, chosen as the possible counterpart with the largest variability in Gaia DR2
    \begin{itemize}[nosep]
    \item $G > 16$
    \item $L > 8$ in at least one of the three eROSITA energy bands
    \item Detections in three eROSITA energy bands
    \item $\log{\left( f_x/f_{opt}\right)} < 2.7$
    \item $\rm radec\_err > 0$ and $\rm s/radec\_err < 2.1$, where s is the separation between the Gaia and eROSITA sources
    \item Neither of the following two conditions are met
        \begin{itemize}[nosep] %logpmdpm = $ \log_{10}(\mu/\sigma_\mu) $
            \item $(\log_{10}(\mu/\sigma_\mu) < 0.301)$ AND $(\varpi/\sigma_\varpi > -1.4996\log_{10}(\mu/\sigma_\mu) – 4.05)$ AND $(\varpi/\sigma_\varpi < 1.4995\log_{10}(\mu/\sigma_\mu) + 4.05)$
            \item $(\log_{10}(\mu/\sigma_\mu) – 0.301)^2/0.39794^2 + (\varpi/\sigma_\varpi)^2/4.5^2 \leq 1$
        \end{itemize}
    \item Is the optical source that meets the above criteria with the largest variability $\log_{10}{\left(\left(\frac{\rm phot\_g\_n\_obs}{\rm phot\_g\_mean\_flux\_over\_error}\right)^{1/2}\right)}$
    \end{itemize}
\item[Data Sources] eROSITA ($f_x$, $L$, radec\_err), Gaia DR2 ($G$, $\mu$, $\sigma_mu$, $\varpi$, $\sigma_\varpi$, $f_{opt}$ (from $G$), phot\_g\_n\_obs, phot\_g\_mean\_flux\_over\_error)
\item[Target priority options] 1900, 2400
\item[Cadence options] bright\_1x1, dark\_1x2, dark\_1x3
\end{description}

\begin{center}\rule{0.5\linewidth}{0.5pt}\end{center}

%mwm_erosita_stars_plan
\subsection{mwm\_erosita\_stars}\label{mwm_erosita_stars_plan}
\begin{description}[nosep]
    \item[Description of selection criteria] Optical or infrared counterpart to a candidate stellar source detected by eROSITA. Counterpart identified using the techniques and data described in \citet{Freund_2022}.
%    \begin{itemize}[nosep]
%        \item Criteria 1
%        \item Criteria 2
%    \end{itemize}
\item[Data Sources] eROSITA, Gaia DR2, 2MASS PSC
\item[Target priority options] 1920, 2400
\item[Cadence options]  bright\_1x1, dark\_1x2, dark\_1x3
\end{description}

\begin{center}\rule{0.5\linewidth}{0.5pt}\end{center}

%mwm_galactic_core_plan
\subsection{mwm\_galactic\_core}\label{mwm_galactic_core_plan}
\begin{description}[nosep]
    \item[Description of selection criteria] A simple color-magnitude cut effectively targets luminous cool giant stars (${\rm median}(\log g) \sim 1.0-1.5$) with little ($<$6\%) contamination from dwarf stars.
    \begin{itemize}[nosep]
        \item $H < 11$
        \item$ G - H > 3.5$ or Gaia non-detection
         \item gal\_contam == 0
         \item cc\_flg == 0
        \item 0 $<$ rd\_flag $<= 3$
         \item ph\_qual flag is A or B
    \end{itemize}
\item[Data Sources] Gaia DR2 (G), 2MASS PSC (H, gal\_contam, cc\_flg, rd\_flag, ph\_qual)
\item[Target priority options] 2710
\item[Cadence options] bright\_1x1
\end{description}

\begin{center}\rule{0.5\linewidth}{0.5pt}\end{center}

%mwm_legacy_ir2opt_plan
\subsection{mwm\_legacy\_ir2opt}\label{mwm_legacy_ir2opt_plan}
\begin{description}[nosep]
    \item[Description of selection criteria] This carton selects all bright Gaia objects that are in APOGEE DR16 (in particular the \texttt{sdss\_apogeeallstarmerge\_r13} file) to be observed with BOSS.
    \begin{itemize}[nosep]
        \item $3 < G < 18$
        \item $G_{\rm BP} > 13$
        \item $G_{\rm RP} > 13$.
%        \item Star in sdss\_apogeeallstarmerge\_r13
    \end{itemize}
\item[Data Sources] Gaia DR2 (G, $G_{\rm BP}$, $G_{\rm RP}$), APOGEE DR16
\item[Target priority options] 6100
\item[Cadence options]  bright\_1x1
\end{description}

\begin{center}\rule{0.5\linewidth}{0.5pt}\end{center}

%mwm_ob_cepheids_plan
\subsection{mwm\_ob\_cepheids}\label{mwm_ob_cepheids_plan}
\begin{description}[nosep]
    \item[Description of selection criteria] Catalog of Cepheids compiled by \citet{Inno2021}
%    \begin{itemize}[nosep]
%        \item Target in Cepheid catalog
%    \end{itemize}
\item[Data Sources] \citet{Inno2021}
\item[Target priority options] 2910
\item[Cadence options] bright\_3x1
\end{description}

\begin{center}\rule{0.5\linewidth}{0.5pt}\end{center}

%mwm_ob_core_plan
\subsection{mwm\_ob\_core}\label{mwm_ob_core_plan}
\begin{description}[nosep]
    \item[Description of selection criteria] This carton uses color and magnitude cuts to select hot, young stars.
    \begin{itemize}[nosep]
        \item $\varpi < 10^{(10 - K - 0.61) / 5)}$
        \item $G < 16$
         \item  $J - K - 0.25  (G - K) < 0.10$
         \item  $J - K - 0.25  (G - K) > -0.30$
         \item  $J - H < 0.15  (G - K) + 0.05$
         \item  $J - H > 0.15  (G - K) - 0.15$
         \item  $J - K < 0.23  (G - K) + 0.03$
         \item  $G > 2(G - K) + 3.0$
         \item  Cross match separation $<1$\arcsec
         \item RUWE $< 1.4$
    \end{itemize}
\item[Data Sources]  Gaia DR2 (G, $\varpi$, RUWE), 2MASS PSC (J, H, K)
\item[Target priority options] 2910
\item[Cadence options]  bright\_3x1
\end{description}

\begin{center}\rule{0.5\linewidth}{0.5pt}\end{center}

%mwm_rv_long_fps_plan
\subsection{mwm\_rv\_long\_fps}\label{mwm_rv_long_fps_plan}
\begin{description}[nosep]
    \item[Description of selection criteria] This carton selects stars previously observed at least 3 times with the APOGEE instrument if it was targeted as part of the main APOGEE sample or the APOGEE-2 binary program.
    \begin{itemize}[nosep]
\item Presence in \texttt{sdss\_apogeeallstarmerge\_r13} file (previously observed with APOGEE 1 and/or 2
\item APOGEE number of visits $\geq3$
\item $H < 11.5$
\item APOGEE TARGFLAGS includes one of the following: APOGEE\_SHORT,
APOGEE\_INTERMEDIATE, APOGEE\_LONG, APOGEE2\_BIN 
\item Gaia-based distance
    \end{itemize}
\item[Data Sources] APOGEE, 2MASS PSC (H), Gaia DR 2
\item[Target priority options] 2510, 2520, 2530, 2540
\item[Cadence options]bright\_6x1
bright\_6x2, bright\_9x1, bright\_9x2, bright\_12x1, bright\_12x2, bright\_15x1, bright\_15x2
\end{description}

\begin{center}\rule{0.5\linewidth}{0.5pt}\end{center}

%mwm_rv_short_fps_plan
\subsection{mwm\_rv\_short\_fps}\label{mwm_rv_short_fps_plan}
\begin{description}[nosep]
    \item[Description of selection criteria] This carton selects stars that were never previously observed with APOGEE and applies the same color and quality cuts as the main APOGEE surveys.
    \begin{itemize}[nosep]
        \item $H< 10.8$
        \item $J – K_s – (1.5 \cdot 0.918 (H – W2 – 0.05)) >= 0.05$
           \item $J\_msigcom,H\_msigcom,K_s\_msigcom   \leq 0.1$
        \item $W2\_sigmpro \leq 0.1$
        \item 2MASS ph\_qual any of the following: AAA, AAB, ABA, BAA, ABB, BAB, BBA, BBB
         \item 2MASS gal\_contam = 0
         \item  2MASS cc\_flg = 0
         \item 2MASS rd\_flg any of the following: 111, 112, 121, 211, 122, 212, 221, 222
         \item 2MASS prox $\geq$ 6
         \item Gaia DR2 parallax exists
         \item Does not match a source in the 2MASS extended catalog
    \end{itemize}
\item[Data Sources] 2MASS PSC (J, H, $K_s$, J\_msigcom, H\_msigcom, K$_s$\_msigcom, ph\_qual, gal\_contam, cc\_flg, rd\_flg, prox), Gaia DR 2, AllWise (W2)
\item[Target priority options] 2515, 2525, 2535, 2545
\item[Cadence options]  bright\_18x1
\end{description}

\begin{center}\rule{0.5\linewidth}{0.5pt}\end{center}

%mwm_snc_100pc_apogee_plan
\subsection{mwm\_snc\_100pc\_apogee}\label{mwm_snc_100pc_apogee_plan}
\begin{description}[nosep]
    \item[Description of selection criteria] Infrared bright targets in the 100 pc volume limited region. 
    \begin{itemize}[nosep]
        \item $\varpi - \sigma_\varpi > 10$
        \item $H<11$
        \item astrometric\_excess\_noise $<2$ if in one of the following regions
        \begin{itemize}[nosep]
        \item $l \leq 180$ AND $ b<-0.139 l + 25$ AND $b>0.139 l-25$
        \item $l>180$ AND $b>-0.139l + 25$ AND $b <0.139l – 25$
        \item $\sqrt{(l-303.2)^2 + 2*(b+44.4)^2 } < 5$ 
        \item $\sqrt{(l-280.3)^2 + 2*(b+33.0)^2} < 8$
        \end{itemize}
    \end{itemize}
\item[Data Sources]  Gaia DR2 ($\varpi$, $\sigma_\varpi$, astrometric\_excess\_noise, $l$, $b$), 2MASS PSC (H)
\item[Target priority options] 1805
\item[Cadence options] bright\_1x1
\end{description}

\begin{center}\rule{0.5\linewidth}{0.5pt}\end{center}

%mwm_snc_100pc_boss_plan
\subsection{mwm\_snc\_100pc\_boss}\label{mwm_snc_100pc_boss_plan}
\begin{description}[nosep]
    \item[Description of selection criteria] Optically bright targets in the 100 pc volume limited region. 
    \begin{itemize}[nosep]
        \item $\varpi - \sigma_\varpi > 10$
        %\item Detected in Gaia
        \item astrometric\_excess\_noise $<2$ if in one of the following regions
        \begin{itemize}[nosep]
        \item $l \leq 180$ AND $ b<-0.139 l + 25$ AND $b>0.139 l-25$
        \item $l>180$ AND $b>-0.139l + 25$ AND $b <0.139l – 25$
        \item $\sqrt{(l-303.2)^2 + 2*(b+44.4)^2 } < 5$ 
        \item $\sqrt{(l-280.3)^2 + 2*(b+33.0)^2} < 8$
        \end{itemize}
    \end{itemize}
\item[Data Sources]  Gaia DR2 ($\varpi$, $\sigma_\varpi$, astrometric\_excess\_noise, $l$, $b$, G)
\item[Target priority options] 1800
\item[Cadence options]  bright\_2x1, dark\_2x1
\end{description}

\begin{center}\rule{0.5\linewidth}{0.5pt}\end{center}

%mwm_snc_250pc_apogee_plan
\subsection{mwm\_snc\_250pc\_apogee}\label{mwm_snc_250pc_apogee_plan}
\begin{description}[nosep]
    \item[Description of selection criteria]  Infrared bright targets in the 250 pc volume limited region. 
    \begin{itemize}[nosep]
       \item $G+5\log_{10}(\varpi / 1000) + 5 < 6$
        \item $\varpi - \sigma_\varpi > 4$
        \item $H<11$
        \item astrometric\_excess\_noise $<2$ if in one of the following regions
        \begin{itemize}[nosep]
        \item $l \leq 180$ AND $ b<-0.139 l + 25$ AND $b>0.139 l-25$
        \item $l>180$ AND $b>-0.139l + 25$ AND $b < 0.139 l–25$
        \item $\sqrt{(l-303.2)^2 + 2*(b+44.4)^2 } < 5$ 
        \item $\sqrt{(l-280.3)^2 + 2*(b+33.0)^2} < 8$
        \end{itemize}
    \end{itemize}
\item[Data Sources]  Gaia DR2 ($\varpi$, $\sigma_\varpi$, astrometric\_excess\_noise, $l$, $b$, G), 2MASS PSC (H)
\item[Target priority options] 1815
\item[Cadence options]  bright\_1x1
\end{description}

\begin{center}\rule{0.5\linewidth}{0.5pt}\end{center}

%mwm_snc_250pc_boss_plan
\subsection{mwm\_snc\_250pc\_boss}\label{mwm_snc_250pc_boss_plan}
\begin{description}[nosep]
    \item[Description of selection criteria]  Optically bright targets in the 250 pc volume limited region. 
    \begin{itemize}[nosep]
       \item $G+5\log_{10}(\varpi / 1000) + 5 < 6$
        \item $\varpi - \sigma_\varpi > 4$
        %\item Detected in Gaia
        \item astrometric\_excess\_noise $<2$ if in one of the following regions
        \begin{itemize}[nosep]
        \item $l \leq 180$ AND $ b<-0.139 l + 25$ AND $b>0.139 l-25$
        \item $l>180$ AND $b>-0.139l + 25$ AND $b <0.139l – 25$
        \item $\sqrt{(l-303.2)^2 + 2*(b+44.4)^2 } < 5$ 
        \item $\sqrt{(l-280.3)^2 + 2*(b+33.0)^2} < 8$
        \end{itemize}
    \end{itemize}
\item[Data Sources]  Gaia DR2 ($\varpi$, $\sigma_\varpi$, astrometric\_excess\_noise, $l$, $b$, G), 2MASS PSC (H)
\item[Target priority options] 1810
\item[Cadence options] bright\_2x1, dark\_2x1
\end{description}

\begin{center}\rule{0.5\linewidth}{0.5pt}\end{center}

%mwm_tess_planet_plan
\subsection{mwm\_tess\_planet}\label{mwm_tess_planet_plan}
\begin{description}[nosep]
    \item[Description of selection criteria] Stars with TESS 2-minute cadence data taken during the nominal 2 year mission, prioritizing those that are either TESS Objects of Interest (TOIs) or Community TESS Objects of Interest (CTOIs)
    \begin{itemize}[nosep]
        \item $7 < H < 12$
    \end{itemize}
\item[Data Sources] 2MASS PSC (H), TESS
\item[Target priority options] 2600, 2605, 2610
\item[Cadence options]  bright\_1x1,  bright\_1x2,  bright\_1x3,  bright\_1x4,  bright\_1x5,  bright\_1x6
\end{description}

\begin{center}\rule{0.5\linewidth}{0.5pt}\end{center}

%mwm_tessrgb_core_plan
\subsection{mwm\_tessrgb\_core}\label{mwm_tessrgb_core_plan}
\begin{description}[nosep]
    \item[Description of selection criteria] This carton selects red giant candidates with a color cut and uses a magnitude cut to limit stars likely to have stellar oscillations detectable in TESS light curve data.
    \begin{itemize}[nosep]
        \item $J - K > 0.5 $
        \item $H < 12$ 
        \item $T < 13$
        \item $H -10 + 5\log_{10}\varpi < 1$
        \item $|b| > 20$
    \end{itemize}
\item[Data Sources]  Gaia DR2 ($\varpi$), 2MASS PSC (J, H, K), TESS (T)
\item[Target priority options] 2800
\item[Cadence options] bright\_1x1,  bright\_1x2,  bright\_1x3,  bright\_1x4,  bright\_1x5,  bright\_1x6
\end{description}

\begin{center}\rule{0.5\linewidth}{0.5pt}\end{center}

%mwm_wd_core_plan
\subsection{mwm\_wd\_core}\label{mwm_wd_core_plan}
\begin{description}[nosep]
    \item[Description of selection criteria] All sufficiently bright white dwarf candidates, where $P_{\rm WD}$ is the probability a target is a white dwarf
    \begin{itemize}[nosep]
        \item $G < 20$
        \item $P_{\rm WD} > 0.5$
    \end{itemize}
\item[Data Sources] Gaia DR2 (G),  \citet[][$P_{\rm WD}$]{GentileFusillo2019}
\item[Target priority options] 1400
\item[Cadence options]  dark\_2x1
\end{description}

\begin{center}\rule{0.5\linewidth}{0.5pt}\end{center}

%mwm_yso_cluster_apogee_plan
\subsection{mwm\_yso\_cluster\_apogee}\label{mwm_yso_cluster_apogee_plan}
\begin{description}[nosep]
    \item[Description of selection criteria] Infrared bright YSO candidates associated with young moving groups, as identified by \cite{Kounkel2020}.
    \begin{itemize}[nosep]
        \item $H < 13$
        \item age $< 10^{7.5}$ years
    \end{itemize}
\item[Data Sources] 2MASS PSC (H), \citet[][age]{Kounkel2020}
\item[Target priority options] 2705
\item[Cadence options]  bright\_3x1
\end{description}

\begin{center}\rule{0.5\linewidth}{0.5pt}\end{center}

%mwm_yso_cluster_boss_plan
\subsection{mwm\_yso\_cluster\_boss}\label{mwm_yso_cluster_boss_plan}
\begin{description}[nosep]
    \item[Description of selection criteria] Optically bright YSO candidates associated with young moving groups, as identified by \cite{Kounkel2020}.
    \begin{itemize}[nosep]
        \item $G_{\rm RP} < 15.5$
        \item  age $< 10^{7.5}$ years
    \end{itemize}
\item[Data Sources]  Gaia DR2 ($G_{\rm RP}$), \citet[][age]{Kounkel2020}
\item[Target priority options] 2705
\item[Cadence options]  bright\_3x1, bright\_4x1, bright\_5x1,  bright\_6x1
\end{description}

\begin{center}\rule{0.5\linewidth}{0.5pt}\end{center}

%mwm_yso_cmz_apogee_plan
\subsection{mwm\_yso\_cmz\_apogee}\label{mwm_yso_cmz_apogee_plan}
\begin{description}[nosep]
    \item[Description of selection criteria] YSOs in the central molecular zone (CMZ), the most extreme star-forming environment in the Milky Way 
    \begin{itemize}[nosep]
        \item $H < 13$
        \item $[8.0]-[24] > 2.5$, in \cite{Gutermuth15a} catalog
        \item $\varpi < 0.2$ or not measured
    \end{itemize}
\item[Data Sources]  Gaia DR2 ($\varpi$), 2MASS PSC (H), Spitzer ([8.0], [24])
\item[Target priority options] 2700
\item[Cadence options]  bright\_3x1
\end{description}

\begin{center}\rule{0.5\linewidth}{0.5pt}\end{center}

%mwm_yso_disk_apogee_plan
\subsection{mwm\_yso\_disk\_apogee}\label{mwm_yso_disk_apogee_plan}
\begin{description}[nosep]
    \item[Description of selection criteria] YSO's expected to have protoplanetary disks, as identified by larger infrared excesses.
    \begin{itemize}[nosep]
        \item $H < 13$
        \item $W1-W2 > 0.25$
        \item $W2-W3 > 0.50$
        \item $W3-W4> 1.50$
        \item $\varpi > 0.3$ mas
    \end{itemize}
\item[Data Sources]  Gaia DR2 ($\varpi$), 2MASS PSC (H), AllWise (W1, W2, W3, W4)
\item[Target priority options] 2705
\item[Cadence options]  bright\_3x1
\end{description}

\begin{center}\rule{0.5\linewidth}{0.5pt}\end{center}

%mwm_yso_disk_boss_plan
\subsection{mwm\_yso\_disk\_boss}\label{mwm_yso_disk_boss_plan}
\begin{description}[nosep]
    \item[Description of selection criteria] Optically bright YSO's expected to have protoplanetary disks, as identified by larger infrared excesses.
    \begin{itemize}[nosep]
        \item $G_{\rm RP} < 8.5$
        \item $W1-W2 > 0.25$
        \item $W2-W3 > 0.50$
        \item $W3-W4> 1.50$
        \item $\varpi > 0.3$ mas
    \end{itemize}
\item[Data Sources] Gaia DR2 ( $G_{\rm RP}$, $\varpi$), AllWise (W1, W2, W3, W4)
\item[Target priority options] 2705
\item[Cadence options]  bright\_3x1, bright\_4x1, bright\_5x1,  bright\_6x1
\end{description}

\begin{center}\rule{0.5\linewidth}{0.5pt}\end{center}

%mwm_yso_embedded_apogee_plan
\subsection{mwm\_yso\_embedded\_apogee}\label{mwm_yso_embedded_apogee_plan}
\begin{description}[nosep]
    \item[Description of selection criteria] Deeply embedded YSO candidates too optically faint for parallax measurements have more stringent infrared color cuts to avoid contamination with reddened field stars.
    \begin{itemize}[nosep]
        \item $H < 13$
        \item $G > 18.5$ or undetected
        \item $J-H > 1$
       \item $H-K > 0.5$
        \item $W1-W2 > 0.5$
        \item $W2-W3 > 1$
        \item $W3-W4 > 1.5$
        \item $W3-W4 > 0.8(W1-W2)+1.1$
    \end{itemize}
\item[Data Sources]  Gaia DR2 (G), 2MASS PSC (J, H, K), AllWise (W1, W2, W3, W4)
\item[Target priority options] 2705
\item[Cadence options]  bright\_3x1
\end{description}

\begin{center}\rule{0.5\linewidth}{0.5pt}\end{center}

%mwm_yso_nebula_apogee_plan
\subsection{mwm\_yso\_nebula\_apogee}\label{mwm_yso_nebula_apogee_plan}
\begin{description}[nosep]
    \item[Description of selection criteria] YSO's with strong nebular emission should saturate in the long AllWise bands and are identified with short wavelength color cuts.
    \begin{itemize}[nosep]
        \item $H < 13$
        \item (no W4 AND $W2-W3 > 4)$ OR (no W3, W4 AND $J-H > 1.1$)
        \item $b < 5^\circ$ 
        \item $b > -5^\circ$ OR $l > 180^\circ$
    \end{itemize}
\item[Data Sources]  2MASS PSC (J, H),  AllWise (W1, W2, W3, W4)
\item[Target priority options] 2705
\item[Cadence options]  bright\_3x1
\end{description}

\begin{center}\rule{0.5\linewidth}{0.5pt}\end{center}

%mwm_yso_pms_apogee_plan
\subsection{mwm\_yso\_pms\_apogee}\label{mwm_yso_pms_apogee_plan}
\begin{description}[nosep]
    \item[Description of selection criteria] Infrared bright YSO candidates based on their location above the main sequence, as identified by \cite{Zari2018} or \cite{McBride2021}.
    \begin{itemize}[nosep]
        \item $H < 13$
    \end{itemize}
\item[Data Sources] 2MASS PSC (H)
\item[Target priority options] 2700
\item[Cadence options]  bright\_3x1
\end{description}

\begin{center}\rule{0.5\linewidth}{0.5pt}\end{center}

%mwm_yso_pms_boss_plan
\subsection{mwm\_yso\_pms\_boss}\label{mwm_yso_pms_boss_plan}
\begin{description}[nosep]
    \item[Description of selection criteria] Optically bright YSO candidates based on their location above the main sequence, as identified by \cite{Zari2018} or \cite{McBride2021}.
    \begin{itemize}[nosep]
        \item $G_{\rm RP} < 15.5$
    \end{itemize}
\item[Data Sources]  Gaia DR2 ($G_{\rm RP}$)
\item[Target priority options] 2700
\item[Cadence options]  bright\_3x1, bright\_4x1, bright\_5x1,  bright\_6x1
\end{description}

\begin{center}\rule{0.5\linewidth}{0.5pt}\end{center}

%mwm_yso_variable_apogee_plan
\subsection{mwm\_yso\_variable\_apogee}\label{mwm_yso_variable_apogee_plan}
\begin{description}[nosep]
    \item[Description of selection criteria] YSO candidates that are variable in the Gaia bands, to be observed with APOGEE. Variability in a given band $X$ is defined by $V_X = \sqrt{N_{obs}\sigma_{\overline{I_X}}/(\overline{I_X})}$, where $\overline{I_X}$ is the mean flux and $\sigma_{\overline{I_X}}$ its associated error. Below, $M_{X} = G_{\rm X}-5(\log_{10}(1000/\varpi)-1)$
    \begin{itemize}[nosep]
        \item $H < 13$
        \item $G < 18.5$
        \item $\varpi > 0.3$
        \item $V_G > 0.02$
        \item $V_{BP} > 0.02$
        \item $V_{RP} > 0.02$
        \item $V_G^{0.75} < V_{BP} < V_G$
        \item $0.75 V_G < V_{RP} < V_G^{0.95}$
        \item $G_{\rm BP}-G_{\rm RP} > 1.3$
        \item $M_{BP} > 5 \log_{10}V_{BP} + 11$
        \item $2.5(G_{\rm BP}-G_{\rm RP}) -1 < M_G < 2.5 * (G_{\rm BP}-G_{\rm RP}) + 2.5 $
    \end{itemize}
\item[Data Sources] Gaia DR2 (G, $G_{\rm RP}$, $G_{\rm BP}$, $\varpi$,  $\overline{I_X}$ ,  $\sigma_{\overline{I_X}}$ ), 2MASS PSC (H)
\item[Target priority options] 2705
\item[Cadence options]  bright\_3x1
\end{description}

\begin{center}\rule{0.5\linewidth}{0.5pt}\end{center}

%mwm_yso_variable_boss_plan
\subsection{mwm\_yso\_variable\_boss}\label{mwm_yso_variable_boss_plan}
\begin{description}[nosep]
    \item[Description of selection criteria] YSO candidates that are variable in the Gaia bands, to be observed with BOSS. Variability in a given band $X$ is defined by $V_X = \sqrt{N_{obs} \sigma_{\overline{I_X}}/(\overline{I_X})}$, where $\overline{I_X}$ is the mean flux and $\sigma_{\overline{I_X}}$ its associated error. Below, $M_{X} = G_{\rm X}-5(\log_{10}(1000/\varpi)-1)$
    \begin{itemize}[nosep]
        \item $H < 13$
        \item $G < 18.5$
        \item $\varpi > 0.3$
        \item $V_G > 0.02$
        \item $V_{BP} > 0.02$
        \item $V_{RP} > 0.02$
        \item $V_G^{0.75} < V_{BP} < V_G$
        \item $0.75 V_G < V_{RP} < V_G^{0.95}$
        \item $G_{\rm BP}-G_{\rm RP} > 1.3$
        \item $M_{BP} > 5 \log_{10}V_{BP} + 11$
        \item $2.5(G_{\rm BP}-G_{\rm RP}) -1 < M_G < 2.5 * (G_{\rm BP}-G_{\rm RP}) + 2.5 $
    \end{itemize}
\item[Data Sources] Gaia DR2 (G, $G_{\rm RP}$, $G_{\rm BP}$, $\varpi$,  $\overline{I_X}$ ,  $\sigma_{\overline{I_X}}$ ), 2MASS PSC (H)
\item[Target priority options] 2705
\item[Cadence options]  bright\_3x1, bright\_4x1, bright\_5x1,  bright\_6x1
\end{description}

\section{Details of BHM \texttt{v0.5.3} target cartons}
\label{sec:bhm_cartons_appendix}
In this appendix we provide a more detailed description of the full set of 
BHM target cartons that are being released as part of SDSS DR18. We describe not only the ``core'' BHM cartons in Table \ref{tab:bhm_cartons}, but also all ancillary/supplementary BHM cartons, which collectively expand the scope, footprint, and depth of the project beyond the core science goals.
We remind the reader that inclusion in a carton means that the target fits the carton selection criteria, not that it is guaranteed to be observed.

For each target carton, we provide the following information: 
The \textbf{target\_selection plan} and \textbf{target\_selection tag} codes 
give the software and configuration versions of the \texttt{target\_selection} 
code that were used in this carton instance.  
In \textbf{Summary}, we provide a short synopsis of the content of the target carton, followed by a \textbf{Simplified description of selection criteria} with a human-readable 
description of the specific selection criteria that have been applied to build the carton, including quantitative limits on colors, magnitudes, and other quantities.
The \textbf{Target priority options} and \textbf{Cadence options} indicate which ``priority'' and ``cadence'' option(s) have been allocated to the targets in this carton. Targets with numerically smaller priority are more likely to be assigned fibers; the cadence describes a target's exposure time requirement. 
Under \textbf{Implementation}, we point to the 
specific section of the \texttt{target\_selection} Python code that implements this carton\footnote{
These can be found at URLs like: \url{https://github.com/sdss/target_selection/blob/0.3.0/python/target_selection/cartons/bhm_aqmes.py}, where `0.3.0' is the \texttt{target\_selection} git tag in this example.}.
Finally, the \textbf{Number of targets} lists the number of targets which pass all of the carton selection criteria. Note that targets are often shared between two or more cartons.

\begin{center}\rule{0.5\linewidth}{0.5pt}\end{center}

\hypertarget{bhm_aqmes_med_plan0.5.0}{%
\subsection{bhm\_aqmes\_med}\label{bhm_aqmes_med_plan0.5.0}}

\noindent\textbf{target\_selection plan:} 0.5.0

\noindent\textbf{target\_selection tag:}
\href{https://github.com/sdss/target_selection/tree/0.3.0/}{0.3.0}

\noindent\textbf{Summary:} Spectroscopically confirmed optically bright SDSS
QSOs, selected from the SDSS QSO catalog (DR16Q,
\citealt{Lyke2020}). Located in 36 mostly disjoint fields within the SDSS QSO
footprint that were pre-selected to contain higher than average numbers
of bright QSOs and CSC targets. The list of field centers can be found
within
\href{https://github.com/sdss/target_selection/blob/0.3.0/python/target_selection/masks/candidate_target_fields_bhm_aqmes_med_v0.3.1.fits}{the
target\_selection repository}.

\noindent\textbf{Simplified description of selection criteria:} Select all
objects from SDSS DR16 QSO catalog that have SDSS
$16.0 < i_{\mathrm{psf}} < 19.1$~AB, and that lie within
1.49~degrees of at least one AQMES-medium field location.

\noindent\textbf{Target priority options:} 1100

\noindent\textbf{Cadence options:} dark\_10x4\_4yr

\noindent\textbf{Implementation:}
\href{https://github.com/sdss/target_selection/blob/0.3.0/python/target_selection/cartons/bhm_aqmes.py}{bhm\_aqmes.py}

\noindent\textbf{Number of targets:} 2663

\begin{center}\rule{0.5\linewidth}{0.5pt}\end{center}

\hypertarget{bhm_aqmes_med_faint_plan0.5.0}{%
\subsection{bhm\_aqmes\_med\_faint}\label{bhm_aqmes_med_faint_plan0.5.0}}

\noindent\textbf{target\_selection plan:} 0.5.0

\noindent\textbf{target\_selection tag:}
\href{https://github.com/sdss/target_selection/tree/0.3.0/}{0.3.0}

\noindent\textbf{Summary:} Spectroscopically confirmed optically faint SDSS QSOs,
selected from the SDSS QSO catalog (DR16Q,
\citealt{Lyke2020}). Located in 36 mostly disjoint fields within the SDSS QSO
footprint that were pre-selected to contain higher than average numbers
of bright QSOs and CSC targets. The list of field centres can be found
within
\href{https://github.com/sdss/target_selection/blob/0.3.0/python/target_selection/masks/candidate_target_fields_bhm_aqmes_med_v0.3.1.fits}{the
target\_selection repository}.

\noindent\textbf{Simplified description of selection criteria:} Select all
objects from SDSS DR16 QSO catalog that have SDSS
$19.1 < i_{\mathrm{psf}} < 21.0$~AB, and that lie within
1.49~degrees of at least one AQMES-medium field location.

\noindent\textbf{Target priority options:} 3100

\noindent\textbf{Cadence options:} dark\_10x4\_4yr

\noindent\textbf{Implementation:}
\href{https://github.com/sdss/target_selection/blob/0.3.0/python/target_selection/cartons/bhm_aqmes.py}{bhm\_aqmes.py}

\noindent\textbf{Number of targets:} 16853

\begin{center}\rule{0.5\linewidth}{0.5pt}\end{center}

\hypertarget{bhm_aqmes_wide2_plan0.5.4}{%
\subsection{bhm\_aqmes\_wide2}\label{bhm_aqmes_wide2_plan0.5.4}}

\noindent\textbf{target\_selection plan:} 0.5.4

\noindent\textbf{target\_selection tag:}
\href{https://github.com/sdss/target_selection/tree/0.3.5/}{0.3.5}

\noindent\textbf{Summary:} Spectroscopically confirmed optically bright SDSS
QSOs, selected from the SDSS QSO catalog (DR16Q,
\citealt{Lyke2020}). Located in 425 fields within the SDSS QSO footprint,
where the choice of survey area prioritized field that overlapped with
the SPIDERS footprint (approx 180\textless{}l\textless{}360~deg), and/or
had higher than average numbers of bright QSOs and CSC targets. The list
of field centers can be found
\href{https://github.com/sdss/target_selection/blob/0.3.0/python/target_selection/masks/candidate_target_fields_bhm_aqmes_wide_v0.3.1.fits}{within
the target\_selection repository}.

\noindent\textbf{Simplified description of selection criteria:} Select all
objects from SDSS DR16 QSO catalog that have SDSS
$16.0 < i_{\mathrm{psf}} < 19.1$~AB, and that lie within
1.49~degrees of at least one AQMES-wide field location.

\noindent\textbf{Target priority options:} 1210, 1211

\noindent\textbf{Cadence options:} dark\_2x4

\noindent\textbf{Implementation:}
\href{https://github.com/sdss/target_selection/blob/0.3.5/python/target_selection/cartons/bhm_aqmes.py}{bhm\_aqmes.py}

\noindent\textbf{Number of targets:} 24142

\begin{center}\rule{0.5\linewidth}{0.5pt}\end{center}

\hypertarget{bhm_aqmes_wide2_faint_plan0.5.4}{%
\subsection{bhm\_aqmes\_wide2\_faint}\label{bhm_aqmes_wide2_faint_plan0.5.4}}

\noindent\textbf{target\_selection plan:} 0.5.4

\noindent\textbf{target\_selection tag:}
\href{https://github.com/sdss/target_selection/tree/0.3.5/}{0.3.5}

\noindent\textbf{Summary:} Spectroscopically confirmed optically faint SDSS QSOs,
selected from the SDSS QSO catalog (DR16Q,
\citealt{Lyke2020}). Located in 425 fields within the SDSS QSO footprint,
where the choice of survey area prioritized field that overlapped with
the SPIDERS footprint (approx 180\textless{}l\textless{}360~deg), and/or
had higher than average numbers of bright QSOs and CSC targets. The list
of field centres can be found
\href{https://github.com/sdss/target_selection/blob/0.3.0/python/target_selection/masks/candidate_target_fields_bhm_aqmes_wide_v0.3.1.fits}{within
the target\_selection repository}.

\noindent\textbf{Simplified description of selection criteria:} Select all
objects from SDSS DR16 QSO catalog that have SDSS
$19.1 < i_{\mathrm{psf}} < 21.0$~AB, and that lie within
1.49~degrees of at least one AQMES-wide field location.

\noindent\textbf{Target priority options:} 3210, 3211

\noindent\textbf{Cadence options:} dark\_2x4

\noindent\textbf{Implementation:}
\href{https://github.com/sdss/target_selection/blob/0.3.5/python/target_selection/cartons/bhm_aqmes.py}{bhm\_aqmes.py}

\noindent\textbf{Number of targets:} 99586

\begin{center}\rule{0.5\linewidth}{0.5pt}\end{center}

\hypertarget{bhm_aqmes_bonus_core_plan0.5.4}{%
\subsection{bhm\_aqmes\_bonus\_core}\label{bhm_aqmes_bonus_core_plan0.5.4}}

\noindent\textbf{target\_selection plan:} 0.5.4

\noindent\textbf{target\_selection tag:}
\href{https://github.com/sdss/target_selection/tree/0.3.5/}{0.3.5}

\noindent\textbf{Summary:} Spectroscopically confirmed optically bright SDSS
QSOs, selected from the SDSS QSO catalog \citep[DR16Q;][]{Lyke2020}. Located anywhere within the SDSS DR16Q footprint.

\noindent\textbf{Simplified description of selection criteria:} Select all
objects from SDSS DR16 QSO catalog that have SDSS
$16.0 < i_{\mathrm{psf}} < 19.1$~AB.

\noindent\textbf{Target priority options:} 3300,3301

\noindent\textbf{Cadence options:} dark\_1x4

\noindent\textbf{Implementation:}
\href{https://github.com/sdss/target_selection/blob/0.3.5/python/target_selection/cartons/bhm_aqmes.py}{bhm\_aqmes.py}

\noindent\textbf{Number of targets:} 83163

\begin{center}\rule{0.5\linewidth}{0.5pt}\end{center}

\hypertarget{bhm_aqmes_bonus_faint_plan0.5.4}{%
\subsection{bhm\_aqmes\_bonus\_faint}\label{bhm_aqmes_bonus_faint_plan0.5.4}}

\noindent\textbf{target\_selection plan:} 0.5.4

\noindent\textbf{target\_selection tag:}
\href{https://github.com/sdss/target_selection/tree/0.3.5/}{0.3.5}

\noindent\textbf{Summary:} Spectroscopically confirmed optically faint SDSS QSOs,
selected from the SDSS QSO catalog \citep[DR16Q;][]{Lyke2020}. Located anywhere within the SDSS DR16Q footprint.

\noindent\textbf{Simplified description of selection criteria:} Select all
objects from SDSS DR16 QSO catalog that have SDSS
$19.1 < i_{\mathrm{psf}} < 21.0$~AB.

\noindent\textbf{Target priority options:} 3302,3303

\noindent\textbf{Cadence options:} dark\_1x4

\noindent\textbf{Implementation:}
\href{https://github.com/sdss/target_selection/blob/0.3.5/python/target_selection/cartons/bhm_aqmes.py}{bhm\_aqmes.py}

\noindent\textbf{Number of targets:} 424163

\begin{center}\rule{0.5\linewidth}{0.5pt}\end{center}

\hypertarget{bhm_aqmes_bonus_bright_plan0.5.4}{%
\subsection{bhm\_aqmes\_bonus\_bright}\label{bhm_aqmes_bonus_bright_plan0.5.4}}

\noindent\textbf{target\_selection plan:} 0.5.4

\noindent\textbf{target\_selection tag:}
\href{https://github.com/sdss/target_selection/tree/0.3.5/}{0.3.5}

\noindent\textbf{Summary:} Spectroscopically confirmed, extremely optically
bright SDSS QSOs, selected from the SDSS QSO catalog \citep[DR16Q;][]{Lyke2020}. Located anywhere within the SDSS DR16Q footprint.

\noindent\textbf{Simplified description of selection criteria:} Select all
objects from SDSS DR16 QSO catalog that have SDSS
$14.0 < i_{\mathrm{psf}} < 18.0$~AB.

\noindent\textbf{Target priority options:} 4040,4041

\noindent\textbf{Cadence options:} bright\_3x1

\noindent\textbf{Implementation:}
\href{https://github.com/sdss/target_selection/blob/0.3.5/python/target_selection/cartons/bhm_aqmes.py}{bhm\_aqmes.py}

\noindent\textbf{Number of targets:} 10848

\begin{center}\rule{0.5\linewidth}{0.5pt}\end{center}

\hypertarget{bhm_rm_ancillary_plan0.5.0}{%
\subsection{bhm\_rm\_ancillary}\label{bhm_rm_ancillary_plan0.5.0}}

\noindent\textbf{target\_selection plan:} 0.5.0

\noindent\textbf{target\_selection tag:}
\href{https://github.com/sdss/target_selection/tree/0.3.0/}{0.3.0}

\noindent\textbf{Summary:} A supporting sample of candidate QSOs that have been
selected by the Gaia-unWISE AGN catalog
(\citealt{Shu2019}) and/or the SDSS XDQSO catalog
(\citealt{Bovy2011}). These targets are located within five (+1 backup) well
known survey fields (SDSS-RM, COSMOS, XMM-LSS, ECDFS, CVZ-S/SEP, and
ELIAS-S1).

\noindent\textbf{Simplified description of selection criteria:} Starting from a
parent catalog of optically selected objects in the RM fields (as
presented by
\citealt{Yang2022}), select candidate QSOs that satisfy all of the
following: i) are identified via external ancillary methods
(photo\_bitmask \& 3 != 0); ii) have
$15 < i_{\mathrm{psf}} < 21.5$~AB
\citep[or $16 < G < 21.7$~Vega in the CVZ-S/SEP field, photometry taken from][]{Yang2022};
iii) do not
have significant detections ($>$3$\sigma$) of parallax and/or proper
motion in Gaia~DR2; iv) are not vetoed due to results of visual
inspections of recent spectroscopy; and v) do not lie in the SDSS-RM
field.

\noindent\textbf{Target priority options:} 900-1050

\noindent\textbf{Cadence options:} dark\_174x8, dark\_100x8

\noindent\textbf{Implementation:}
\href{https://github.com/sdss/target_selection/blob/0.3.0/python/target_selection/cartons/bhm_rm.py}{bhm\_rm.py}

\noindent\textbf{Number of targets:} 943

\begin{center}\rule{0.5\linewidth}{0.5pt}\end{center}

\hypertarget{bhm_rm_core_plan0.5.0}{%
\subsection{bhm\_rm\_core}\label{bhm_rm_core_plan0.5.0}}

\noindent\textbf{target\_selection plan:} 0.5.0

\noindent\textbf{target\_selection tag:}
\href{https://github.com/sdss/target_selection/tree/0.3.0/}{0.3.0}

\noindent\textbf{Summary:} A sample of candidate QSOs selected via the methods
presented by
\citet{Yang2022}. These targets are located within five (+1 backup) well
known survey fields (SDSS-RM, COSMOS, XMM-LSS, ECDFS, CVZ-S/SEP, and
ELIAS-S1).

\noindent\textbf{Simplified description of selection criteria:} Starting from a
parent catalog of optically selected objects in the RM fields (as
presented by
\citealt{Yang2022}), select candidate QSOs that satisfy all of the
following: i) are identified via the Skew-T algorithm (skewt\_qso == 1);
ii) have $17 < i_{\mathrm{psf}} < 21.5$~AB
\citep[or $16 < G < 21.7$~Vega in the CVZ-S/SEP field, photometry taken from][]{Yang2022};
iii) do not
have significant detections ($>3\sigma$) of parallax and/or proper
motion in Gaia DR2; iv) are not vetoed due to results of visual
inspections of recent spectroscopy; v) have detections in all of the
$gri$ bands (a Gaia detection is sufficient in the CVZ-S/SEP field); and
vi) do not lie in the SDSS-RM field.

\noindent\textbf{Target priority options:} 900-1050

\noindent\textbf{Cadence options:} dark\_174x8, dark\_100x8

\noindent\textbf{Implementation:}
\href{https://github.com/sdss/target_selection/blob/0.3.0/python/target_selection/cartons/bhm_rm.py}{bhm\_rm.py}

\noindent\textbf{Number of targets:} 3721

\begin{center}\rule{0.5\linewidth}{0.5pt}\end{center}

\hypertarget{bhm_rm_var_plan0.5.0}{%
\subsection{bhm\_rm\_var}\label{bhm_rm_var_plan0.5.0}}

\noindent\textbf{target\_selection plan:} 0.5.0

\noindent\textbf{target\_selection tag:}
\href{https://github.com/sdss/target_selection/tree/0.3.0/}{0.3.0}

\noindent\textbf{Summary:} A sample of candidate QSOs selected via their optical
variability properties, as presented by
\citet{Yang2022}. These targets are located within five (+1 backup) well
known survey fields (SDSS-RM, COSMOS, XMM-LSS, ECDFS, CVZ-S/SEP, and
ELIAS-S1).

\noindent\textbf{Simplified description of selection criteria:} Starting from a
parent catalog of optically selected objects in the RM fields (as
presented by
\citealt{Yang2022}), select candidate QSOs that satisfy all of the
following: i) have significant variability in the DES or PanSTARRS1
multi-epoch photometry
($\mathrm{var\_sn[}g\mathrm{]}>3$ and $\mathrm{var\_rms[}g\mathrm{]}>0.05$);
%(var\_sn{[}g{]}\textgreater{}3 and %var\_rms{[}g{]}\textgreater{}0.05);
ii) have $17 < i_{\mathrm{psf}} < 20.5$~AB
\citep[or $16 < G < 21.7$~Vega in the CVZ-S/SEP field, photometry taken from][]{Yang2022};
%17\textless{}psfmag\_i\textless{}20.5~AB
%($16 < G < 21.7$~Vega in the CVZ-S/SEP field);
iii) do not
have significant detections ($>3\sigma$) of parallax and/or proper
motion in Gaia DR2; iv) are not vetoed due to results of visual
inspections of recent spectroscopy; and v) do not lie in the SDSS-RM
field.

\noindent\textbf{Target priority options:} 900-1050

\noindent\textbf{Cadence options:} dark\_174x8, dark\_100x8

\noindent\textbf{Implementation:}
\href{https://github.com/sdss/target_selection/blob/0.3.0/python/target_selection/cartons/bhm_rm.py}{bhm\_rm.py}

\noindent\textbf{Number of targets:} 934

\begin{center}\rule{0.5\linewidth}{0.5pt}\end{center}

\hypertarget{bhm_rm_known_spec_plan0.5.0}{%
\subsection{bhm\_rm\_known\_spec}\label{bhm_rm_known_spec_plan0.5.0}}

\noindent\textbf{target\_selection plan:} 0.5.0

\noindent\textbf{target\_selection tag:}
\href{https://github.com/sdss/target_selection/tree/0.3.0/}{0.3.0}

\noindent\textbf{Summary:} A sample of known QSOs identified through optical
spectroscopy from various projects, as collated by
\citet{Yang2022}. These targets are located within five (+1 backup) well
known survey fields (SDSS-RM, COSMOS, XMM-LSS, ECDFS, CVZ-S/SEP, and
ELIAS-S1).

\noindent\textbf{Simplified description of selection criteria:} Starting from a
parent catalog of optically selected objects in the RM fields (as
presented by
\citealt{Yang2022}), select targets which satisfy all of the following: i)
are flagged as having a spectroscopic identification (in the parent
catalog or in the bhm\_rm\_tweaks table); ii) have
$15 < i_{\mathrm{psf}} < 21.7$~AB (SDSS-RM, CDFS, ELIAS-S1 fields),
$15 < i_{\mathrm{psf}} < 21.5$~AB (COSMOS and XMM-LSS fields),
$16 < G < 21.7$~Vega in the CVZ-S/SEP field
\citep[photometry taken from][]{Yang2022};
%15\textless{}psfmag\_i\textless{}21.7~AB (SDSS-RM, CDFS, ELIAS-S1
%field), 16\textless{}G\textless{}21.7~Vega (CVZ-S/SEP field),
%15\textless{}psfmag\_i\textless{}21.5~AB (COSMOS and XMM-LSS fields);
iii) have a spectroscopic redshift in the range
0.005\textless{}z\textless{}7; and iv) are not vetoed due to results of
visual inspections of recent spectroscopy.

\noindent\textbf{Target priority options:} 900-1050

\noindent\textbf{Cadence options:} dark\_174x8, dark\_100x8

\noindent\textbf{Implementation:}
\href{https://github.com/sdss/target_selection/blob/0.3.0/python/target_selection/cartons/bhm_rm.py}{bhm\_rm.py}

\noindent\textbf{Number of targets:} 3022

\begin{center}\rule{0.5\linewidth}{0.5pt}\end{center}

\hypertarget{bhm_csc_apogee_plan0.5.15}{%
\subsection{bhm\_csc\_apogee}\label{bhm_csc_apogee_plan0.5.15}}

\noindent\textbf{target\_selection plan:} 0.5.15

\noindent\textbf{target\_selection tag:}
\href{https://github.com/sdss/target_selection/tree/0.3.14/}{0.3.14}

\noindent\textbf{Summary:} X-ray sources from the CSC2.0 source catalog with
NIR counterparts in 2MASS PSC

\noindent\textbf{Simplified description of selection criteria:} Starting from the
parent catalog of CSC2.0 sources associated with optical/IR counterparts
(\texttt{bhm\_csc\_v2}). Select entries satisfying the following criteria: i) NIR
counterpart is from the 2MASS catalog, ii) 2MASS $H$-band magnitude
measurement is not null and in the accepted range for SDSS-V:
$7.0 < H < 14.0$. Allocate cadence (exposure time)
requests based on $H$ magnitude.

\noindent\textbf{Target priority options:} 2930-2939

\noindent\textbf{Cadence options:} bright\_1x1,bright\_3x1

\noindent\textbf{Implementation:}
\href{https://github.com/sdss/target_selection/blob/0.3.14/python/target_selection/cartons/bhm_csc.py}{bhm\_csc.py}

\noindent\textbf{Number of targets:} 48928

\begin{center}\rule{0.5\linewidth}{0.5pt}\end{center}

\hypertarget{bhm_csc_boss_plan0.5.15}{%
\subsection{bhm\_csc\_boss}\label{bhm_csc_boss_plan0.5.15}}

\noindent\textbf{target\_selection plan:} 0.5.15

\noindent\textbf{target\_selection tag:}
\href{https://github.com/sdss/target_selection/tree/0.3.14/}{0.3.14}

\noindent\textbf{Summary:} X-ray sources from the CSC2.0 source catalog with
counterparts in Panstarrs1-DR1 or Gaia DR2.

\noindent\textbf{Simplified description of selection criteria:} Starting from the
parent catalog of CSC2.0 sources associated with optical/IR counterparts
(\texttt{bhm\_csc\_v2}). Select entries satisfying the following criteria: i)
optical counterpart is from the PanSTARRS1 or Gaia DR2 catalogs, ii)
optical flux/magnitude is in the accepted range for SDSS-V:
PanSTARRS1 $g_{\mathrm{psf}},r_{\mathrm{psf}},i_{\mathrm{psf}},z_{\mathrm{psf}} > 13.5$~AB, and
%psfmag\_{[}g,r,i,z{]}\textgreater{}13.5~AB and
non-Null $i_{\mathrm{psf}}$ (objects with PanSTARRS1 counterparts); $G,G_{RP} > 13.0$~Vega
(Gaia DR2 counterparts). Deprioritize targets which already have good
quality SDSS spectroscopy. Allocate cadence (exposure time) requests
based on optical brightness (PanSTARRS1 $i_{\mathrm{psf}}$ or Gaia $G$).

\noindent\textbf{Target priority options:} 1920-1939, 2920-2939

\noindent\textbf{Cadence options:} bright\_1x1,dark\_1x2,dark\_1x4

\noindent\textbf{Implementation:}
\href{https://github.com/sdss/target_selection/blob/0.3.14/python/target_selection/cartons/bhm_csc.py}{bhm\_csc.py}

\noindent\textbf{Number of targets:} 122731

\begin{center}\rule{0.5\linewidth}{0.5pt}\end{center}

\hypertarget{bhm_gua_bright_plan0.5.0}{%
\subsection{bhm\_gua\_bright}\label{bhm_gua_bright_plan0.5.0}}

\noindent\textbf{target\_selection plan:} 0.5.0

\noindent\textbf{target\_selection tag:}
\href{https://github.com/sdss/target_selection/tree/0.3.0/}{0.3.0}

\noindent\textbf{Summary:} A sample of optically bright candidate AGN lacking
spectroscopic confirmations, derived from the parent sample presented by
\citet{Shu2019}, who applied a machine-learning approach to select QSO
candidates from a combination of the Gaia DR2 and unWISE catalogs.

\noindent\textbf{Simplified description of selection criteria:} Starting with the
\citet{Shu2019} catalog, select targets that satisfy the following
criteria: i) have a Random Forest probability of being a QSO
of \textgreater{}0.8, ii) are in the magnitude range suitable for BOSS
spectroscopy in bright time ($G_{\mathrm{dered}} > 13.0$ and
$G_{RP,\mathrm{dered}} > 13.5$, as well as $G_{\mathrm{dered}} < 18.5$ or
$G_{RP,\mathrm{dered}} < 18.5$\, Vega), and iii) do not have good optical
spectroscopic measurements from a previous iteration of SDSS.

\noindent\textbf{Target priority options:} 4040

\noindent\textbf{Cadence options:} bright\_2x1

\noindent\textbf{Implementation:}
\href{https://github.com/sdss/target_selection/blob/0.3.0/python/target_selection/cartons/bhm_gua.py}{bhm\_gua.py}

\noindent\textbf{Number of targets:} 254601

\begin{center}\rule{0.5\linewidth}{0.5pt}\end{center}

\hypertarget{bhm_gua_dark_plan0.5.0}{%
\subsection{bhm\_gua\_dark}\label{bhm_gua_dark_plan0.5.0}}

\noindent\textbf{target\_selection plan:} 0.5.0

\noindent\textbf{target\_selection tag:}
\href{https://github.com/sdss/target_selection/tree/0.3.0/}{0.3.0}

\noindent\textbf{Summary:} A sample of optically faint candidate AGN lacking
spectroscopic confirmations, derived from the parent sample presented by
\citet{Shu2019}, who applied a machine-learning approach to select QSO
candidates from a combination of the Gaia DR2 and unWISE catalogs.

\noindent\textbf{Simplified description of selection criteria:} Starting with the
\citet{Shu2019} catalog, select targets which satisfy the following
criteria: i) have a Random Forest probability of being a QSO
of \textgreater{}0.8, ii) are in the magnitude range suitable for BOSS
spectroscopy in dark time ($G_{\mathrm{dered}} > 16.5$ and
$G_{RP,\mathrm{dered}} > 16.5$,
as well as $G_{\mathrm{dered}} < 21.2$ or
$G_{RP,\mathrm{dered}} < 21.0$\, Vega), and iii) do not
have good optical spectroscopic measurements from a previous iteration
of SDSS.

\noindent\textbf{Target priority options:} 3400

\noindent\textbf{Cadence options:} dark\_1x4

\noindent\textbf{Implementation:}
\href{https://github.com/sdss/target_selection/blob/0.3.0/python/target_selection/cartons/bhm_gua.py}{bhm\_gua.py}

\noindent\textbf{Number of targets:} 2156582

\begin{center}\rule{0.5\linewidth}{0.5pt}\end{center}

\hypertarget{bhm_colr_galaxies_lsdr8_plan0.5.16}{%
\subsection{bhm\_colr\_galaxies\_lsdr8}\label{bhm_colr_galaxies_lsdr8_plan0.5.16}}

\noindent\textbf{target\_selection plan:} 0.5.16

\noindent\textbf{target\_selection tag:}
\href{https://github.com/sdss/target_selection/tree/0.3.13/}{0.3.13}

\noindent\textbf{Summary:} A supplementary magnitude limited sample of optically
bright galaxies selected from the DESI Legacy Survey DR8 optical/IR
imaging catalog. Selection is based on optical morphology, lack of
Gaia DR2 parallax, and several magnitude cuts.

\noindent\textbf{Simplified description of selection criteria:} Starting from the
DESI Legacy Survey DR8 catalog (lsdr8), select entries satisfying all of
the following criteria: i) lsdr8 morphological type != `PSF', ii) zero
or Null parallax in Gaia DR2, 
iii) $z_{\mathrm{model,dered}} < 19.0$~AB,
and $z_{\mathrm{fiber,dered}} < 19.5$~AB,
and $z_{\mathrm{fiber}} < 19.0$~AB,
and $r_{\mathrm{fiber}} > 16.0$~AB,
%iii) dereddened z-band model
%mag\textless{}19.0~AB, and dereddened z-band fiber
%mag\textless{}19.5~AB, and 16\textless{}apparent z-band fiber
%mag\textless{}19.0~AB, and apparent r-band fiber
%mag\textgreater{}16.0~AB,
and $G>15.0$~Vega, and $G_{RP}>15.0$~Vega
(photometry from lsdr8).

\noindent\textbf{Target priority options:} 7100

\noindent\textbf{Cadence options:} bright\_1x1, dark\_1x1, dark\_1x4

\noindent\textbf{Implementation:}
\href{https://github.com/sdss/target_selection/blob/0.3.13/python/target_selection/cartons/bhm_galaxies.py}{bhm\_galaxies.py}

\noindent\textbf{Number of targets:} 7320203

\begin{center}\rule{0.5\linewidth}{0.5pt}\end{center}

\hypertarget{bhm_spiders_agn_lsdr8_plan0.5.0}{%
\subsection{bhm\_spiders\_agn\_lsdr8}\label{bhm_spiders_agn_lsdr8_plan0.5.0}}

\noindent\textbf{target\_selection plan:} 0.5.0

\noindent\textbf{target\_selection tag:}
\href{https://github.com/sdss/target_selection/tree/0.3.0/}{0.3.0}

\noindent\textbf{Summary:} This is the highest priority carton for SPIDERS AGN
wide area follow up. The carton provides optical counterparts to
point-like (unresolved) X-ray sources detected in early reductions of
the first 6-months of eROSITA all sky survey data (eRASS:1). The sample
is expected to contain a mixture of QSOs, AGN, stars and compact
objects. The X-ray sources have been cross-matched by the eROSITA-DE
team to DESI Legacy Survey DR8 (lsdr8)
optical/IR counterparts. All targets are located in the sky hemisphere
where MPE controls the data rights (approx.
180\textless{}l\textless{}360~deg). Due to the footprint of lsdr8,
nearly all targets in this carton are located at high Galactic latitudes
($|b|>15$~deg).

\noindent\textbf{Simplified description of selection criteria:} Starting from a
parent catalog of eRASS:1 point source $\rightarrow$ legacysurvey.org/dr8
associations (method: NWAY assisted by optical/IR priors computed via a
pre-trained Random Forest, building on
\citealt{Salvato2022}), select targets which meet all of the following criteria:
i) have eROSITA detection likelihood\textgreater{}6.0, ii) have an X-ray
$\rightarrow$ optical/IR cross-match probability (NWAY) of p\_any\textgreater{}0.1,
iii) have $13.5 < r_{\mathrm{fibertot}} < 22.5$ or
$13.5 < z_{\mathrm{fibertot}} < 21.0$~AB, iv) are not saturated in
Legacy Survey imaging, v) have at least one observation in r-band and at
least one observation in g- or z-band, and vi) if detected by Gaia DR2 then
have $G>13.5$ and $G_{RP} > 13.5$~Vega (photometry from lsdr8).
%iii) have 13.5\textless{}fibertotmag\_r\textless{}22.5 or
%13.5\textless{}fibertotmag\_z\textless{}21.0, iv) are not saturated in
%legacysurvey imaging, v) have at least one observation in r-band and at
%least one observation in g- or z-band, and vi) if detected by Gaia DR2 then
%have G\textgreater{}13.5 and RP\textgreater{}13.5~Vega.
We deprioritize
targets if any of the following criteria are met: a) the target already
has existing good quality SDSS spectroscopy, b) the X-ray detection
likelihood is \textless{}8.0, or c) the target is a secondary
X-ray$\rightarrow$optical/IR association. We assign cadences (exposure time
requests) based on optical brightness.

\noindent\textbf{Target priority options:} 1520-1523, 1720-1723

\noindent\textbf{Cadence options:} bright\_2x1, dark\_1x2, dark\_1x4

\noindent\textbf{Implementation:}
\href{https://github.com/sdss/target_selection/blob/0.3.0/python/target_selection/cartons/bhm_spiders_agn.py}{bhm\_spiders\_agn.py}

\noindent\textbf{Number of targets:} 235745

\begin{center}\rule{0.5\linewidth}{0.5pt}\end{center}

\hypertarget{bhm_spiders_agn_ps1dr2_plan0.5.0}{%
\subsection{bhm\_spiders\_agn\_ps1dr2}\label{bhm_spiders_agn_ps1dr2_plan0.5.0}}

\noindent\textbf{target\_selection plan:} 0.5.0

\noindent\textbf{target\_selection tag:}
\href{https://github.com/sdss/target_selection/tree/0.3.0/}{0.3.0}

\noindent\textbf{Summary:} This is the second highest priority carton for SPIDERS
AGN wide area follow up, included to expand the survey footprint
beyond Legacy Survey DR8. The carton provides optical counterparts to
point-like (unresolved) X-ray sources detected in early reductions of
the first 6-months of eROSITA all sky survey data (eRASS:1). The sample
is expected to contain a mixture of QSOs, AGN, stars and compact
objects. The X-ray sources have been cross-matched by the eROSITA-DE
team, first to CatWISE2020  mid-IR sources \citep{Marocco2021}, 
and then to optical counterparts from the
\href{https://outerspace.stsci.edu/display/PANSTARRS/}{PanSTARRS1 DR2}
catalog. All targets are located in the sky hemisphere where MPE
controls the data rights (approx. 180\textless{}l\textless{}360~deg),
and at Dec\textgreater{}$-30$~deg, spanning a wide range of Galactic
latitudes. Targets at low Galactic latitudes
($|b|<15$~deg) do not drive survey strategy.

\noindent\textbf{Simplified description of selection criteria:} Starting from a
parent catalog of eRASS:1 point source $\rightarrow$ CatWISE2020 $\rightarrow$ PanSTARRS1
associations (method: NWAY assisted by IR priors computed via a
pre-trained Random Forest, building on
\citealt{Salvato2022}), select targets which meet all of the following criteria:
i) have eROSITA detection likelihood\textgreater{}6.0, ii) have an
X-ray$\rightarrow$IR cross-match probability of p\_any\textgreater{}0.1, iii) have
PanSTARRS1 $g_{\mathrm{psf}},r_{\mathrm{psf}},i_{\mathrm{psf}},z_{\mathrm{psf}} > 13.5$~AB and at
%psfmag\_g, psfmag\_r, psfmag\_i, psfmag\_z\textgreater{}13.5~AB and at
least one of PanSTARRS1 $g_{\mathrm{psf}} < 22.5$, $r_{\mathrm{psf}}< 22.0$, $i_{\mathrm{psf}} < 21.5$ or $z_{\mathrm{psf}} <20.5$~AB,
%psfmag\_g\textless{}22.5, psfmag\_r\textless{}22.0,
%psfmag\_i\textless{}21.5 or psfmag\_z\textless{}20.5,
iv) are not
associated with a bad PanSTARRS1 image stack, v) have non-null
measurements of $g_{\mathrm{psf}},r_{\mathrm{psf}}$ and $i_{\mathrm{psf}}$, and vi) if detected by
Gaia DR2 then have $G>13.5$ and $G_{RP} > 13.5$~Vega. We
deprioritize targets if any of the following criteria are met: a) the
target already has existing good quality SDSS spectroscopy, b) the X-ray
detection likelihood is \textless{}8.0, or c) the target is a secondary
X-ray$\rightarrow$IR association. We assign cadences (exposure time requests) based
on optical brightness.

\noindent\textbf{Target priority options:}
1530-1533,1730-1733,3530-3533,3730-3732

\noindent\textbf{Cadence options:} bright\_2x1, dark\_1x2, dark\_1x4

\noindent\textbf{Implementation:}
\href{https://github.com/sdss/target_selection/blob/0.3.0/python/target_selection/cartons/bhm_spiders_agn.py}{bhm\_spiders\_agn.py}

\noindent\textbf{Number of targets:} 200681

\begin{center}\rule{0.5\linewidth}{0.5pt}\end{center}

\hypertarget{bhm_spiders_agn_gaiadr2_plan0.5.0}{%
\subsection{bhm\_spiders\_agn\_gaiadr2}\label{bhm_spiders_agn_gaiadr2_plan0.5.0}}

\noindent\textbf{target\_selection plan:} 0.5.0

\noindent\textbf{target\_selection tag:}
\href{https://github.com/sdss/target_selection/tree/0.3.0/}{0.3.0}

\noindent\textbf{Summary:} This is the third highest priority carton for SPIDERS
AGN wide area follow up, included to expand the survey footprint
to the full hemisphere where X-ray sources are available (beyond
Legacy Survey DR8 and PanSTARRS1). The carton provides optical
counterparts to point-like (unresolved) X-ray sources detected in early
reductions of the first 6-months of eROSITA all sky survey data
(eRASS:1). The sample is expected to contain a mixture of QSOs, AGN,
stars and compact objects. The X-ray sources have been cross-matched by
the eROSITA-DE team, first to
CatWISE2020 mid-IR sources \citep{Marocco2021}, 
and then to optical counterparts from the Gaia-DR2
catalog. All targets are located in the sky hemisphere where MPE
controls the data rights (approx. 180\textless{}l\textless{}360~deg).
The targets in this carton distributed over a wide range of Galactic
latitudes, but targets at low Galactic latitudes
($|b|<15$~deg) do not drive survey strategy.

\noindent\textbf{Simplified description of selection criteria:} Starting from a
parent catalog of eRASS:1 point source $\rightarrow$ CatWISE2020 $\rightarrow$ Gaia DR2
associations (method: NWAY assisted by IR priors computed via a
pre-trained Random Forest, building on
\citealt{Salvato2022}), select targets which meet all of the following criteria:
i) have eROSITA detection likelihood\textgreater{}6.0, ii) have an
X-ray$\rightarrow$IR cross-match probability of p\_any\textgreater{}0.1, and iii) have
$G>13.5$ and $G_{RP} > 13.5$~Vega. We deprioritize targets
if any of the following criteria are met: a) the target already has
existing good quality SDSS spectroscopy, b) the X-ray detection
likelihood is \textless{}8.0, or c) the target is a secondary X-ray$\rightarrow$IR
association. We assign cadences (exposure time requests) based on
optical brightness.

\noindent\textbf{Target priority options:}
1540-1543,1740-1743,3540-3543,3740-3742

\noindent\textbf{Cadence options:} bright\_2x1, dark\_1x2, dark\_1x4

\noindent\textbf{Implementation:}
\href{https://github.com/sdss/target_selection/blob/0.3.0/python/target_selection/cartons/bhm_spiders_agn.py}{bhm\_spiders\_agn.py}

\noindent\textbf{Number of targets:} 324576

\begin{center}\rule{0.5\linewidth}{0.5pt}\end{center}

\hypertarget{bhm_spiders_agn_skymapperdr2_plan0.5.0}{%
\subsection{bhm\_spiders\_agn\_skymapperdr2}\label{bhm_spiders_agn_skymapperdr2_plan0.5.0}}

\noindent\textbf{target\_selection plan:} 0.5.0

\noindent\textbf{target\_selection tag:}
\href{https://github.com/sdss/target_selection/tree/0.3.0/}{0.3.0}

\noindent\textbf{Summary:} This is a lower ranked carton for SPIDERS AGN wide
area follow up, which supplements the survey in areas that rely on
Gaia~DR2 (beyond Legacy Survey DR8 and PanSTARRS1) by recovering
extended targets (galaxies) that are missed by Gaia. The carton provides
optical counterparts to point-like (unresolved) X-ray sources detected
in early reductions of the first 6-months of eROSITA all sky survey data
(eRASS:1). The sample is expected to contain a mixture of QSOs, AGN,
stars and compact objects. The X-ray sources have been cross-matched by
the eROSITA-DE team, first to CatWISE2020 mid-IR sources \citep{Marocco2021},
and then to optical counterparts from the
SkyMapper-DR2 catalog \citep{Onken2019}. All targets are located in the sky hemisphere where MPE
controls the data rights (approx. 180\textless{}l\textless{}360~deg) and
at Dec\textless{}0~deg, spanning a wide range of Galactic latitudes.
Targets at low Galactic latitudes ($|b|<15$~deg)
do not drive survey strategy.

\noindent\textbf{Simplified description of selection criteria:} Starting from a
parent catalog of eRASS:1 point source $\rightarrow$ CatWISE2020 $\rightarrow$ SkyMapper-dr2
associations (method: NWAY assisted by IR priors computed via a
pre-trained Random Forest, building on
\citealt{Salvato2022}), select targets which meet all of the following criteria:
i) have eROSITA detection likelihood\textgreater{}6.0, ii) have an
X-ray$\rightarrow$IR cross-match probability of p\_any\textgreater{}0.1, iii) have
SkyMapper $g_{\mathrm{psf}},r_{\mathrm{psf}},i_{\mathrm{psf}},z_{\mathrm{psf}} > 13.5$~AB and at
least one of SkyMapper $g_{\mathrm{psf}} < 22.5$, $r_{\mathrm{psf}}< 22.0$, $i_{\mathrm{psf}} < 21.5$ or $z_{\mathrm{psf}} <20.5$~AB,
iv) are not
associated with a bad SkyMapper source detection (flags\textless{}4),
v) have non-null
measurements of $g_{\mathrm{psf}},r_{\mathrm{psf}}$ and $i_{\mathrm{psf}}$, and vi) if detected by
Gaia DR2 then have $G>13.5$ and $G_{RP} > 13.5$~Vega.
We deprioritize targets if any of the
following criteria are met: a) the target already has existing good
quality SDSS spectroscopy, b) the X-ray detection likelihood is
\textless{}8.0, or c) the target is a secondary X-ray$\rightarrow$IR association. We
assign cadences (exposure time requests) based on optical brightness.

\noindent\textbf{Target priority options:}
1550-1553,1750-1753,3550-3553,3750-3752

\noindent\textbf{Cadence options:} bright\_2x1, dark\_1x2, dark\_1x4

\noindent\textbf{Implementation:}
\href{https://github.com/sdss/target_selection/blob/0.3.0/python/target_selection/cartons/bhm_spiders_agn.py}{bhm\_spiders\_agn.py}

\noindent\textbf{Number of targets:} 82683

\begin{center}\rule{0.5\linewidth}{0.5pt}\end{center}

\hypertarget{bhm_spiders_agn_supercosmos_plan0.5.0}{%
\subsection{bhm\_spiders\_agn\_supercosmos}\label{bhm_spiders_agn_supercosmos_plan0.5.0}}

\noindent\textbf{target\_selection plan:} 0.5.0

\noindent\textbf{target\_selection tag:}
\href{https://github.com/sdss/target_selection/tree/0.3.0/}{0.3.0}

\noindent\textbf{Summary:} This is a lower ranked carton for SPIDERS AGN wide
area follow up, which supplements the survey in areas which rely on
Gaia-DR2 (beyond DESI Legacy Survey DR8 and PanSTARRS1) by recovering
extended targets (galaxies) that are missed by Gaia. The carton provides
optical counterparts to point-like (unresolved) X-ray sources detected
in early reductions of the first 6-months of eROSITA all sky survey data
(eRASS:1). The sample is expected to contain a mixture of QSOs, AGN,
stars and compact objects. The X-ray sources have been cross-matched by
the eROSITA-DE team, first to CatWISE2020 mid-IR sources \citep{Marocco2021}, 
and then to optical counterparts from the
\href{http://www-wfau.roe.ac.uk/sss/intro.html}{SuperCosmos Sky Surveys}
catalog (derived from scans of photographic plates). All targets are
located in the sky hemisphere where MPE controls the data rights
(approx. 180\textless{}l\textless{}360~deg), spanning a wide range of
Galactic latitudes. Targets at low Galactic latitudes
($|b|<15$~deg) do not drive survey strategy.

\noindent\textbf{Simplified description of selection criteria:} Starting from a
parent catalog of eRASS:1 point source $\rightarrow$ CatWISE2020 $\rightarrow$ SuperCosmos
associations (method: NWAY assisted by IR priors computed via a
pre-trained Random Forest, building on
\citealt{Salvato2022}), select targets which meet all of the following criteria:
i) have eROSITA detection likelihood\textgreater{}6.0, ii) have an
X-ray$\rightarrow$IR cross-match probability of p\_any\textgreater{}0.1, iii) have
$B_{\mathrm{J,psf}}, R_{\mathrm{psf}}, I_{\mathrm{psf}} > 13.5$~Vega and at least one of
$B_{\mathrm{J,psf}} < 22.5$, $R_{\mathrm{psf}} < 22.0$, or
$I_{\mathrm{psf}} < 21.5$~Vega (photometry from SuperCosmos),
and iv) if detected by Gaia DR2 then have $G>13.5$ and $G_{RP} > 13.5$~Vega.
We deprioritize
targets if any of the following criteria are met: a) the target already
has existing good quality SDSS spectroscopy, b) the X-ray detection
likelihood is \textless{}8.0, or c) the target is a secondary X-ray$\rightarrow$IR
association. We assign cadences (exposure time requests) based on
optical brightness.

\noindent\textbf{Target priority options:}
1560-1563,1760-1763,3560-3563,3760-3763

\noindent\textbf{Cadence options:} bright\_2x1, dark\_1x2, dark\_1x4

\noindent\textbf{Implementation:}
\href{https://github.com/sdss/target_selection/blob/0.3.0/python/target_selection/cartons/bhm_spiders_agn.py}{bhm\_spiders\_agn.py}

\noindent\textbf{Number of targets:} 430780

\begin{center}\rule{0.5\linewidth}{0.5pt}\end{center}

\hypertarget{bhm_spiders_agn_efeds_stragglers_plan0.5.0}{%
\subsection{bhm\_spiders\_agn\_efeds\_stragglers}\label{bhm_spiders_agn_efeds_stragglers_plan0.5.0}}

\noindent\textbf{target\_selection plan:} 0.5.0

\noindent\textbf{target\_selection tag:}
\href{https://github.com/sdss/target_selection/tree/0.3.0/}{0.3.0}

\noindent\textbf{Summary:} This is an opportunistic supplementary carton for
SPIDERS AGN follow up, aiming, where fiber resources allow, to gather a
small additional number of spectra for targets in the eFEDS field (which
has been repeatedly surveyed in earlier iterations of SDSS). This carton
provides optical counterparts to point-like (unresolved) X-ray sources
detected in early reductions of the eROSITA/eFEDS performance validation
field. The sample is expected to contain a mixture of QSOs, AGN, stars
and compact objects. The X-ray sources have been cross-matched by the
eROSITA-DE team to
DESI Legacy Survey DR8
optical/IR counterparts. All targets in this carton are located within
the eFEDS field (approx 126\textless{}RA\textless{}146,
-3\textless{}Dec\textless{}+6~deg). These targets do not drive survey
strategy.

\noindent\textbf{Simplified description of selection criteria:} Starting from a
parent catalog of eFEDS point source $\rightarrow$ legacysurvey.org/dr8
associations (method: NWAY assisted by optical/IR priors computed via a
pre-trained Random Forest, see
\citealt{Salvato2022}), select targets which meet all of the following criteria:
i) have eROSITA detection likelihood\textgreater{}6.0, ii) have an X-ray
$\rightarrow$ optical/IR cross-match probability (NWAY) of p\_any\textgreater{}0.1,
iii) have $13.5 < r_{\mathrm{fibertot}} < 22.5$ or
$13.5 < z_{\mathrm{fibertot}} < 21.0$~AB,
iv) are not saturated in
Legacy Survey imaging, v) have at least one observation in $r$-band and at
least one observation in $g$- or $z$-band, and vi) if detected by Gaia DR2 then
have $G>13.5$ and $G_{RP} > 13.5$~Vega. We deprioritize
targets if any of the following criteria are met: a) the target already
has existing good quality SDSS spectroscopy, b) the X-ray detection
likelihood is \textless{}8.0, or c) the target is a secondary
X-ray$\rightarrow$optical/IR association. We assign cadences (exposure time
requests) based on optical brightness.

\noindent\textbf{Target priority options:} 1510-1514, 1710-1714

\noindent\textbf{Cadence options:} bright\_2x1, dark\_1x2, dark\_1x4

\noindent\textbf{Implementation:}
\href{https://github.com/sdss/target_selection/blob/0.3.0/python/target_selection/cartons/bhm_spiders_agn.py}{bhm\_spiders\_agn.py}

\noindent\textbf{Number of targets:} 15926

\begin{center}\rule{0.5\linewidth}{0.5pt}\end{center}

\hypertarget{bhm_spiders_agn_sep_plan0.5.0}{%
\subsection{bhm\_spiders\_agn\_sep}\label{bhm_spiders_agn_sep_plan0.5.0}}

\noindent\textbf{target\_selection plan:} 0.5.0

\noindent\textbf{target\_selection tag:}
\href{https://github.com/sdss/target_selection/tree/0.3.0/}{0.3.0}

\noindent\textbf{Summary:} This special carton is dedicated to SPIDERS AGN follow
up in the CVZ-S/SEP field. The carton provides optical counterparts to
point-like (unresolved) X-ray sources detected in a dedicated analysis
of the first 6-months of eROSITA scanning data near the SEP. The X-ray
sources have been cross-matched by the eROSITA-DE team, first to
CatWISE2020 mid-IR sources \citep{Marocco2021}, 
and then to optical counterparts from the Gaia~DR2
catalog, using additional filtering (including Gaia EDR3 astrometric
information) to reduce the contamination from foreground stars located in the
LMC. All targets are located within 1.5~deg of the South Ecliptic Pole (RA,Dec =
90.0,-66.56~deg).

\noindent\textbf{Simplified description of selection criteria:} Starting from a
parent catalog of eRASS:1/SEP point source $\rightarrow$ CatWISE2020 $\rightarrow$ Gaia DR2
associations (method: NWAY assisted by IR priors computed via a
pre-trained Random Forest, building on
\citealt{Salvato2022}), select targets which meet all of the following criteria:
i) have eROSITA detection likelihood\textgreater{}6.0, ii) have an
X-ray$\rightarrow$IR cross-match probability of p\_any\textgreater{}0.1, and iii) have
Gaia $G > 13.5$ and $G_{RP} > 13.5$~Vega. We deprioritize targets
if any of the following criteria are met: a) the X-ray detection
likelihood is \textless{}8.0, or b) the target is a secondary X-ray$\rightarrow$IR
association. We assign cadences (exposure time requests) based on
optical brightness.

\noindent\textbf{Target priority options:} 1510,1512

\noindent\textbf{Cadence options:} bright\_2x1, dark\_1x2, dark\_1x4

\noindent\textbf{Implementation:}
\href{https://github.com/sdss/target_selection/blob/0.3.0/python/target_selection/cartons/bhm_spiders_agn.py}{bhm\_spiders\_agn.py}

\noindent\textbf{Number of targets:} 697

\begin{center}\rule{0.5\linewidth}{0.5pt}\end{center}

\hypertarget{bhm_spiders_clusters_lsdr8_plan0.5.0}{%
\subsection{bhm\_spiders\_clusters\_lsdr8}\label{bhm_spiders_clusters_lsdr8_plan0.5.0}}

\noindent\textbf{target\_selection plan:} 0.5.0

\noindent\textbf{target\_selection tag:}
\href{https://github.com/sdss/target_selection/tree/0.3.0/}{0.3.0}

\noindent\textbf{Summary:} This is the highest priority carton for SPIDERS
Clusters wide area follow up. The carton provides a list of galaxies
which are candidate members of clusters selected from early reductions
of the first 6-months of eROSITA all sky survey data (eRASS:1). The
X-ray clusters have been associated by the eROSITA-DE team to
DESI Legacy Survey DR8 (lsdr8)
optical/IR counterparts using the eROMaPPeR red-sequence finder
algorithm
(\citealt{Rykoff2014};
\citealt{IderChitham2020}). All targets are located in the sky hemisphere
where MPE controls the data rights (approx.
180\textless{}l\textless{}360~deg). Due to the footprint of lsdr8,
nearly all targets in this carton are located at high Galactic latitudes
(\textbar{}b\textbar{}\textgreater{}15~deg).

\noindent\textbf{Simplified description of selection criteria:} Starting from a
parent catalog of eRASS:1 $\rightarrow$ lsdr8 eROMaPPeR cluster
associations, select targets which meet all of the following criteria:
i) have $13.5 < r_{\mathrm{fibertot}} < 21.0$ or
$13.5 < z_{\mathrm{fibertot}} < 20.0$~AB, ii) if detected by Gaia
DR2 then have $G > 13.5$ and $G_{RP} > 13.5$~Vega (photometry from lsdr8), and iii) do
not have existing good quality SDSS spectroscopy. We assign a range of
priorities to targets in this carton, with BCGs top ranked, followed by
candidate member galaxies according their probability of membership. We
assign cadences (exposure time requests) based on optical brightness.

\noindent\textbf{Target priority options:} 1501,1630-1659

\noindent\textbf{Cadence options:} bright\_2x1, dark\_1x2, dark\_1x4

\noindent\textbf{Implementation:}
\href{https://github.com/sdss/target_selection/blob/0.3.0/python/target_selection/cartons/bhm_spiders_clusters.py}{bhm\_spiders\_clusters.py}

\noindent\textbf{Number of targets:} 87490

\begin{center}\rule{0.5\linewidth}{0.5pt}\end{center}

\hypertarget{bhm_spiders_clusters_ps1dr2_plan0.5.0}{%
\subsection{bhm\_spiders\_clusters\_ps1dr2}\label{bhm_spiders_clusters_ps1dr2_plan0.5.0}}

\noindent\textbf{target\_selection plan:} 0.5.0

\noindent\textbf{target\_selection tag:}
\href{https://github.com/sdss/target_selection/tree/0.3.0/}{0.3.0}

\noindent\textbf{Summary:} This is the second highest priority carton for SPIDERS
Clusters wide area follow up, designed to expand the survey area
beyond the Legacy Survey DR8 footprint. The carton provides a list of
galaxies that are candidate members of clusters selected from early
reductions of the first 6-months of eROSITA all sky survey data
(eRASS:1). The X-ray clusters have been associated by the eROSITA-DE
team to the
\href{https://outerspace.stsci.edu/display/PANSTARRS/}{PanSTARRS1-DR2}
catalog using the eROMaPPeR red-sequence finder algorithm
(\citealt{Rykoff2014};
\citealt{IderChitham2020}). All targets are located in the sky hemisphere
where MPE controls the data rights (approx.
180\textless{}l\textless{}360~deg). Nearly all targets in this carton
are located at high Galactic latitudes
(\textbar{}b\textbar{}\textgreater{}15~deg).

\noindent\textbf{Simplified description of selection criteria:} Starting from a
parent catalog of eRASS:1 $\rightarrow$ PanSTARRS1-DR2 eROMaPPeR cluster
associations, select targets which meet all of the following criteria:
i) have PanSTARRS1 $r_{\mathrm{psf}},i_{\mathrm{psf}},z_{\mathrm{psf}} > 13.5$~AB and at
least one of $r_{\mathrm{psf}}< 21.5$, $i_{\mathrm{psf}} < 21.0$ or $z_{\mathrm{psf}} <20.5$~AB,
and ii) do not have existing good quality SDSS
spectroscopy. We assign a range of priorities to targets in this carton,
with BCGs top ranked, followed by candidate member galaxies according
their probability of membership. We assign cadences (exposure time
requests) based on optical brightness.

\noindent\textbf{Target priority options:} 1502,1660-1689

\noindent\textbf{Cadence options:} bright\_2x1,dark\_1x2,dark\_1x4

\noindent\textbf{Implementation:}
\href{https://github.com/sdss/target_selection/blob/0.3.0/python/target_selection/cartons/bhm_spiders_clusters.py}{bhm\_spiders\_clusters.py}

\noindent\textbf{Number of targets:} 86179

\begin{center}\rule{0.5\linewidth}{0.5pt}\end{center}

\hypertarget{bhm_spiders_clusters_efeds_stragglers_plan0.5.0}{%
\subsection{bhm\_spiders\_clusters\_efeds\_stragglers}\label{bhm_spiders_clusters_efeds_stragglers_plan0.5.0}}

\noindent\textbf{target\_selection plan:} 0.5.0

\noindent\textbf{target\_selection tag:}
\href{https://github.com/sdss/target_selection/tree/0.3.0/}{0.3.0}

\noindent\textbf{Summary:} This is an opportunistic supplementary carton for
SPIDERS Clusters follow up, aiming, where fiber resources allow, to
gather a small additional number of spectra for targets in the eFEDS
field (which has been repeatedly surveyed in earlier iterations of
SDSS). The carton provides a list of galaxies which are candidate
members of clusters selected from early reductions of the eROSITA
performance verification survey in the eFEDS field. The X-ray clusters
have been associated by the eROSITA-DE team to
DESI Legacy Survey DR8 (lsdr8)
optical/IR counterparts. All targets in this carton are located within
the eFEDS field (approx 126\textless{}RA\textless{}146,
-3\textless{}Dec\textless{}+6~deg).

\noindent\textbf{Simplified description of selection criteria:} Starting from a
parent catalog of eFEDS $\rightarrow$
\href{https://www.legacysurvey.org/dr8}{legacysurvey.org/dr8} cluster
associations (eROMaPPeR,
\citealt{Rykoff2014};
\citealt{IderChitham2020};
\citealt{Liu2022}), select targets which meet all of the following criteria:
 have $13.5 < r_{\mathrm{fibertot}} < 21.0$ or
$13.5 < z_{\mathrm{fibertot}} < 20.0$~AB, ii) if detected by Gaia
DR2 then have $G > 13.5$ and $G_{RP} > 13.5$~Vega (photometry from lsdr8),
and iii) do not have existing good quality SDSS spectroscopy. We assign a range of
priorities to targets in this carton, with BCGs top ranked, followed by
candidate member galaxies according their probability of membership. We
assign cadences (exposure time requests) based on optical brightness.

\noindent\textbf{Target priority options:} 1500,1600-1629

\noindent\textbf{Cadence options:} dark\_1x2,dark\_1x4

\noindent\textbf{Implementation:}
\href{https://github.com/sdss/target_selection/blob/0.3.0/python/target_selection/cartons/bhm_spiders_clusters.py}{bhm\_spiders\_clusters.py}

\noindent\textbf{Number of targets:} 3060

\begin{center}\rule{0.5\linewidth}{0.5pt}\end{center}

\hypertarget{bhm_spiders_agn-efeds_plan0.1.0}{%
\subsection{bhm\_spiders\_agn-efeds}\label{bhm_spiders_agn-efeds_plan0.1.0}}

\noindent\textbf{target\_selection plan:} 0.1.0

\noindent\textbf{target\_selection tag:}
\href{https://github.com/sdss/target_selection/tree/0.1.0/}{0.1.0}

\noindent\textbf{Summary:} A carton used during SDSS-V plate-mode observations
that contains candidate AGN targets found in the eROSITA/eFEDS X-ray
survey field. This carton provides optical counterparts to point-like
(unresolved) X-ray sources detected in early reductions (`c940/V2T') of
the eROSITA/eFEDS performance validation field. The sample is expected
to contain a mixture of QSOs, AGN, stars and compact objects. The X-ray
sources have been cross-matched by the eROSITA-DE team to
DESI Legacy Survey DR8 (lsdr8)
optical/IR counterparts. All targets in this carton are located within
the eFEDS field (approx 126\textless{}RA\textless{}146,
-3\textless{}Dec\textless{}+6~deg).

\noindent\textbf{Simplified description of selection criteria:} Starting from a
parent catalog of eFEDS point source $\rightarrow$
\href{https://www.legacysurvey.org/dr8}{lsdr8}
associations (primarily via NWAY assisted by optical/IR priors computed
via a pre-trained Random Forest, see
\citealt{Salvato2022}, supplemented by counterparts selected via a Likelihood
Ratio using $r$-band magnitudes), select targets which meet all of the
following criteria: i) have eROSITA detection
likelihood\textgreater{}6.0, ii) have an X-ray $\rightarrow$ optical/IR cross-match
probability of either p\_any\textgreater{}0.1 (NWAY associations) or
LR\textgreater{}0.2 (Likelihood Ratio associations), 
iii) have $r_{\mathrm{fiber}} > 16.5$ and either $r_{\mathrm{fiber}} < 22.0$
or $z_{\mathrm{fiber}} < 21.0$~AB (photometry from lsdr8),
and iv) did not receive high quality
spectroscopy during the SDSS-IV observations of the eFEDS field
(\citealt{Abdurrouf_2021_sdssDR17}). We deprioritize targets if any of the following criteria
are met: a) the target already has existing good quality SDSS
spectroscopy in SDSS DR16, b) the X-ray detection likelihood is
\textless{}8.0, c) the target is a secondary X-ray$\rightarrow$optical/IR
association, or d) the optical/IR counterpart was only chosen by the
LR method. All targets were assigned a nominal cadence of:
bhm\_spiders\_1x8 (8x15mins dark time).

\noindent\textbf{Target priority options:} 1510-1519

\noindent\textbf{Cadence options:} bhm\_spiders\_1x8

\noindent\textbf{Implementation:}
\href{https://github.com/sdss/target_selection/blob/0.1.0/python/target_selection/cartons/bhm_spiders_agn.py}{bhm\_spiders\_agn.py}

\noindent\textbf{Number of targets:} 12459

\begin{center}\rule{0.5\linewidth}{0.5pt}\end{center}

\hypertarget{bhm_spiders_clusters-efeds-ls-redmapper_plan0.1.0}{%
\subsection{bhm\_spiders\_clusters-efeds-ls-redmapper}\label{bhm_spiders_clusters-efeds-ls-redmapper_plan0.1.0}}

\noindent\textbf{target\_selection plan:} 0.1.0

\noindent\textbf{target\_selection tag:}
\href{https://github.com/sdss/target_selection/tree/0.1.0/}{0.1.0}

\noindent\textbf{Summary:} A carton used during SDSS-V plate-mode observations
that contains galaxy cluster targets found in the eROSITA/eFEDS X-ray
survey field. The carton provides a list of galaxies which are candidate
members of clusters selected from early reductions (`c940') of the
eROSITA performance verification survey in the eFEDS field. The parent
sample of galaxy clusters and their member galaxies have been selected
via a joint analysis of X-ray and (several) optical/IR datasets using
the eROMaPPeR red-sequence finder algorithm
(\citealt{Rykoff2014};
\citealt{IderChitham2020}). This particular carton relies on optical/IR data
from DESI Legacy Survey DR8 (lsdr8). All
targets in this carton are located within the eFEDS field (approx
126\textless{}RA\textless{}146, -3\textless{}Dec\textless{}+6~deg).

\noindent\textbf{Simplified description of selection criteria:} Starting from a
parent catalog of eFEDS $\rightarrow$ optical/IR cluster associations, select
targets which meet all of the following criteria: i) are selected by
eROMaPPeR applied to
\href{https://www.legacysurvey.org/dr8}{lsdr8}
photometric data, ii) have eROSITA X-ray detection likelihood
\textgreater{} 8.0,
iii)  have $r_{\mathrm{fiber}} > 16.5$ and either $r_{\mathrm{fiber}} < 21.0$
or $z_{\mathrm{fiber}} < 20.0$~AB (photometry from lsdr8), and iv) do not
have existing good quality (SDSS or external) spectroscopy. We assign a
range of priorities to targets in this carton, with BCGs top ranked,
followed by candidate member galaxies according their probability of
membership. All targets were assigned a nominal cadence of:
bhm\_spiders\_1x8 (8x15mins dark time).

\noindent\textbf{Target priority options:} 1500, 1511-\/-1610

\noindent\textbf{Cadence options:} dark\_1x8

\noindent\textbf{Implementation:}
\href{https://github.com/sdss/target_selection/blob/0.1.0/python/target_selection/cartons/bhm_spiders_clusters.py}{bhm\_spiders\_clusters.py}

\noindent\textbf{Number of targets:} 4432

\begin{center}\rule{0.5\linewidth}{0.5pt}\end{center}

\hypertarget{bhm_spiders_clusters-efeds-sdss-redmapper_plan0.1.0}{%
\subsection{bhm\_spiders\_clusters-efeds-sdss-redmapper}\label{bhm_spiders_clusters-efeds-sdss-redmapper_plan0.1.0}}

\noindent\textbf{target\_selection plan:} 0.1.0

\noindent\textbf{target\_selection tag:}
\href{https://github.com/sdss/target_selection/tree/0.1.0/}{0.1.0}

\noindent\textbf{Summary:} A carton used during SDSS-V plate-mode observations
that contains galaxy cluster targets found in the eROSITA/eFEDS X-ray
survey field. The carton provides a list of galaxies which are candidate
members of clusters selected from early reductions (`c940') of the
eROSITA performance verification survey in the eFEDS field. The parent
sample of galaxy clusters and their member galaxies have been selected
via a joint analysis of X-ray and (several) optical/IR datasets using
the eROMaPPeR red-sequence finder algorithm
(\citealt{Rykoff2014};
\citealt{IderChitham2020}). This particular carton relies on optical photometric data
from SDSS~DR13. All targets in this
carton are located within the eFEDS field (approx
126\textless{}RA\textless{}146, -3\textless{}Dec\textless{}+6~deg).

\noindent\textbf{Simplified description of selection criteria:} Starting from a
parent catalog of eFEDS $\rightarrow$ optical/IR cluster associations, select
targets which meet all of the following criteria: i) are selected by
eROMaPPeR applied to \href{https://www.sdss.org/dr13/}{SDSS dr13}
photometric data, ii) have eROSITA X-ray detection likelihood
\textgreater{} 8.0,
iii) have $r_{\mathrm{fiber}} > 16.5$ and either $r_{\mathrm{fiber}} < 21.0$
or $z_{\mathrm{fiber}} < 20.0$~AB (photometry from DESI Legacy Survey DR8), and iv) do not
have existing good quality (SDSS or external) spectroscopy. We assign a
range of priorities to targets in this carton, with BCGs top ranked,
followed by candidate member galaxies according their probability of
membership. All targets were assigned a nominal cadence of:
bhm\_spiders\_1x8 (8x15mins dark time).

\noindent\textbf{Target priority options:} 1500, 1511-1610

\noindent\textbf{Cadence options:} dark\_1x8

\noindent\textbf{Implementation:}
\href{https://github.com/sdss/target_selection/blob/0.1.0/python/target_selection/cartons/bhm_spiders_clusters.py}{bhm\_spiders\_clusters.py}

\noindent\textbf{Number of targets:} 4304

\begin{center}\rule{0.5\linewidth}{0.5pt}\end{center}

\hypertarget{bhm_spiders_clusters-efeds-hsc-redmapper_plan0.1.0}{%
\subsection{bhm\_spiders\_clusters-efeds-hsc-redmapper}\label{bhm_spiders_clusters-efeds-hsc-redmapper_plan0.1.0}}

\noindent\textbf{target\_selection plan:} 0.1.0

\noindent\textbf{target\_selection tag:}
\href{https://github.com/sdss/target_selection/tree/0.1.0/}{0.1.0}

\noindent\textbf{Summary:} A carton used during SDSS-V plate-mode observations
that contains galaxy cluster targets found in the eROSITA/eFEDS X-ray
survey field. The carton provides a list of galaxies which are candidate
members of clusters selected from early reductions (`c940') of the
eROSITA performance verification survey in the eFEDS field. The parent
sample of galaxy clusters and their member galaxies have been selected
via a joint analysis of X-ray and (several) optical/IR datasets using
the eROMaPPeR red-sequence finder algorithm
(\citealt{Rykoff2014};
\citealt{IderChitham2020}). This particular carton relies on optical/IR data
from the \href{https://hsc.mtk.nao.ac.jp/ssp/}{Hyper Suprime-Cam Subaru
Strategic Program (HSC-SSP)}. All targets in this carton are located
within the eFEDS field (approx 126\textless{}RA\textless{}146,
-3\textless{}Dec\textless{}+6~deg).

\noindent\textbf{Simplified description of selection criteria:} Starting from a
parent catalog of eFEDS $\rightarrow$ optical/IR cluster associations, select
targets which meet all of the following criteria: i) are selected by
eROMaPPeR applied to \href{https://hsc.mtk.nao.ac.jp/ssp/}{HSC-SSP}
photometric data, ii) have eROSITA X-ray detection likelihood
\textgreater{} 8.0, iii) have $r_{\mathrm{fiber}} > 16.5$ and either $r_{\mathrm{fiber}} < 21.0$
or $z_{\mathrm{fiber}} < 20.0$~AB (photometry from DESI Legacy Survey DR8), and iv) do not
have existing good quality (SDSS or external) spectroscopy. We assign a
range of priorities to targets in this carton, with BCGs top ranked,
followed by candidate member galaxies according their probability of
membership. All targets were assigned a nominal cadence of:
bhm\_spiders\_1x8 (8x15mins dark time).

\noindent\textbf{Target priority options:} 1500, 1511-1610

\noindent\textbf{Cadence options:} dark\_1x8

\noindent\textbf{Implementation:}
\href{https://github.com/sdss/target_selection/blob/0.1.0/python/target_selection/cartons/bhm_spiders_clusters.py}{bhm\_spiders\_clusters.py}

\noindent\textbf{Number of targets:} 924

\begin{center}\rule{0.5\linewidth}{0.5pt}\end{center}

\hypertarget{bhm_spiders_clusters-efeds-erosita_plan0.1.0}{%
\subsection{bhm\_spiders\_clusters-efeds-erosita}\label{bhm_spiders_clusters-efeds-erosita_plan0.1.0}}

\noindent\textbf{target\_selection plan:} 0.1.0

\noindent\textbf{target\_selection tag:}
\href{https://github.com/sdss/target_selection/tree/0.1.0/}{0.1.0}

\noindent\textbf{Summary:} A carton used during SDSS-V plate-mode observations
that contains galaxy cluster targets found in the eROSITA/eFEDS X-ray
survey field. The carton provides a list of galaxies which are candidate
members of clusters selected from early reductions (`c940') of the
eROSITA performance verification survey in the eFEDS field. The parent
sample of galaxy clusters and their member galaxies have been selected
via a joint analysis of X-ray and (several) optical/IR datasets. This
particular carton includes counterparts to X-ray extended sources that
were not selected by the eROMaPPeR red sequence finder algorithm when
applied to any of the DESI Legacy Survey DR8, SDSS DR13, or HSC-SSP datasets
(i.e., complementary to the cartons:
bhm\_spiders\_clusters-efeds-ls-redmapper,
bhm\_spiders\_clusters-efeds-sdss-redmapper and
bhm\_spiders\_clusters-efeds-hsc-redmapper). All targets in this carton
are located within the eFEDS field (approx
126\textless{}RA\textless{}146, -3\textless{}Dec\textless{}+6~deg).

\noindent\textbf{Simplified description of selection criteria:} Starting from a
parent catalog of eFEDS $\rightarrow$ optical/IR cluster associations, select
targets which meet all of the following criteria: i) are identified as
being X-ray extended but not selected via the eROMaPPeR algorithm, ii)
have eROSITA X-ray detection likelihood \textgreater{} 8.0, iii) have $r_{\mathrm{fiber}} > 16.5$ and either $r_{\mathrm{fiber}} < 21.0$
or $z_{\mathrm{fiber}} < 20.0$~AB (photometry from DESI Legacy Survey DR8),
and iv) do not have existing good quality (SDSS
or external) spectroscopy. We assign a range of priorities to targets in
this carton, with BCGs top ranked, followed by candidate member galaxies
according their probability of membership. All targets were assigned a
nominal cadence of: bhm\_spiders\_1x8 (8x15mins dark time).

\noindent\textbf{Target priority options:} 1500, 1511-1535

\noindent\textbf{Cadence options:} dark\_1x8

\noindent\textbf{Implementation:}
\href{https://github.com/sdss/target_selection/blob/0.1.0/python/target_selection/cartons/bhm_spiders_clusters.py}{bhm\_spiders\_clusters.py}

\noindent\textbf{Number of targets:} 15

\end{document}